\documentclass[twocolumn,times,tighten]{aastex631}

\usepackage{graphicx} 
\usepackage{latexsym} 
\usepackage{bm}       
\usepackage[english]{babel}
\usepackage{booktabs} 
\usepackage{multirow} 
\usepackage{epstopdf}
\usepackage{hyperref}
\usepackage{cleveref}
\usepackage{float}

\newcommand{\angstrom}{\textup{\AA}}

\usepackage{color} 

\newcommand{\etal}{et~al.\thinspace}

\newcommand{\U}{\emph{U}}
\newcommand{\Ub}{\emph{U}-band }%
\newcommand{\Uab}{$U_{AB}$ }

\newcommand{\B}{\emph{B}}

\newcommand{\mab}{$m_{\rm AB}$ }

\newcommand{\sbabr}{$\mu^{AB}_{r}$ }

\newcommand{\rab}{$r_{AB}$}

\newcommand{\V}{\emph{V}}

\newcommand{\hst}{\emph{HST}}
\newcommand{\Uz}{\emph{U}}
\newcommand{\R}{\emph{R}}
\newcommand{\rb}{$r$-band }%
\newcommand{\rs}{Sloan $r$-band}

\newlength{\txw}
\newlength{\txh}

\newcommand{\sextractor}{\textsc{SExtractor}}
\newcommand{\idl}{\textsc{IDL} }

\newcommand{\swarp}{\textsc{swarp} }

\newcommand{\del}[1]{\relax}%
\newcommand{\DELETED}[1]{\relax}%
{\relax}%

\shorttitle{LBT \rb{} GOODS-N}
\shortauthors{Ashcraft \etal}

\begin{document}

\title{Deep Large Binocular Camera \rb Observations of the GOODS-N Field}

\correspondingauthor{Tyler McCabe}
\email{tyler.mccabe@asu.edu}

\author[0000-0003-4439-6003]{Teresa A. Ashcraft}
\affiliation{Department of Physics, Eckerd College, St. Petersburg, FL USA}
\author[0000-0002-5506-3880]{Tyler McCabe}
\affiliation{School of Earth and Space Exploration, Arizona State University, Tempe, AZ 85287-1404, USA}
\author[0000-0002-9961-2984]{Caleb Redshaw}
\affiliation{School of Earth and Space Exploration, Arizona State University, Tempe, AZ 85287-1404, USA}
\author[0000-0001-8156-6281]{Rogier A. Windhorst}
\affiliation{School of Earth and Space Exploration, Arizona State University, Tempe, AZ 85287-1404, USA}
\author[0000-0003-1268-5230]{Rolf A. Jansen}
\affiliation{School of Earth and Space Exploration, Arizona State University, Tempe, AZ 85287-1404, USA}
\author[0000-0003-3329-1337]{Seth H. Cohen}
\affiliation{School of Earth and Space Exploration, Arizona State University, Tempe, AZ 85287-1404, USA}
\author[0000-0001-6650-2853]{Timothy Carleton}
\affiliation{School of Earth and Space Exploration, Arizona State University, Tempe, AZ 85287-1404, USA}
\author[0000-0003-3030-1763]{Kris Ganzel}
\affiliation{School of Earth and Space Exploration, Arizona State University, Tempe, AZ 85287-1404, USA}

\author[0000-0002-6610-2048]{Anton M. Koekemoer}
\affiliation{Space Telescope Science Institute, 3700 San Martin Dr., Baltimore, MD 21218, USA}

\author[0000-0003-0894-1588]{Russell E. Ryan}
\affiliation{Space Telescope Science Institute, Baltimore, MD 21218, USA}

\author[0000-0001-6342-9662]{Mario Nonino}
\affiliation{INAF - Osservatorio Astronomico di Trieste, Via Bazzoni 2, 34124 Trieste, Italy}
\author[0000-0002-7409-8114]{Diego Paris}
\affiliation{INAF - Osservatorio Astronomico di Roma, Via Frascati 33, I-00078 Monte Porzio Catone, Italy}
\author[0000-0002-5688-0663]{Andrea Grazian}
\affiliation{INAF - Osservatorio Astronomico di Padova Vicolo dell'Osservatorio, 5 Padova (PD) I-35122, Italy}
\author[0000-0003-3820-2823]{Adriano Fontana} 
\affiliation{INAF - Osservatorio Astronomico di Roma, Via Frascati 33, I-00078 Monte Porzio Catone, Italy}
\author[0000-0003-0734-1273]{Emanuele Giallongo}
\affiliation{INAF - Osservatorio Astronomico di Roma, Via Frascati 33, I-00078 Monte Porzio Catone, Italy}
\author[0000-0003-3754-387X]{Roberto Speziali} 
\affiliation{INAF - Osservatorio Astronomico di Roma, Via Frascati 33, I-00078 Monte Porzio Catone, Italy}
\author[0000-0003-1033-1340]{Vincenzo Testa}
\affiliation{INAF - Osservatorio Astronomico di Roma, Via Frascati 33, I-00078 Monte Porzio Catone, Italy}
 
\author[0000-0003-4432-5037]{Konstantina Boutsia}
\affiliation{Carnegie Observatories, Las Campanas Observatory, Colina El Pino, Casilla 601, La Serena, Chile}

\author[0000-0002-8190-7573]{Robert W. O'Connell} 
\affiliation{Department of Astronomy, University of Virginia, Charlottesville, VA 22904-4325, USA}

\author[0000-0001-7016-5220]{Michael J. Rutkowski} 
\affiliation{Department of Physics \& Astronomy, Minnesota State University, Mankato, Mankato, MN 56001}

\author[0000-0002-9136-8876]{Claudia Scarlata}
\affiliation{Minnesota Institute for Astrophysics, University of Minnesota, 116 Church Street SE, Minneapolis, MN 55455, USA}

\author[0000-0002-7064-5424]{Harry I. Teplitz}
\affiliation{Infrared Processing and Analysis Center, MS 100-22, Caltech, Pasadena, CA 91125, USA}

\author[0000-0002-9373-3865]{Xin Wang}
\affiliation{Infrared Processing and Analysis Center, MS 100-22, Caltech, Pasadena, CA 91125, USA}

\author[0000-0002-9946-4731]{Marc Rafelski}
\affiliation{Space Telescope Science Institute, Baltimore, MD 21218, USA}
\affiliation{Department of Physics \& Astronomy, Johns Hopkins University, Baltimore, MD 21218, USA}
 
\author[0000-0001-9440-8872]{Norman A. Grogin}
\affiliation{Space Telescope Science Institute, Baltimore, MD 21218, USA}

\begin{abstract}
We obtained 838 Sloan $r$-band images ($\sim$28 hrs) of the GOODS-North field with the Large Binocular Camera (LBC) on the Large Binocular Telescope in order to study the presence of extended, low surface brightness features in galaxies and investigate the trade-off between image depth and resolution. The individual images were sorted by effective seeing, which allowed for optimal resolution and optimal depth mosaics to be created with all images with seeing FWHM$<$ $0\farcs 9$ and FWHM$<$ $2\farcs 0$, respectively. Examining bright galaxies and their substructure as well as accurately deblending overlapping objects requires the optimal resolution mosaic, while detecting the faintest objects possible (to a limiting magnitude of \mab$\sim$\,29.2\,mag) requires the optimal depth mosaic. The better surface brightness sensitivity resulting from the larger LBC pixels, compared to those of extant WFC3/UVIS and ACS/WFC cameras aboard the Hubble Space Telescope (\hst) allows for unambiguous detection of {\it both} diffuse flux {\it and} very faint tidal tails. We created azimuthally-averaged radial surface brightness profiles were created for the 360 brightest galaxies in the mosaics. We find little difference in the majority of the light profiles from the optimal resolution and optimal depth mosaics. However, $\lesssim$15\% of the profiles show excess flux in the galaxy outskirts down to surface brightness levels of \sbabr$\!\simeq\,$31\,mag arcsec$^{-2}$. This is relevant to Extragalactic Background Light (EBL) studies as diffuse light in the outer regions of galaxies are thought to be a major contribution to the EBL. While some additional diffuse light exists in the optimal depth profiles compared to the shallower, optimal resolution profiles, we find that diffuse light in galaxy outskirts is a minor contribution to the EBL overall in the $r$-band. 
\end{abstract}


\section{INTRODUCTION}

Galaxy mergers and interactions play a critical role in galaxy evolution and are
observed across cosmic time \citep{Barnes1992,Bundy2005,Lotz2008,Lotz2011}. In the nearby Universe, mergers and interactions are able to be observed with high resolution using the \textit{Hubble Space Telescope} (\hst{}). These observations have shown that as redshift increases, galaxies appear more irregular, have closer neighbors, exhibit features of recent interactions, or appear to be in the process of merging \citep[and references therein]{Burkey1994,Duncan2019}. Various studies have visually identified and classified these merging systems based upon appearance \citep{Darg2010} and features such as tidal tails, streams, and other diffuse/extended flux regions \citep{Elmegreen2007, Mohamed2011, Elmegreen2021}. However, these features can be missed by high-resolution \hst{} imaging due to their intrinsically low surface brightness. 

Tidal tails and bridges of matter between galaxies are clear signatures of past or on-going
interactions \citep{Toomre1972}. These interactions are known to trigger star formation and play a critical role in galaxy evolution throughout the Universe \citep[and references therein]{Hernquist1989, Conselice2014}. A few studies have looked for interacting
systems within various extragalactic deep fields. For example \citet{Elmegreen2007}
examined the Galaxy Evolution from Morphologies and SEDs (GEMS) survey \citep{Rix2004} and the Great Observatories Origins Deep Survey (GOODS) South field \citep{Giavalisco2004} for mergers and galaxy interactions to
z$\simeq$1.4. They defined a sample of 100 objects, and measured properties of
the galaxies and star-forming clumps within the interacting galaxies. Similarly, \citet{Wen2016}
 identified a sample of 461 merging galaxies with long tidal tails
in the COSMOS field \citep{Scoville2007} using \hst/ACS F814W (I-band). They only included galaxies
in 0.2$\lesssim$ z$\lesssim$ 1 which corresponds to rest frame optical light sampled by the F814W filter with a
surface brightness limit of $\sim$25.1 mag arcsec$^{-2}$. However, most of
their sample have intrinsic surface brightness $\gtrsim$ 23.1\,mag
arcsec$^{-2}$.

\citet{Straughn2006} and \citet{Straughn2015} identified ''tadpole" galaxies, based on their
asymmetric knot-plus-tail morphologies visible in \hst/ACS F775W at intermediate
redshifts (0.3$\lesssim$ z$\lesssim$ 3.2) in the Hubble Ultra Deep Field \citep[HUDF;][]{Beckwith2006}. Using
multi-wavelength data, they studied rest frame UV/optical properties of these
galaxies in comparison with other field galaxies. They measured the star formation histories and ages of these galaxies and concluded that ''tadpole" galaxies are still actively assembling either through late-stage merging or cold gas accretion.

The Large Binocular Telescope (LBT) is able to obtain imaging for 4 of the 5 CANDELS \citep{Grogin2011,Koekemoer2011} fields that are in the northern hemisphere or around the celestial equator \citep[McCabe et al. 2022 in prep]{Ashcraft2018, Otteson2021, Redshaw2022}. 
\citet{Ashcraft2018} presented ultra deep \Ub imaging of GOODS-N \citep{Giavalisco2004} and created optimal resolution and optimal depth mosaics, which represent the best \Ub imaging that can be achieved from the ground. Each mirror of the LBT is equipped with a Large Binocular Camera (LBC), which allowed for parallel \Ub and Sloan $r$-band imaging. With the large field of view (FOV) of the LBC, our GOODS-N observations encompass the \hst{} footprint, which makes the complementary, very deep \rb{} data especially useful for larger survey volumes. 

Deep imaging of the CANDELS fields also allows for investigations into the amount to which galaxies contribute to the Extragalactic Background Light (EBL) \citep[and references therein]{McVittie1959, Driver2016, Windhorst2022, Carleton2022}. Currently, a discrepancy exists between EBL predictions from integrated galaxy counts \citep{Driver2011,Driver2016, Andrews2018} and from direct measurements \citep{Puget1996, Hauser1998, Matsumoto2005, Matsumoto2011, Matsumoto2018, Lauer2021, Lauer2022, Korngut2022}. Ultra-Diffuse Galaxies, diffuse intragroup or intracluster light, as well as faint light in the outskirts of galaxies have all been proposed as sources that would be capable of closing the discrepancy between galaxy counts and direct EBL measurements. 

Using the capabilities of the LBT/LBC for deep \rb{} imaging allows the detection of faint flux in galaxy outskirts in the form of star forming clumps, tidal tails/mergers, and diffuse light. This paper builds upon the previous \Ub work of \citet{Ashcraft2018, Otteson2021, Redshaw2022} and McCabe et al. 2022 (in prep) by utilizing the seeing sorted stacking procedure to create optimal depth and optimal resolution mosaics of GOODS-N for the $r$-band obtained simultaneously. Using these mosaics, we attempt to address the level to which faint, extended light in the outer regions of galaxies can contribute to the total observed EBL. 
 
This paper is organized as follows. In \S$\,2$, we describe the acquired data and the creation of optimal depth/resolution mosaics and corresponding object catalogs. \S$\,3$ describes the surface brightness profiles for the 360 brightest galaxies in the mosaic FOV and the implications for any excess diffuse flux. \S$\,4$ describes a collection of galaxies with signatures of interactions out to redshifts of z$\lesssim$ 0.9. Lastly, \S$\,$5 summarizes and discusses these results. Unless stated otherwise, all magnitudes presented in this paper are in the AB system \citep{Oke1983}.

\section{OBSERVATIONS}
\label{sec:obsr}
\deleted{\subsection{Large Binocular Camera Sloan $r$ Filter}\label{sec:lbcr}}

The LBCs are two wide field, prime focus cameras, one for each of the 8.4\,m primary mirrors of the LBT. The LBCs are unique as one camera is red optimized (LBC-Red; LBCR), while the other is blue optimized (LBC-Blue; LBCB). We utilized the LBCR camera along with the \rs{} filter, which has a central wavelength of $\lambda_c$\,=\,6195\,$\angstrom$, a bandwidth of 1300\,$\angstrom$ (full width at half maximum; FWHM), and a peak CCD quantum efficiency of $\sim\!96\%$ within the \rb{} filter. The LBC focal planes are composed of 4 EEV42-90 CCD detectors each, which have an average plate scale of $\sim0\farcs225$ \citep{Giallongo2008}.

\subsection{\rs{} Observations of the GOODS-N Field}

Using binocular imaging mode, LBC observations of the GOODS-N field were carried out in dark time from January 2013 through March 2014. This mode allowed for \Ub observations to be collected with the LBCB camera simultaneous with \rb{} observations with the LBCR camera. The \Ub data were presented in \citet{Ashcraft2018}, where the seeing sorted stacking technique and the associated trade off between optimal resolution and optimal depth mosaics were discussed. Utilizing a combination of US and Italian partner institutions, 838 \rb{} total exposures were collected using the LBCR with a total integration time of $\sim28$\,hr (100,727\,s). The individual exposures in this data set each had an exposure time of 120.2\,s. 

We utilized a dither pattern around a common center position for all images taken over many nights, which included a minimal shift to fill in the gaps between detectors and for removal of detector defects and cosmic ray rejection. The \hst{} GOODS-N field covers an area of $\sim$\,0.021 deg $^2$, which was easily contained inside the LBC's large FOV of $\sim$\,0.16 deg$^2$. Calibration data, bias frames, and twilight sky-flats were taken on most nights, and were used with the LBC pipeline for the standard data reduction steps \citep[see][for details]{Giallongo2008}.


\noindent\begin{figure}[t!]
\includegraphics[width=0.48\txw]{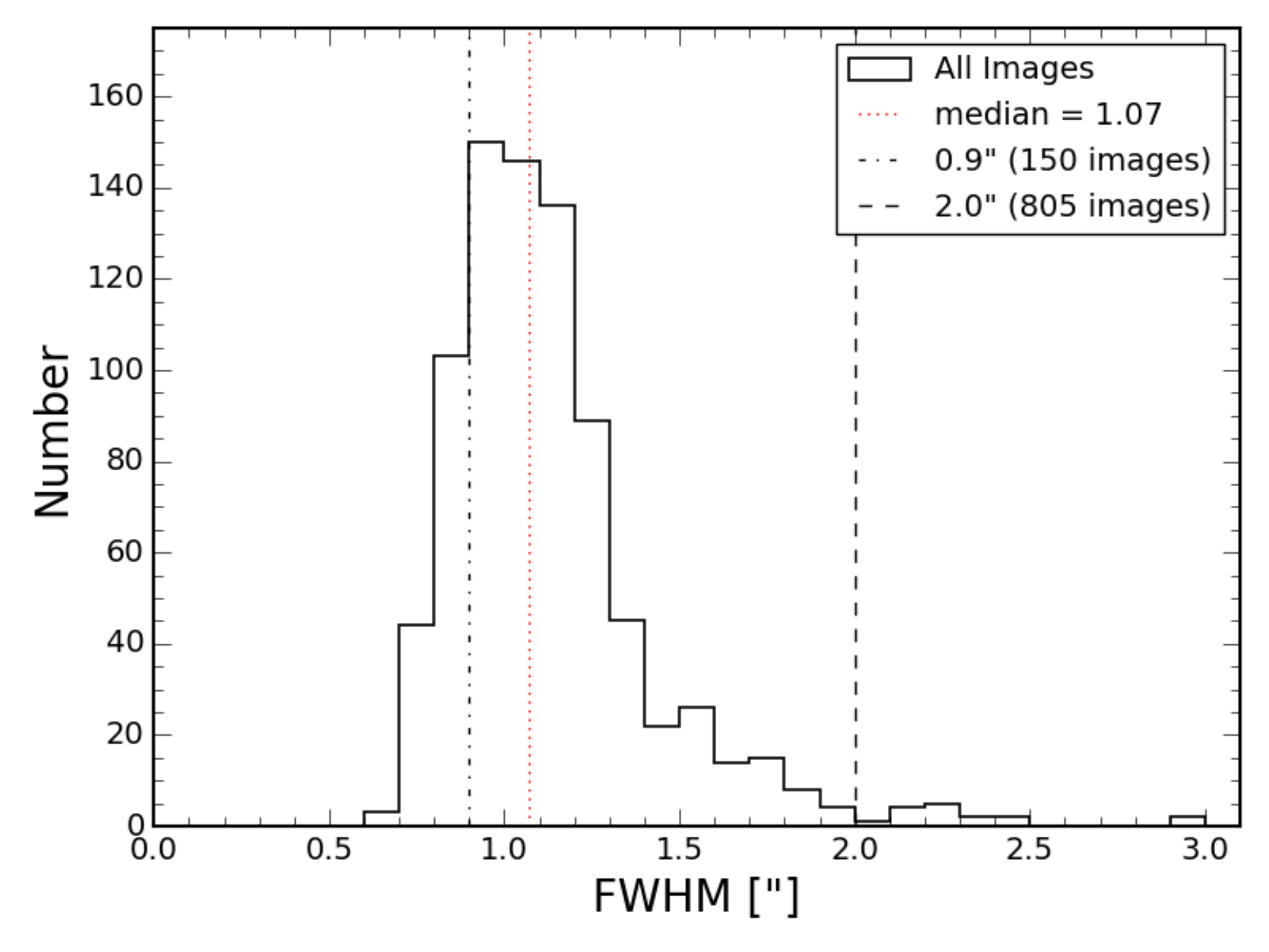}
\caption{\noindent\small
A histogram of the FWHM measured from unsaturated stars for the 838 individual \rb{} exposures taken in the GOODS-N field with the LBC-Red camera. The dot-dashed line represents the largest FWHM ($0\farcs9$) included in the optimal resolution mosaic. The vertical dashed line represents the cut-off of FWHM$\,=2\farcs0$ used for the optimal depth mosaic, which includes 805 exposures. The median value of the FWHM distribution, FWHM$\,=1\farcs07$, is highlighted by the red, dotted line. }
\label{figure:histr}
\end{figure}

\subsection{Creating \rb{} Mosaics}
\label{sec:mosaicsr}

For each of the 838 individual exposures, the Gaussian FWHM was measured for unsaturated stars in the FOV, with the median value corresponding to the seeing of the entire exposure. As described in \citet{Ashcraft2018},  \citet{Otteson2021} and \citet{Redshaw2022}, this allows for a seeing distribution to be created for the entire dataset as shown in Figure \ref{figure:histr}. The median FWHM for all images is $\sim$1$\farcs07$, which is marginally larger than the typical seeing conditions on Mt. Graham for \rb{} of $0\farcs97 \pm 0\farcs06$ \citep{Taylor2004}. Following the prescription from \citet{Ashcraft2018}, the optimal depth \rb{} stack was created with all exposures with FWHM$\,\lesssim\!2\farcs0$, which excludes only the 33 images with the worst seeing ($\sim\!4\%$ of the dataset). The optimal resolution \rb{} stack was created using all images with FWHM $\lesssim$ $0\farcs9$, which corresponds to the best 150 exposures ($\sim\!18\%$ of the data). 

\begin{figure}[h]
\centering
\includegraphics[width=0.48\txw]{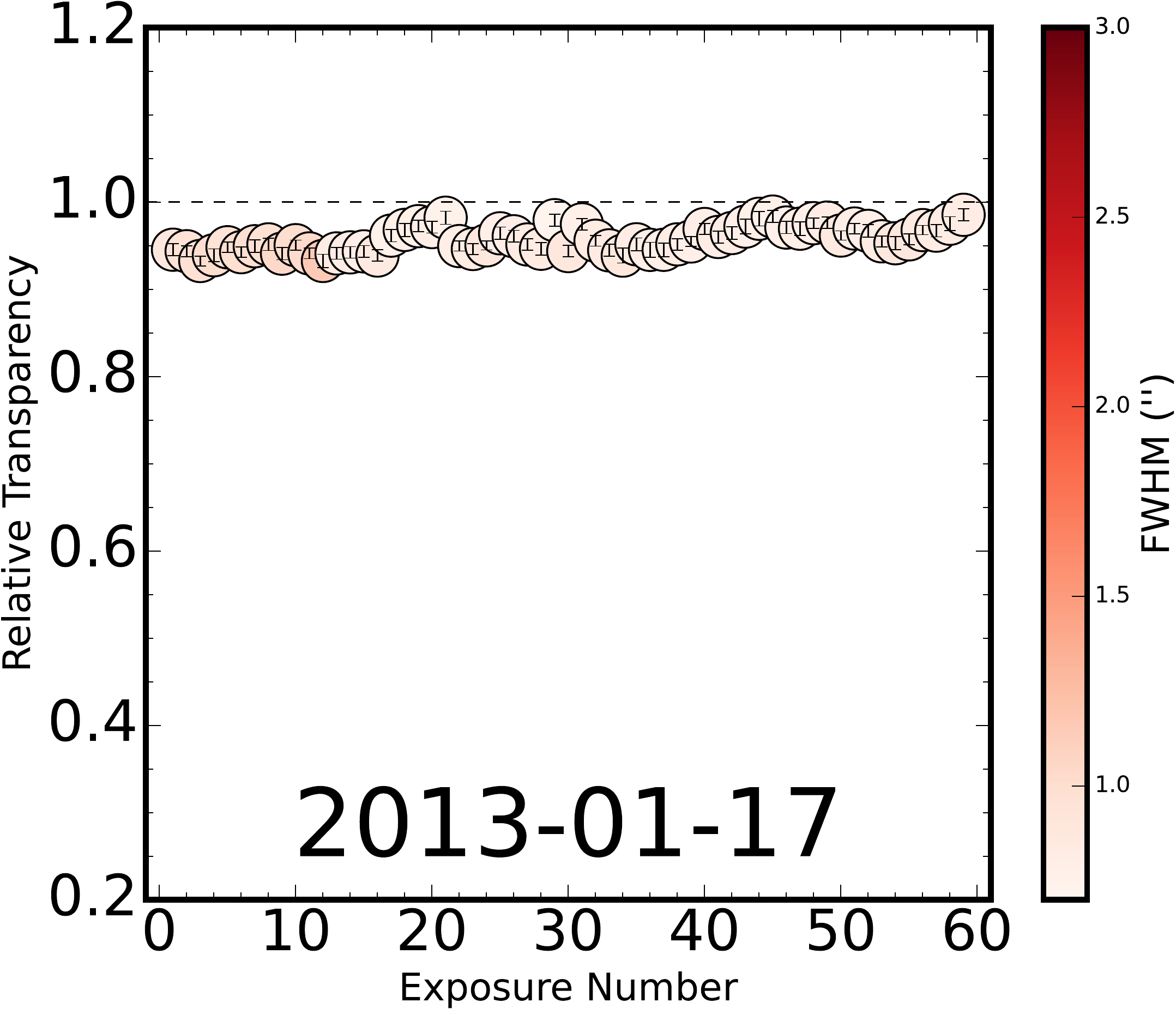}
\caption{Relative transparency values for the first 59 exposures in the sample from January 17, 2013. The color of each data point represents the median seeing value of $\sim$100 unsaturated stars identified in the FOV. The conditions on this particular night allowed for particularly high transparency. On average, across the entire dataset, the relative transparency is $\sim$80\% with some nights between 40-60\%. The flux across each image was scaled so that the relative transparency values were equal to unity and uniform from image to image and therefore, night to night. The complete figure set (15 images) for the full dataset is available in the online journal. }
\label{figure:transp}
\end{figure}

Prior to creating the mosaics, the relative transparency was calculated for each of the 838 exposures following the prescription in \citet{Otteson2021,Redshaw2022} and McCabe et al. (2022; in prep.). In order to account for the night to night differences in relative atmospheric transparency, the flux ratio between $\sim$100 unsaturated stars was taken with respect to the flux values from the Sloan Digital Sky Survey (SDSS) Data Release 16 \citep{Blanton2017,Ahumada2020} in the \rb{}. In this case, a relative transparency value of 1 indicates that the median flux from the $\sim$100 matched SDSS stars in the \rb{} is equal to the flux from the same stars in the individual exposure. Relative transparency values less than 1 indicate that there is less flux received from these stars than expected based upon SDSS \rb{} values. Figure \ref{figure:transp} shows the relative transparency for the night of January 17, 2013, where the median transparency varied from $\sim$0.94 to $\sim$0.98. We corrected for these night-by-night transparency differences by simply scaling images by the median transparency offset to achieve a transparency of 1.0 in the affected images. 

The optimal resolution and optimal depth mosaics were created by combining individual exposures with \swarp \citep{Bertin2002,Bertin2010}. Table \ref{table:swarpr} summarizes some of the key \swarp parameters used for creating these mosaics. The choice of parameters is almost identical to those used previously in \citet{Ashcraft2018}, except for using a ``BACK\_SIZE" parameter of 280 pixels for the mesh size, and a ``BACK\_FILTERSIZE" of 7. The background parameters were increased to compensate for the increased number of bright saturated stars in individual images compared to the \Ub data of \citet{Ashcraft2018}.

\figsetstart
\figsetnum{2}
\figsettitle{Relative transparency of the GOODS-N LBT \rb{} images from January 2013 to March 2014.}

\figsetgrpstart
\figsetgrpnum{2.1}
\figsetgrptitle{Relative transparency values from January 17, 2013.}
\figsetplot{Fig2.1.pdf}
\figsetgrpnote{Relative transparency values from January 17, 2013. The color of each data point represents the median seeing value of $\sim$100 unsaturated stars identified in the FOV. The flux across each image was then scaled so that the relative transparency values were equal to unity and uniform from image to image and night to night.}
\figsetgrpend

\figsetgrpstart
\figsetgrpnum{2.2}
\figsetgrptitle{Relative transparency values from February 15, 2013.}
\figsetplot{Fig2.2.pdf}
\figsetgrpnote{Relative transparency values from February 15, 2013. The color of each data point represents the median seeing value of $\sim$100 unsaturated stars identified in the FOV. The flux across each image was then scaled so that the relative transparency values were equal to unity and uniform from image to image and night to night.}
\figsetgrpend

\figsetgrpstart
\figsetgrpnum{2.3}
\figsetgrptitle{Relative transparency values from February 16, 2013.}
\figsetplot{Fig2.3.pdf}
\figsetgrpnote{Relative transparency values from February 16, 2013. The color of each data point represents the median seeing value of $\sim$100 unsaturated stars identified in the FOV. The flux across each image was then scaled so that the relative transparency values were equal to unity and uniform from image to image and night to night.}
\figsetgrpend

\figsetgrpstart
\figsetgrpnum{2.4}
\figsetgrptitle{Relative transparency values from February 17, 2013.}
\figsetplot{Fig2.4.pdf}
\figsetgrpnote{Relative transparency values from February 17, 2013. The color of each data point represents the median seeing value of $\sim$100 unsaturated stars identified in the FOV. The flux across each image was then scaled so that the relative transparency values were equal to unity and uniform from image to image and night to night.}
\figsetgrpend

\figsetgrpstart
\figsetgrpnum{2.5}
\figsetgrptitle{Relative transparency values from February 18, 2013.}
\figsetplot{Fig2.5.pdf}
\figsetgrpnote{Relative transparency values from February 18, 2013. The color of each data point represents the median seeing value of $\sim$100 unsaturated stars identified in the FOV. The flux across each image was then scaled so that the relative transparency values were equal to unity and uniform from image to image and night to night.}
\figsetgrpend

\figsetgrpstart
\figsetgrpnum{2.6}
\figsetgrptitle{Relative transparency values from April 3, 2013.}
\figsetplot{Fig2.6.pdf}
\figsetgrpnote{Relative transparency values from April 3, 2013. The color of each data point represents the median seeing value of $\sim$100 unsaturated stars identified in the FOV. The flux across each image was then scaled so that the relative transparency values were equal to unity and uniform from image to image and night to night.}
\figsetgrpend

\figsetgrpstart
\figsetgrpnum{2.7}
\figsetgrptitle{Relative transparency values from April 4, 2013.}
\figsetplot{Fig2.7.pdf}
\figsetgrpnote{Relative transparency values from April 4, 2013. The color of each data point represents the median seeing value of $\sim$100 unsaturated stars identified in the FOV. The flux across each image was then scaled so that the relative transparency values were equal to unity and uniform from image to image and night to night.}
\figsetgrpend

\figsetgrpstart
\figsetgrpnum{2.8}
\figsetgrptitle{Relative transparency values from April 6, 2013.}
\figsetplot{Fig2.8.pdf}
\figsetgrpnote{Relative transparency values from April 6, 2013. The color of each data point represents the median seeing value of $\sim$100 unsaturated stars identified in the FOV. The flux across each image was then scaled so that the relative transparency values were equal to unity and uniform from image to image and night to night.}
\figsetgrpend

\figsetgrpstart
\figsetgrpnum{2.9}
\figsetgrptitle{Relative transparency values from April 7, 2013.}
\figsetplot{Fig2.9.pdf}
\figsetgrpnote{Relative transparency values from April 7, 2013. The color of each data point represents the median seeing value of $\sim$100 unsaturated stars identified in the FOV. The flux across each image was then scaled so that the relative transparency values were equal to unity and uniform from image to image and night to night.}
\figsetgrpend

\figsetgrpstart
\figsetgrpnum{2.10}
\figsetgrptitle{Relative transparency values from May 8, 2013.}
\figsetplot{Fig2.10.pdf}
\figsetgrpnote{Relative transparency values from May 8, 2013. The color of each data point represents the median seeing value of $\sim$100 unsaturated stars identified in the FOV. The flux across each image was then scaled so that the relative transparency values were equal to unity and uniform from image to image and night to night.}
\figsetgrpend

\figsetgrpstart
\figsetgrpnum{2.11}
\figsetgrptitle{Relative transparency values from May 10, 2013.}
\figsetplot{Fig2.11.pdf}
\figsetgrpnote{Relative transparency values from May 10, 2013. The color of each data point represents the median seeing value of $\sim$100 unsaturated stars identified in the FOV. The flux across each image was then scaled so that the relative transparency values were equal to unity and uniform from image to image and night to night.}
\figsetgrpend

\figsetgrpstart
\figsetgrpnum{2.12}
\figsetgrptitle{Relative transparency values from May 11, 2013.}
\figsetplot{Fig2.12.pdf}
\figsetgrpnote{Relative transparency values from May 11, 2013. The color of each data point represents the median seeing value of $\sim$100 unsaturated stars identified in the FOV. The flux across each image was then scaled so that the relative transparency values were equal to unity and uniform from image to image and night to night.}
\figsetgrpend

\figsetgrpstart
\figsetgrpnum{2.13}
\figsetgrptitle{Relative transparency values from May 13, 2013.}
\figsetplot{Fig2.13.pdf}
\figsetgrpnote{Relative transparency values from May 13, 2013. The color of each data point represents the median seeing value of $\sim$100 unsaturated stars identified in the FOV. The flux across each image was then scaled so that the relative transparency values were equal to unity and uniform from image to image and night to night.}
\figsetgrpend

\figsetgrpstart
\figsetgrpnum{2.14}
\figsetgrptitle{Relative transparency values from June 10, 2013.}
\figsetplot{Fig2.14.pdf}
\figsetgrpnote{Relative transparency values from June 10, 2013. The color of each data point represents the median seeing value of $\sim$100 unsaturated stars identified in the FOV. The flux across each image was then scaled so that the relative transparency values were equal to unity and uniform from image to image and night to night.}
\figsetgrpend

\figsetgrpstart
\figsetgrpnum{2.15}
\figsetgrptitle{Relative transparency values from March 7, 2014.}
\figsetplot{Fig2.15.pdf}
\figsetgrpnote{Relative transparency values from March 7, 2014. The color of each data point represents the median seeing value of $\sim$100 unsaturated stars identified in the FOV. The flux across each image was then scaled so that the relative transparency values were equal to unity and uniform from image to image and night to night.}
\figsetgrpend

\figsetend

\noindent\begin{deluxetable}{lc}
\tabletypesize{\footnotesize} 
\tablewidth{0pt}
\tablecolumns{2}
\tablehead{
\colhead{Keyword}   & \colhead{Value}}
\startdata
COMBINE\_TYPE       & CLIPPED \\
WEIGHT\_TYPE        & MAP\_WEIGHT \\
PIXELSCALE\_TYPE    & Median \\
CENTER (J2000)      & 12:36:54.5, +62:15:41.1 \\
IMAGE\_SIZE (pix)   & 6351, 6751  \\
RESAMPLING\_TYPE    & LANCZOS3 \\
CLIP\_SIGMA         & 5.0  \\
CLIP\_AMPFRAC       & 0.5 \\
BACK\_SIZE          & 280 \\
BACK\_FILTERSIZE    & 7 \\
\enddata
\caption{SWARP Configuration Parameters adopted for the \rb{} Mosaics.}
\label{table:swarpr}
\end{deluxetable}


\subsection{LBC \rb{} Catalogs}
\label{sec:catalogsr}

\sextractor{} \citep{Bertin1996} was used to identify objects and create photometric catalogs. We developed a \sextractor{} parameter set which adequately balanced the unique source separation without removing extended, low surface brightness features from their host galaxies. Beginning with the \sextractor{} parameters used in the \citet{Ashcraft2018} analysis, various parameters were tweaked to optimize object detection and deblending. For the \rb{} catalogs, the deblending parameters ``DEBLEND\_NTHRESH" and ``DEBLEND\_MINCONT" were adjusted to more accurately separate objects, especially in the more dense regions of the field. For object detection, a Gaussian filter was used to smooth the image with a convolving kernel (FWHM 2.0\,pixels) and a convolution image size of $5\times5$\,pixels. It was found that changing the deblending parameters affected the number of objects detected, but the choice of the smoothing filter did not have a significant impact on number of extracted objects by \sextractor{}. However, failure to use a smoothing filter resulted in a large amount of spurious detections. The major \sextractor{} parameters used to create the final \rb{} catalogs are listed in Table \ref{table:sextr}.


\noindent\begin{deluxetable}{lcc}[b!]
\tabletypesize{\footnotesize} 
\tablewidth{0pt}
\tablecolumns{3}
\tablehead{
\colhead{Keyword} & \colhead{Optimized Resolution} & \colhead{Optimized Depth} }
\startdata
DETECT\_MINAREA     & 6         & 6\\
DETECT\_THRESH      & 1.0       & 1.0\\
ANALYSIS\_THRESH    & 1.0       & 1.0\\
DEBLEND\_NTHRESH    & 64        & 64\\
DEBLEND\_MINCONT    & 0.008     & 0.004\\
WEIGHT\_TYPE        & MAP\_RMS  & MAP\_RMS \\
\enddata
\caption{\textsc{SExtractor} Configuration Parameters adopted for the $r$-sloan Mosaics }
\label{table:sextr}
\end{deluxetable}


A mask image was generated to discard any bright stars and their surrounding corrupted areas during the \rb{} object detection process. The mask was created from the optimal depth image, which had the largest Gaussian wings resulting from the larger FWHM of included exposures. For consistency, the same mask was used for both mosaics. The final object catalogs excluded all objects with the \sextractor{} \texttt{FLAGS} value larger than 3, and magnitude errors greater than $\sigma$\mab $>$\,0.4 mag. Lastly, photometric zero points were determined by identifying unsaturated stars with AB magnitudes between \rab{}$\simeq\!18$ and \rab{}$\simeq\!22$ mag and matching them to SDSS \rb{} magnitudes. Approximately 170 stars within this magnitude range were verified in the LBC images. Stars with nearby neighboring objects were excluded in order to prevent potentially biased flux measurements. We measured photometric zero points of 28.06 and 28.05 mag for the optimal resolution and optimal depth mosaics, respectively. 


\section{ANALYSIS} \label{sec:analysisr}

\begin{figure*}[t!]
\includegraphics[width=0.96\txw]{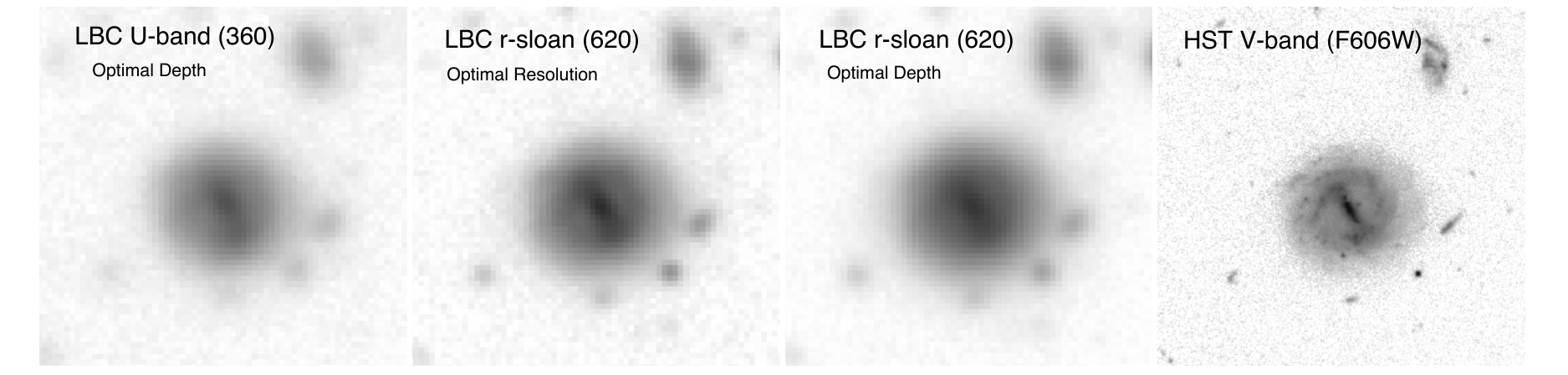}
\includegraphics[width=0.96\txw]{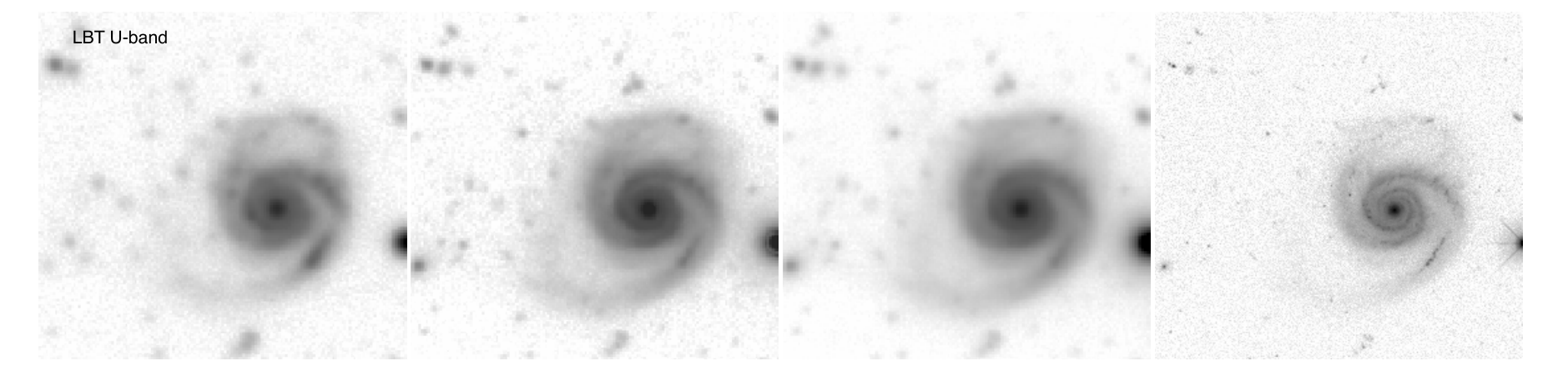}
\includegraphics[width=0.96\txw]{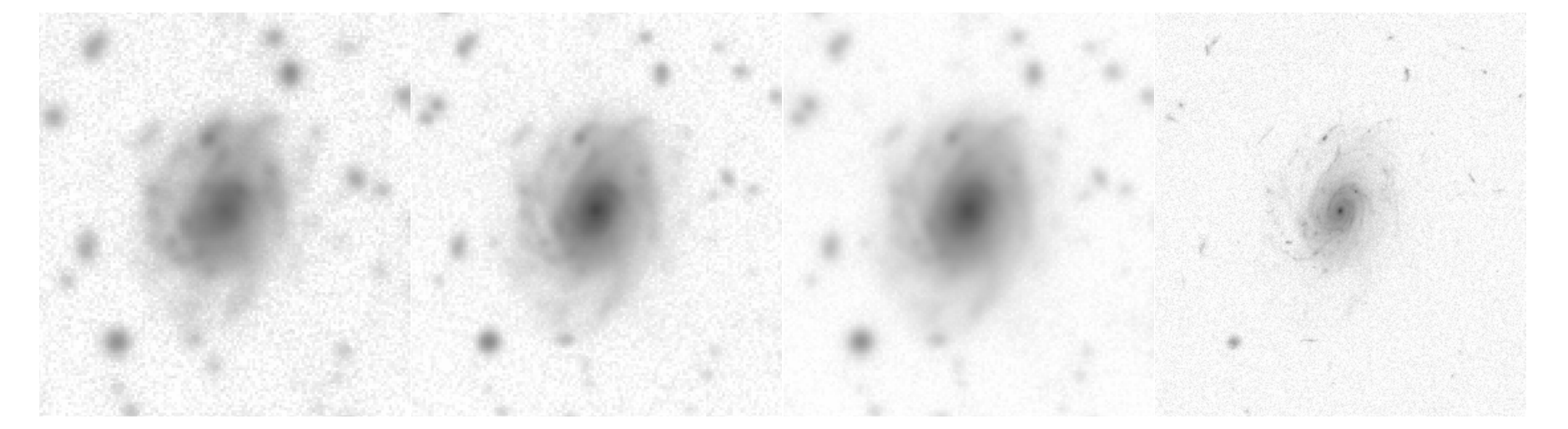}
\includegraphics[width=0.96\txw]{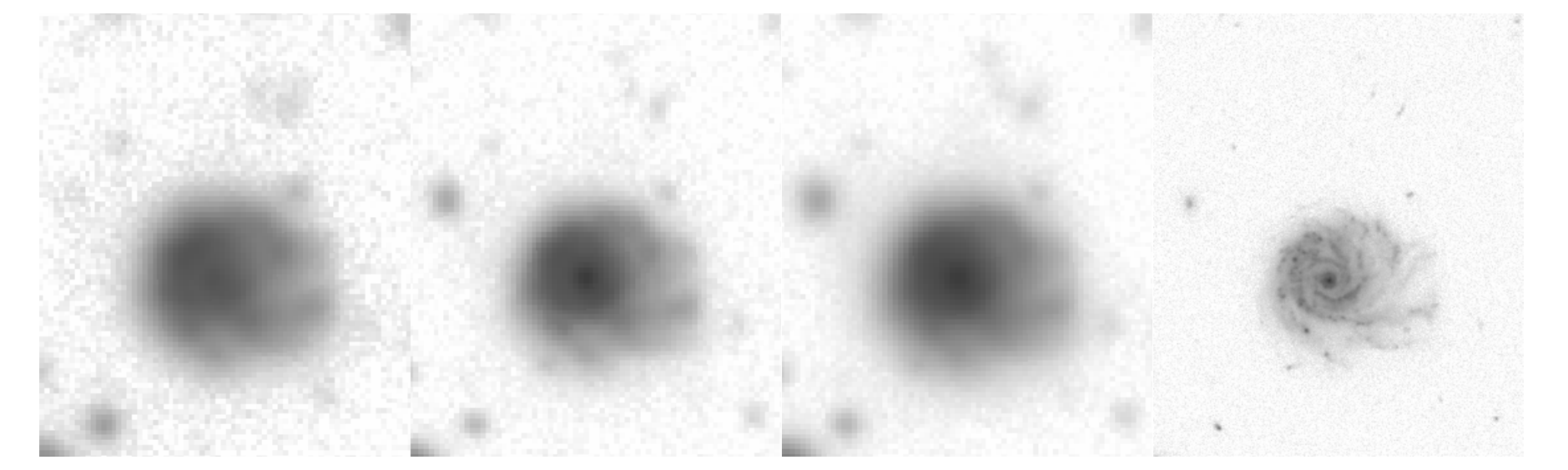}
\caption{\noindent\small 
Four bright face-on spiral galaxies in the GOODS-N field, which are also
observed in \hst{} CANDELS. The top galaxy has \Uab=\,20.8\,mag and
\rab{}=\,19.5\,mag, the second galaxy has \Uab=\,18.0\,mag and
\rab{}=\,17.5\,mag, the third galaxy has \Uab=\,19.9\,mag and
\rab{}=\,18.5\,mag, and the bottom galaxy has \Uab=\,19.5\,mag and
\rab{}=\,18.5\,mag. The LBC optimal depth \Ub image, LBT \rb{} optimal
resolution image, LBT \rb{} optimal depth image, and the \hst{} ACS $V$-band
(F606W; \citet{Giavalisco2004}) image are shown from the left columns to the
right columns, respectively. \deleted{{\bf Please magnify the PDF versions of Fig. 2--4
and Fig. 9--22 as needed to see the details.}}
}
\label{figure:brightgalsHSTr}
\end{figure*}

\begin{figure*}[t!]
\includegraphics[width=0.96\txw]{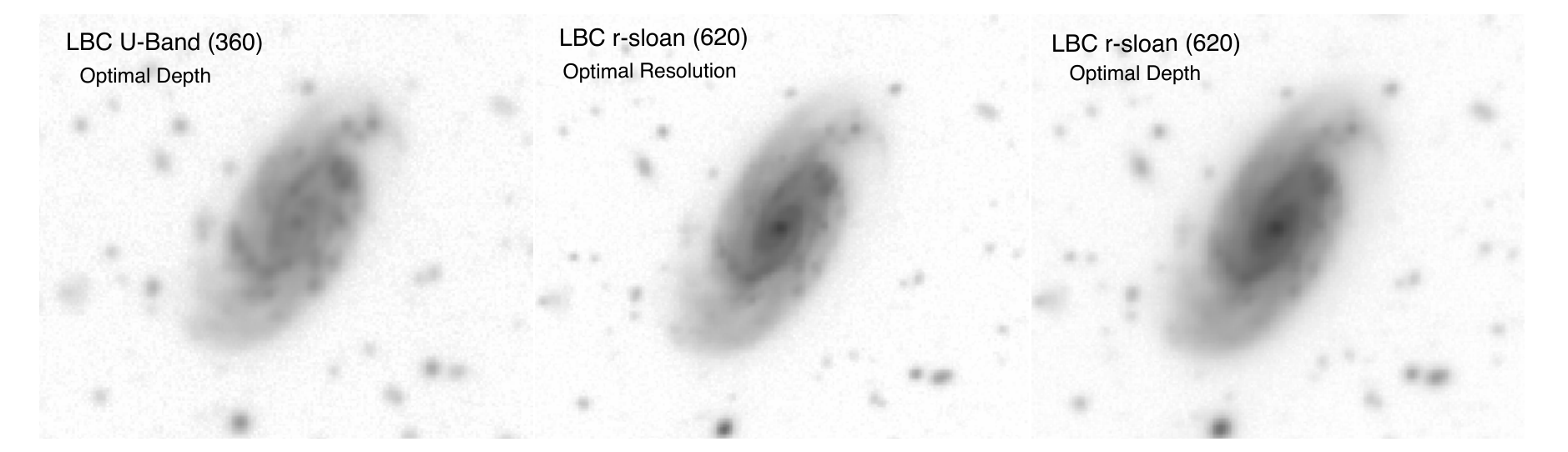}
\includegraphics[width=0.96\txw]{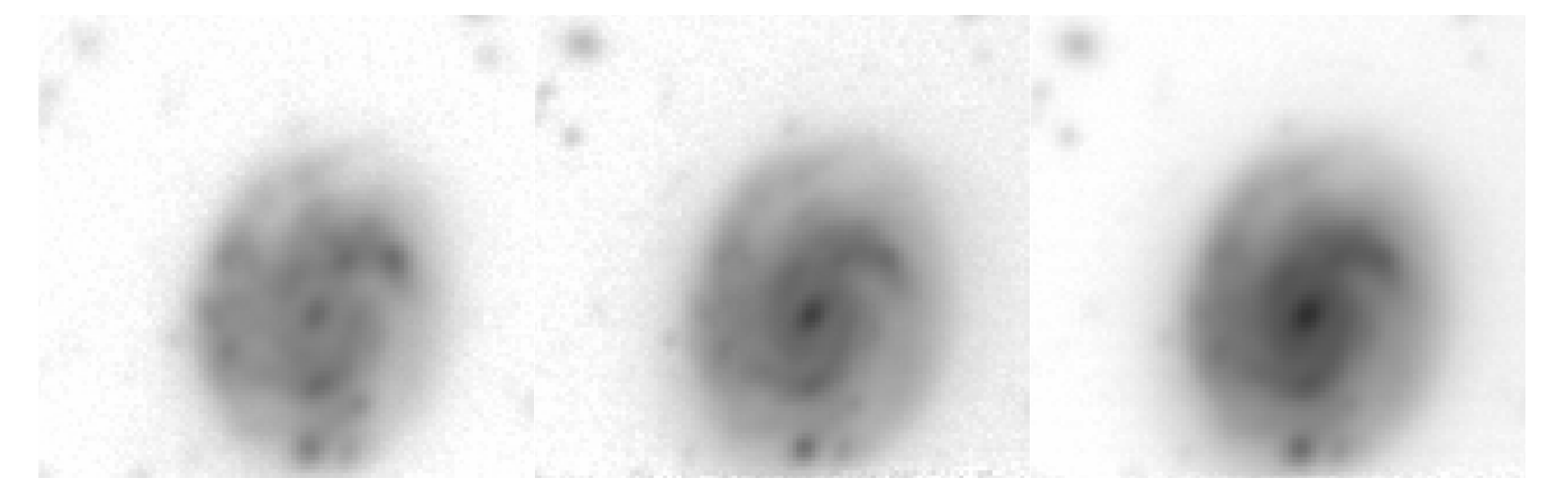}
\caption{\noindent\small 
Two of the brightest galaxies in the LBC FOV, but outside the \hst{} CANDELS area. The
top galaxy has \Uab = 18.3 mag and \rab{} = 16.7 mag, and the bottom galaxy has
\Uab = 19.2 mag and \rab{} = 18.3 mag. From left to right, the columns show the
optimal depth \Ub image, the optimal resolution \rb{} image, and the optimal
depth \rb{} image, respectively. Both of these face-on spirals are sufficiently well-resolved spatially that several sub-structure features including clumps and spiral arms.
Most of these features can be seen in both the optimal resolution and
optimal depth mosaics. However, the features are sharper and easier to
distinguish in the optimal resolution mosaic, while some features blur together
in the optimal depth mosaic.}
\label{figure:brightgalsr}
\end{figure*}

The trade off between the optimal resolution and optimal depth mosaics is clear when looking at the larger and brighter galaxies in the LBT/LBC mosaics. When lower resolution images (FWHM $\gtrsim1\farcs0$) are included, the light from galaxies smooths out and substructures are lost within brighter and larger galaxies (see Fig.\,\ref{figure:brightgalsHSTr} and Fig.\,\ref{figure:brightgalsr}). This phenomena is most apparent when comparing bright face-on spiral galaxies (\rab{}$\simeq 17.5-19.5$\,mag) from the \rb{} optimal resolution ($0\farcs9$ FWHM) and optimal depth ($2\farcs0$ FWHM) mosaics and the \Ub optimal depth mosaic ($1\farcs8$ FWHM; \citet{Ashcraft2018}) to \textit{HST}-ACS $V_{606}$ images of \citet{Giavalisco2004}.

Figure \ref{figure:brightgalsHSTr} clearly illustrates the power of having both ground based optimal resolution and optimal depth images in addition to high resolution \hst{} imaging. In the top panel, a bar is clearly present in the \hst{} image and the \rb{} optimal resolution mosaic, but is much less discernible in the optimal depth mosaic. This feature is also discernible in the LBT mosaics, especially in the optimal resolution \rb{} mosaic. Of the detectable
features in the LBT mosaics, the smallest scale galaxy features are easier to
identify in the optimal resolution mosaic. Most notably, we note that in the deeper LBT \rb{} mosaics, extended low-surface brightness flux in the outer parts of the galaxy is present, which is not always apparent in the \hst{} images.

Similar to Figure \ref{figure:brightgalsHSTr}, Figure \ref{figure:brightgalsr} shows two additional bright galaxies that fall outside of the \hst{} footprint. These galaxies exhibit signatures of star forming clumps within their spiral arms, which are particularly prominent in the optimal depth images. While clumps are observed to be more prevalent at higher redshifts, these optimal depth and resolution mosaics may be useful in determining the origin and ages of the clumps in addition to serving as analogs for high redshift galaxies \citep[and references therein]{Elmegreen2009Bruce, Elmegreen2009, Overzier2009, Fisher2014, Adams2022}.

\begin{figure*}
\includegraphics[width=0.96\txw]{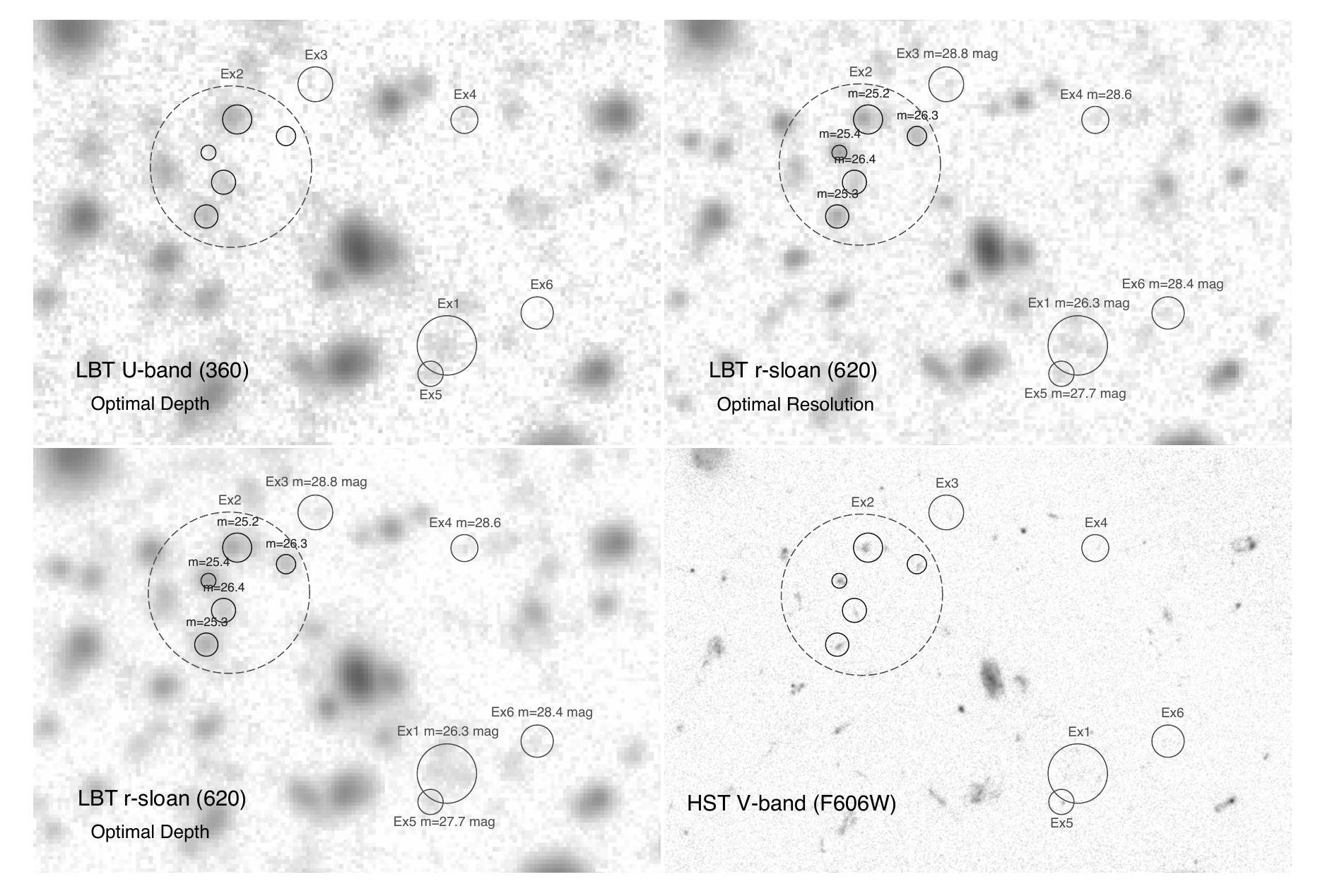}
\caption{\noindent\small The LBC optimal depth \Ub (top left), optimal
resolution \rb{} (top right), optimal depth \rb{} (bottom left), and \hst{} ACS
$V$-band F606W (bottom right)images are shown for the same region of the
GOODS-N field. In this region, the faintest flux detected in the LBT \rb{}
mosaics (\rab{}$\simeq$\,28.5 mag for the optimal depth image) is not always
discernible in the \hst{} images (see example regions 3, 4 and 6). The smallest
objects in the \hst{} V-band (F606W; FWHM$\lesssim$ $0\farcs68$) are not always
detected as single objects in the LBC mosaic catalogs, especially for the
lower-resolution optimal depth LBT mosaics. The dashed circle region of example
2 demonstrates an area where \sextractor{} detected a few separate objects as
one in the LBT optimal depth \rb{} mosaic. However, \sextractor{} was able to
deblend objects in the LBT optimal resolution \rb{} mosaic shown by the solid
circles within the example region 2 dashed circle. The circle of example 1
shows a low-surface brightness region, which is hard to detect in the \hst{}
$V$-band. In contrast, the compact object in the example 5 circle is fainter
than the low-surface brightness region of example 1, yet its small size makes
it easy to detect in the high-resolution \hst{} images. All magnitudes are
measured in the optimal depth LBT $r$-band except for the objects within the
example 2 dashed circle, which come from the optimal resolution $r$-band
catalog.}
\label{figure:faintgalsr}
\end{figure*}

The lower-resolution images also make it more difficult to deblend neighboring
objects (see Fig.\,\ref{figure:faintgalsr}). Example Region 2 in
Fig.\,\ref{figure:faintgalsr} shows a large region, which is detected as one
object by \sextractor{} in the optimal depth, yet lower-resolution mosaic,
indicated by a dashed circle, while \sextractor{} is able to separate the
objects within the dashed circle in the higher-resolution mosaic, indicated
by solid circles\footnote{Note that within region 2, the galaxy farthest to the right does not
appear in the LBT \Ub image, as it has been redshifted beyond
detection ($z_{spec}$\,=\,3.52)}. All magnitudes presented in
Fig.\,\ref{figure:faintgalsr} are measured in the optimal depth LBT $r$-band,
except for the objects within the example 2 dashed circle, which come from the
optimal resolution $r$-band catalog. Objects 3, 4, and 6 in
Fig.\,\ref{figure:faintgalsr} are examples of faint objects detected in the
optimal depth LBT image \rab{}\,$\gtrsim$\,28.4 mag. Only object 6 was detected
in the optimal resolution catalog, which is near its limit for reliable
detections, and all three regions show almost no measurable flux above the
background levels in the \hst{} F606W filter. 

Figure\,\ref{figure:faintgalsr} shows a low-surface brightness region in the example of circle 1, which is barely detectable in the \hst{} F606W mosaic. In contrast, the compact object in the example
of circle 5 is fainter (\mab$\simeq$\,27.7 mag) than the low-surface brightness
region of example 1 (\mab$\simeq$\,26.3 mag), yet its small size and
higher-surface brightness makes it easy to detect in the high-resolution \hst{}
images.


\noindent
\vspace*{-\baselineskip}



\noindent
\vspace*{-\baselineskip}


\subsection{Optimal Resolution versus Optimal Depth LBT \rb{} Mosaics}

\begin{figure}
\includegraphics[width=\columnwidth]{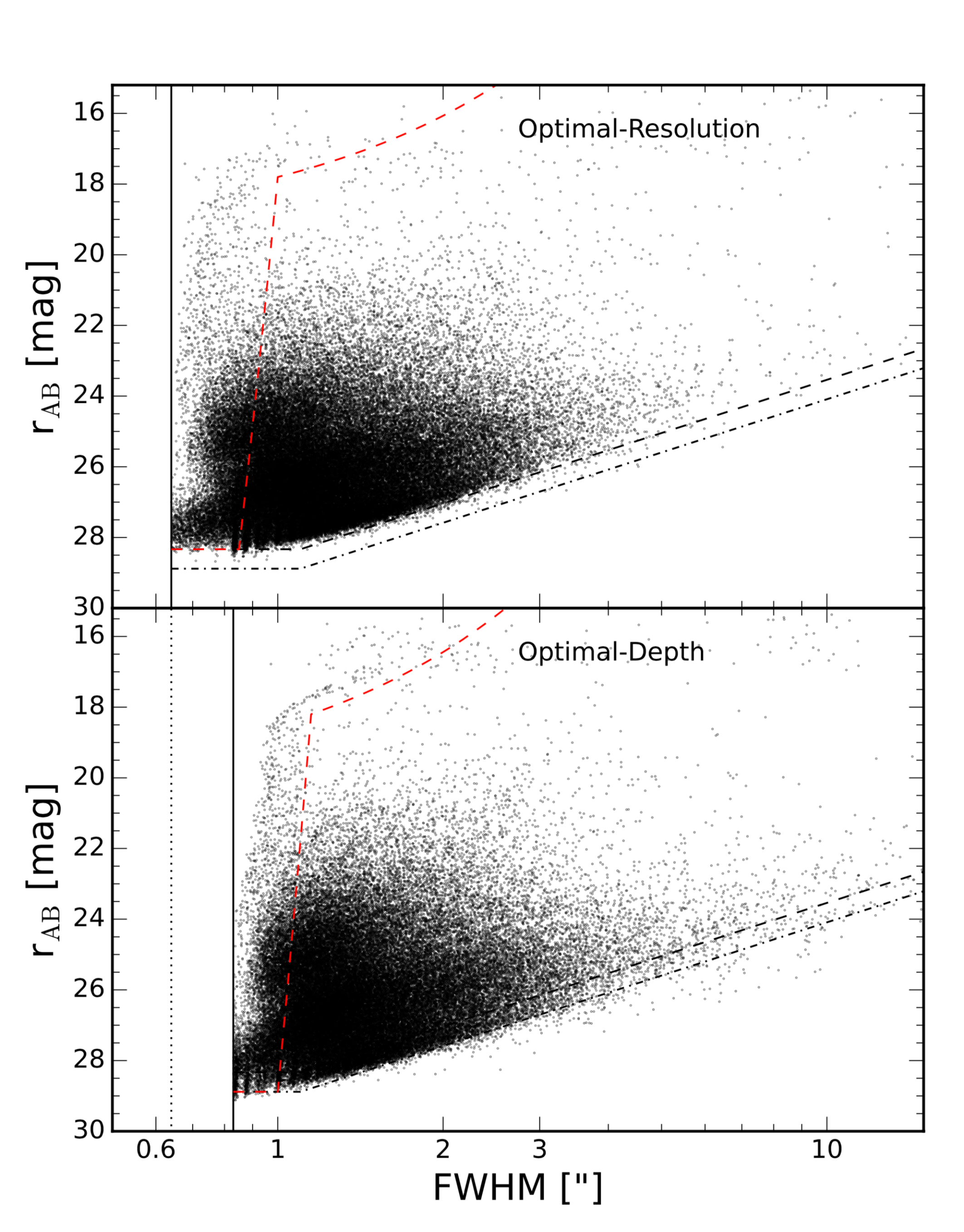}
\caption{\noindent\small 
Total \sextractor{} object magnitude vs.\ FWHM as measured with \sextractor{}
for the optimal resolution mosaic (top) and the deepest-lower-resolution mosaic
(bottom). The solid lines represent the FWHM-limit used in each image, and the black dotted and dotted-dashed lines
represent the effective surface-brightness limits for the optimal resolution and optimal depth mosaics, respectively.}
\label{figure:magradr}
\end{figure}

In order to compare the optimal resolution and optimal depth mosaics, the catalogs described in \S \ref{sec:catalogsr} were used to look at object magnitude as a function of FWHM (Figure \ref{figure:magradr}). The solid lines represent the FWHM limits of $\sim$ $0\farcs65$ and $\sim$ $0\farcs90$ for the optimal resolution and optimal depth mosaics, respectively. Objects with \mab$\gtrsim$\,29 mag generally have sizes ``smaller" than the FWHM limit of the Point Spread Function (PSF), and therefore are not reliable detections. The black dashed and dot-dashed lines represent the effective surface brightness limits for the optimal resolution and optimal depth mosaics, respectively, while the red dashed line denotes the star/galaxy separation. 


\begin{figure}
\includegraphics[width=\columnwidth]{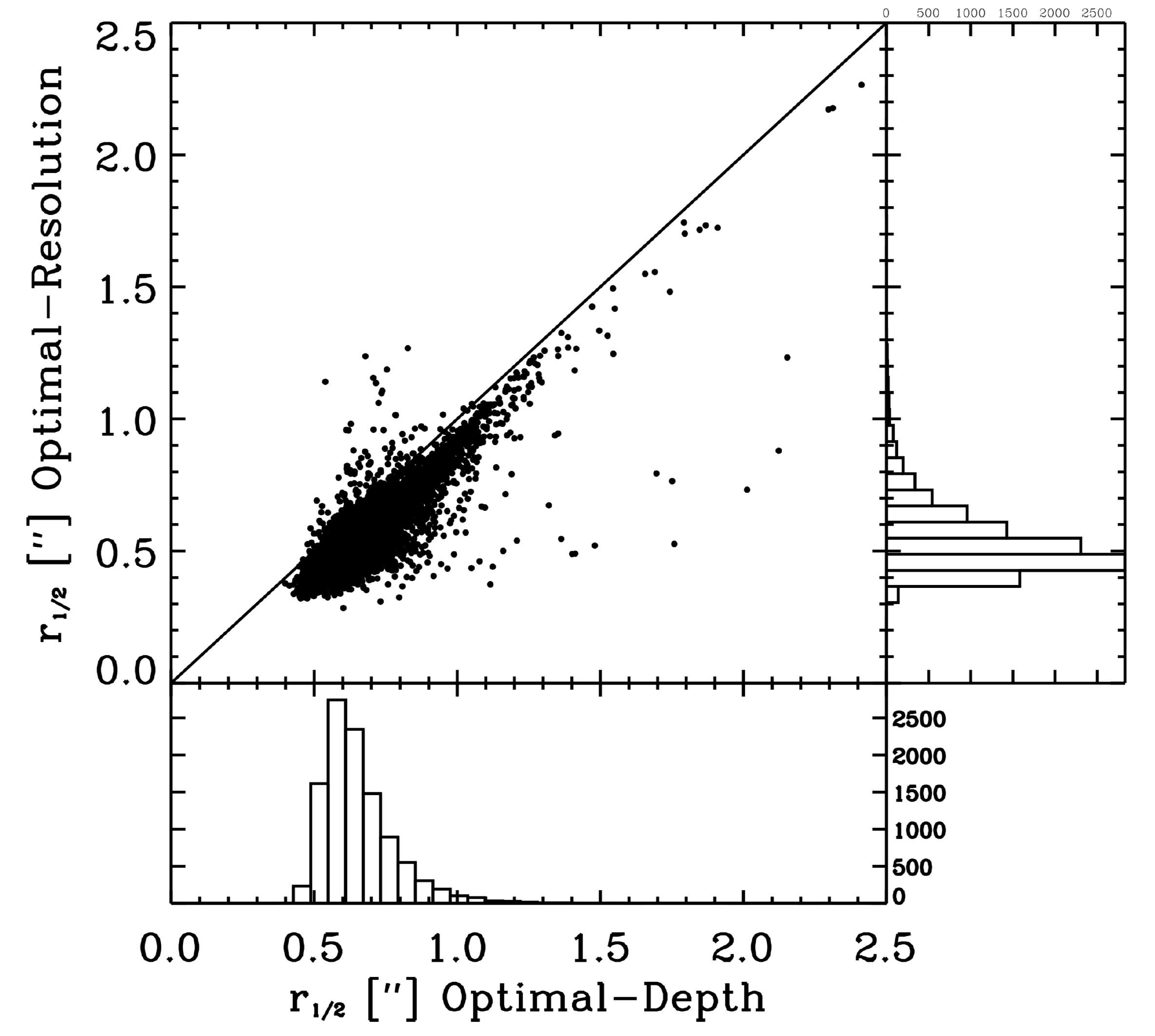}
\caption{\noindent\small 
Half-light radius for galaxies with \rab{}$\lesssim$ 27 mag were measured
using SExtractor. The optimal resolution image indeed yields consistently
smaller half-light radii compared to the optimal depth image.}
\deleted{, although the best fit regression still has a slope close to unity:
r$_{\textrm{optimal resolution}}$ $\simeq$ 1.00\ 
r$_{\textrm{optimal depth}}-0\farcs15$.}
\label{figure:rvsr2r}
\end{figure}

For further comparison, the half-light radii measured by \sextractor{} for both the optimal resolution and optimal depth images for galaxies brighter than \rab{}$\lesssim$ 27 mag is shown in Fig.\,\ref{figure:rvsr2r}. Unsurprisingly, the half-light radius of the optimal resolution image is consistently smaller than the optimal depth image. The half-light radii distributions peak at $\sim0\farcs45$ and $\sim0\farcs58$ for the optimal resolution image and the optimal depth image, respectively.


\begin{figure}
\includegraphics[width=\columnwidth]{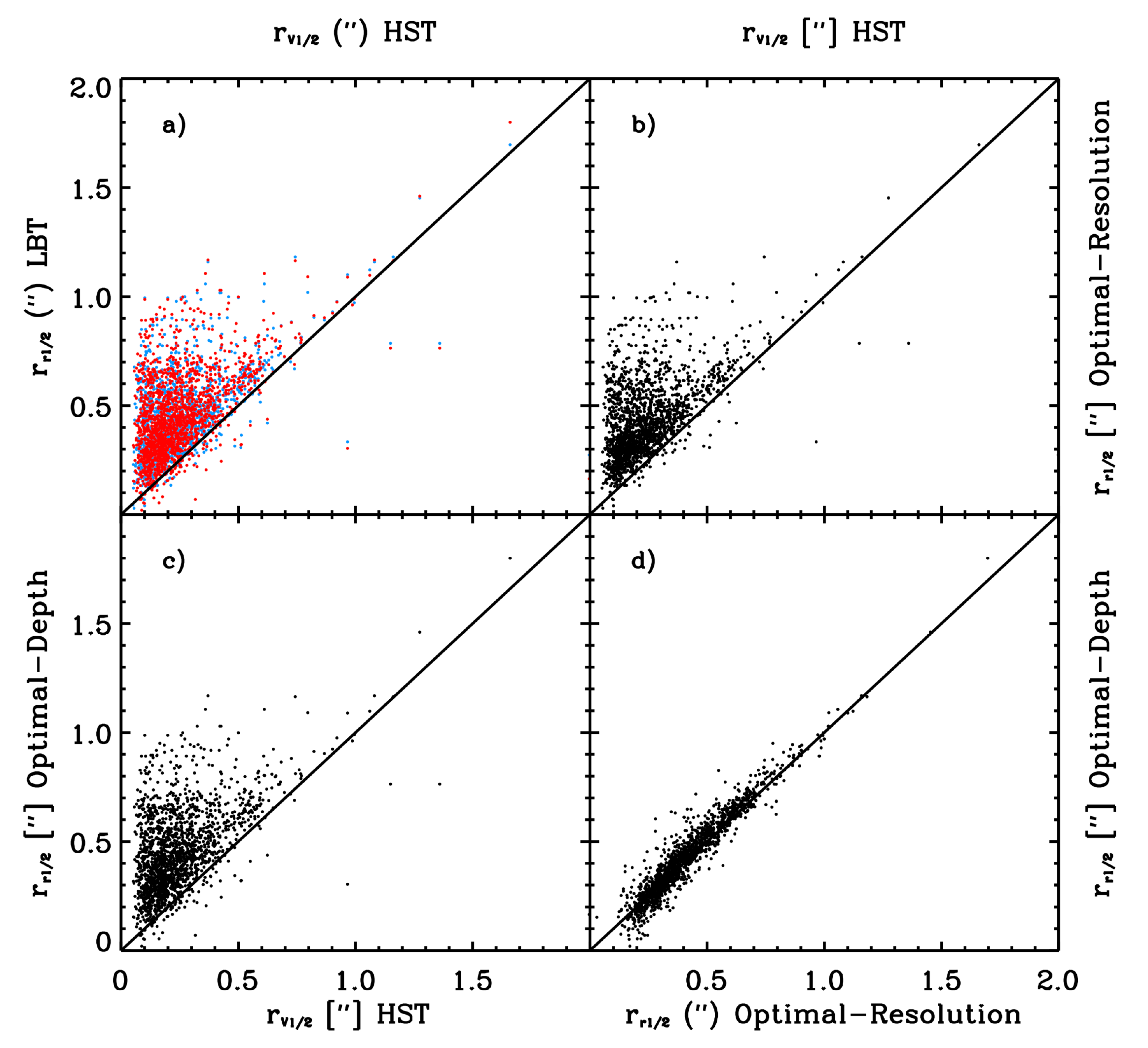}
\caption{\noindent\small
Comparison plots of half-light radius measured with \sextractor{} for two LBT
mosaics, optimal resolution and optimal depth, and the \hst{} \V--band. We only
included galaxies with 18 mag $\leq$\,\rab{} $\leq 27$ mag. All half-light
radii plotted have been corrected using
$r_{corr}=\sqrt{r^{2}-(\textrm{FWHM}/2)^{2}}$. Half-light radius for the
optimal resolution (blue) and lower-resolution optimal depth (red) images are
compared to the half-light radius for the \hst{} \V--band (a; \citet{Giavalisco2004}).
The optimal resolution (b) or the optimal depth (c) half-light radius
are compared to the \hst{} \V--band. The bottom right panel (d) compares the
optimal depth to the optimal resolution half-light radius measurements.}
\label{figure:rhstr}
\end{figure}

The \sextractor{} half-light radii were compared to the equivalent \hst{} \emph{V}--band catalog of the GOODS-N field from \citet{Giavalisco2004}. For this analysis, only galaxies with magnitudes between 18\,$\leq$\,\mab$\!\leq$\,27 mag as measured in the \hst{} \emph{V}--band were selected. Figure \ref{figure:rhstr} shows that the radii measured in the optimal resolution image (blue dots) agree slightly better with the \hst{} size measurements with less scatter than the sizes measured in the lower resolution image (red dots). In order to accurately represent intrinsic object sizes, we subtracted the PSF FWHM-values of of $0\farcs67$ and, $0\farcs98$ in quadrature from the optimal resolution and depth measurements, respectively ($r_{corr}=\sqrt{r^{2}-(\textrm{FWHM}/2)^{2}}$). For consistency, the PSF-size was subtracted in quadrature for the \emph{V}--band \hst{} images as well, but since the \hst/ACS PSF is so narrow ($0\farcs08$ FWHM; see Fig. 10a of \citet{Windhorst2011}), this correction had almost no effect except for the very smallest and most faint objects.


\section{Analysis}

\subsection{Tidal Tails and Merging Galaxies}
\label{sec:tidaltails}

We visually inspected the LBT GOODS-N \rb{} mosaics for signs of galaxy interactions such as tidal tails, diffuse plumes, and bridges. Approximately, 60 galaxies were found that were candidates for interactions, the 30 most obvious galaxies were selected for the sample of interacting galaxies. These galaxies are listed in Table \ref{table:intgals} along with their classification, redshift, and position. Interacting systems can resemble multiple types, especially when looking at both the \hst{} and LBT images of the same systems. Each of these classifications is defined as follows, following \citet{Elmegreen2007}:
\begin{itemize}
\vspace{-0.25cm}
\item Diffuse: Interactions are defined by diffuse plumes with small, star forming regions.
\vspace{-0.25cm}
\item Antennae: Antennae type interactions resemble the local Antennae system with long tidal tails and signs of possible double nuclei.
\vspace{-0.25cm}
\item M51: M51 interactions consist of a main spiral galaxy with a tidal arm that forms a bridge towards a companion.
\vspace{-0.25cm}
\item Assembly: Assembly types are irregular galaxies that appear to be merging and are made up of small pieces.
\vspace{-0.25cm}
\item Equal mass: Equal mass types appear as though two similar sized galaxies are in the process of merging.
\vspace{-0.25cm}
\item ``Shrimp'': `Shrimp'' types are characterized by a single prominent arm and uniformly distributed star forming clumps without signs of a central nucleus or core. 
\vspace{-0.25cm}
\item Tidal tail: Tidal tail interactions exist when both interacting galaxies still resemble their pre-merger shape, but have an extended tidal tail which indicates a recent interaction.
\vspace{-0.25cm}
\end{itemize}

Since this sample of interacting galaxies was selected with the goal of observing extended flux in the outer regions of the galaxies, it is inherently not a complete sample of interacting galaxies in the GOODS-N field. After the galaxies were selected, \hst{} imaging was examined if available. For a complete sample, deeper \hst{} imaging and spectroscopic redshifts would be required to confidently confirm galaxy morphology and interacting systems. However, since the \rb{} mosaics are $\gtrsim$90\% complete to SB levels of \mab$\lesssim$ 31 mag arcsec$^{-2}$, these 30 galaxies constitute a representative sample to this SB limit. While \hst{} imaging is not available for each galaxy in our sample, Appendix \ref{ap:mergers} shows the \rb{} and \Ub optimal depth images and \hst{} (F275W, F336W, F435W, F606W, F775W, F814W, F850LP, F105W, F125W, F160W, which correspond to the NUV, u, b, v, i, I, z, Y, J, H filters) images when available. Our sample covers objects with tidal features and redshifts in the range of $z\,\simeq$\,0.03--1.0. 

Below we briefly comment on a small subset of the 30 interacting galaxies from Table \ref{table:intgals}. Full discussion and LBT/\hst{} imaging of these galaxies can be found in Appendix \ref{ap:mergers}. 
\begin{itemize}
    \item Galaxy 1: This galaxy appears to have both a plume and a tidal tail in the LBT $r$-band (Figure \ref{figure:gal1}). The plume appears above the galaxy, while the tidal tail extends towards the lower left. These features are not easily identifiable in \hst{} imaging, but become apparent after smoothing the F160W data. When looking at high resolution imaging with \hst{} F336W and F275W, this galaxy appears to be composed of multiple clumps. 

    \item Galaxy 3: This galaxy has a diffuse tail above the galaxy that is present in both the LBT and \hst{} imaging (Figure \ref{figure:gal3}). However, the LBT \rb{} image also shows a second tidal tail below the galaxy that is not present in the \hst{} imaging. 

    \item Galaxy 6: LBT \rb{} imaging of this galaxy shows diffuse, tidal debris to the right side of the galaxy, which is not easily identifiable in the \hst{} imaging (Figure \ref{figure:gal5}). The tidal debris appears to make a loop around the galaxy in the deep \rb{} image.
\end{itemize}
These examples show that deep \rb{} imaging with the LBT/LBC is critical for finding low surface brightness regions such as tidal tails, plumes, and streams that are not otherwise identifiable with \hst{} imaging. These data, along with multiwavelength \hst{} imaging and the \Ub data from \citet{Ashcraft2018} should be used in future studies to study the ages, colors, and properties of these galaxy interactions. 


\begin{table*}[t]
\begin{center}
\caption{List of Interacting and Merging Galaxies in the GOODS-N  \label{table:intgals}} 
\begin{tabular}{ccccc|ccccc} \tableline \tableline
Number & Interaction Type & \emph{z} & RA & DEC & Number & Interaction Type & \emph{z} & RA & DEC\\      
\tableline
 1   &  Antennae        & 0.457$^1$    & 189.143651& 62.203670& 16   & Diffuse &  0.48$^7$    & 189.546198& 62.073688\\
 2   & Tidal Tail       &  0.375$^2$   & 188.991511& 62.260236& 17   & Equal &  0.034$^1$     & 189.046012& 62.275677\\
 3   &  Diffuse         &    0.530$^3$ & 189.308326& 62.343553& 18   & Diffuse & 0.560$^8$    & 189.531696& 62.088388\\
 4   & Shrimp/Equal     &   0.299$^1$  & 189.399092& 62.302951& 19   & Tidal Tail &   --      & 188.878151& 62.105181\\
 5   &  M51             &    0.253$^1$ & 189.220288& 62.302170& 20   & Diffuse &  --          & 188.849864& 62.105759\\
 6   &  Tidal Tail      & 0.306$^1$    & 189.469709& 62.274555& 21   & Diffuse/Tidal Tail& -- & 189.316382& 62.440579\\
 7   & Equal            &   0.637$^4$  & 189.027615& 62.164315& 22   &  Equal/Diffuse &   --  & 189.279264& 62.454440\\
 8   & Diffuse          &  0.440$^1$   & 189.446132& 62.275576& 23   & Tidal Tail &   --      & 189.270045& 62.458354\\
 9   &  Tidal Tail/M51  &   0.377$^1$  & 189.420590& 62.254833& 24   & Tidal Tail  &   --     & 189.495589& 62.399123\\
10   & Shrimp           & --           & 189.080363& 62.301563& 25   & Assembly &   --        & 189.183626& 62.357549\\
11   & Tidal Tail/Diffuse & 0.798$^5$  & 189.191198& 62.107990& 26   & Diffuse &   --         & 189.208981& 62.458493\\
12   & Diffuse          &   --         & 189.222929& 62.103161& 27   & Tidal Tail & 0.334$^8$ & 189.216085& 62.427977\\
13   & Assembly         &   0.9374$^4$ & 189.195308& 62.116697& 28   & Antennae  &   --       & 189.153633& 62.414972\\
14   & Antennae         & 0.277$^6$    & 189.622461& 62.280396& 29   & Diffuse &   --         & 189.086340& 62.419340\\
15   & Equal/Tidal Tail &  0.456       & 189.547345& 62.175356& 30   & M51 &  --              & 189.145952& 62.453802\\
\tableline
\end{tabular}
\end{center}
\vspace{-0.5cm}
\tablecomments{(1) \citet{Wirth2004}; (2) \citet{Barger2002}; (3) \citet{Moran2007}; 
(4) \citet{Barger2008}; (5) \citet{Treu2005}; (6) \citet{Casey2012}; (7) \citet{Rafferty2011}; (8) \citet{Albareti2017} }
\end{table*}


\subsection{Implications for the Extragalactic Background Light}
\label{sec:sbprofilesr}

The powerful ability of the optimal depth mosaics to highlight extended, low surface brightness (SB) features in the outskirts of galaxies can be utilized to investigate the contribution of diffuse galaxy light to the EBL. For galaxies brighter than \mab$\simeq$21.5 mag, the azimuthally averaged radial surface brightness (SB) profiles were measured using the custom \idl program \texttt{galprof\footnote{http://www.public.asu.edu/~rjansen/idl/galprof1.0/galprof.pro}} for both the optimal resolution and optimal depth \rb{} mosaics. This left a sample of 360 galaxies suitable for the surface brightness analysis after eliminating galaxies in close proximity to bright stars or the edge of the FOV. The 360 galaxy profiles were analyzed and the excess light in the optimal depth profile was ranked as ``confident'', ``potential'', or ``identical''. Surface brightness profiles ranked as ``confident'' exhibited a $\sim$1.0 mag arcsec$^{-2}$ difference in the two profiles over multiple radial points or had multiple data points between the two profiles that were separated by more than the 1$\sigma$ uncertainty ranges plotted. Profiles ranked as ``potential'' were classified by a $\sim$0.5 mag arcsec$^{-2}$ difference between the two profiles. However, typically the uncertainty ranges in the optimal depth and resolution data points encompassed the corresponding data point, which did not allow for as confident of a classification. The majority of the surface brightness profiles were identified as ``identical'', where there was no apparent difference between the optimal depth and optimal resolution profiles. 

Prior to analyzing the surface brightness profiles, we measured the surface brightness sky limits at which the two mosaics began to significantly differ. In order to accomplish this, model galaxies with pure exponential disk profiles matching the size of actual galaxies with \mab$\simeq$19--21.5 mag were created. Approximately 250 non-saturated stars
were used to create a model PSF-star for both the optimal resolution and optimal depth mosaics. The model galaxies were then convolved with the corresponding PSF and random background pixels were sampled to create a background map. Then, \texttt{galprof} was run to create surface brightness profiles of the model galaxies. In this analysis, we find the optimal resolution and depth profiles deviated at surface brightness levels of \sbabr$\sim$31 mag arcsec$^{-2}$.

\begin{figure*}[t!]
\centerline{
 \includegraphics[width=0.9\txw]{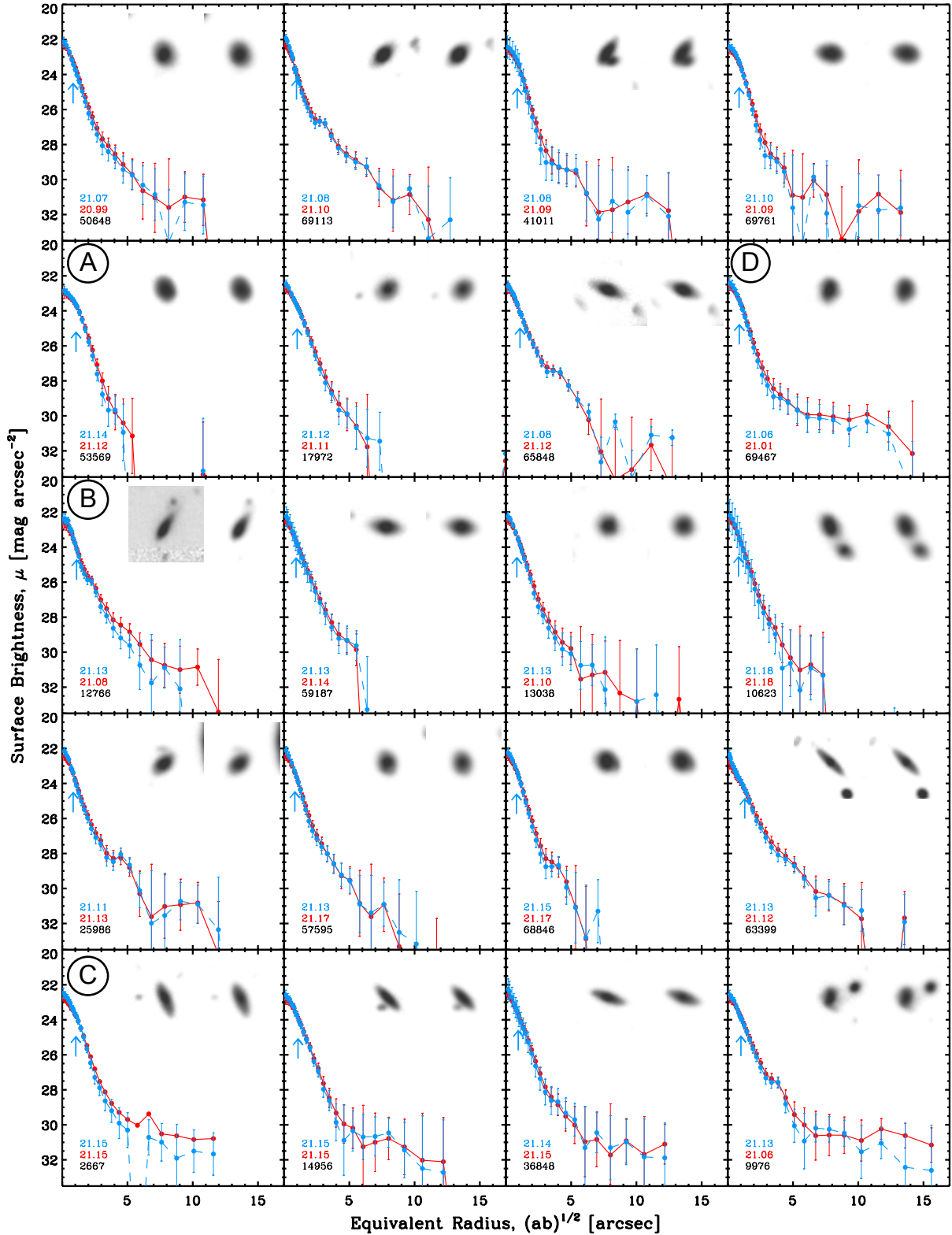}
 }
\caption{\small\noindent 
Radial surface brightness profiles for 20 of the 360 brightest objects with \rab{}$\lesssim$21.5~mag. The blue data points show the surface brightness profile for the optimal resolution image, while the red points show the corresponding profile for the optimal depth image. The total integrated \rab magnitudes from the profiles are listed in the lower left corner. The blue arrow represents the half-light radius measured with \sextractor{}. The galaxy inset images show the optimal resolution (left) and optimal depth (right) images. Galaxies A, B, C, and D clearly show excess flux in the outskirts of the optimal depth surface brightness profiles at levels brighter than \rab{}$\lesssim$31~mag. The complete figure set showing all 360 surface brightness profiles is available in the online journal.}
\label{figure:sbpr}
\end{figure*}

Figure \ref{figure:sbpr} shows a representative selection of 20 out of 360 surface brightness profiles for galaxies in the optimal resolution (blue) and optimal depth (red) mosaics. 16/20 do not exhibit any distinguishable difference between the two profiles to \sbabr$\lesssim$31~mag~arcsec$^{-2}$ between the profiles and were categorized as ``identical''. Galaxies A, B, C, and D are four examples of galaxies that were ranked as ``confident'' or ``potential'' candidates for having significant additional light in the optimal depth surface brightness profiles. Galaxies A and D were categorized as ``potential'' since they exhibited $\sim$0.5 mag arcsec$^{-2}$ differences in surface brightness with both profiles being within the uncertainty ranges of the other. However, galaxies B and C show larger, more robust differences between the optimal depth and resolution profiles\footnote{The bottom right profiles in Figure \ref{figure:sbpr} were not labeled as ``potential'' or ``confident'' as the optimal depth profile was not consistently brighter than the optimal depth.}.  

\citet[Fig. 2]{Driver2016} showed that the number density of galaxies in the \rb{} peaks at \mab$\simeq$\,19--24 mag. Thus, this subset of galaxies constitutes a representative sample of galaxies which significantly contributes to the EBL to surface brightness levels of \sbabr$\lesssim$ 31 mag arcsec$^{-2}$, where the background levels of the optimal depth and optimal resolution mosaics begin to differ. Of the 360 galaxies with surface brightness profiles, only 19 were labeled as ``confident'' and 32 galaxies were labeled as ``potential''. Therefore, 5-14\% of galaxies in this sample have excess flux in their outskirts out to surface brightness levels of \sbabr$\lesssim$ 31 mag arcsec$^{-2}$, which could contribute to missing, diffuse EBL light as summarized by \citet{Driver2010, Driver2016} and \citet{Windhorst2018}. However, the possibility of more uniform, missing flux from all galaxies cannot be ruled out as it would be diffuse enough across the LBT FOV where it would be removed during the \swarp{} background subtraction process. This excess \rb{} light could be the result of an older population of stars in the galaxy outskirts, star formation being traced by H$\alpha$ out to redshifts of $z \lesssim$~0.2, or tidal tails from galaxy interactions.

\begin{figure}
\includegraphics[width=\columnwidth]{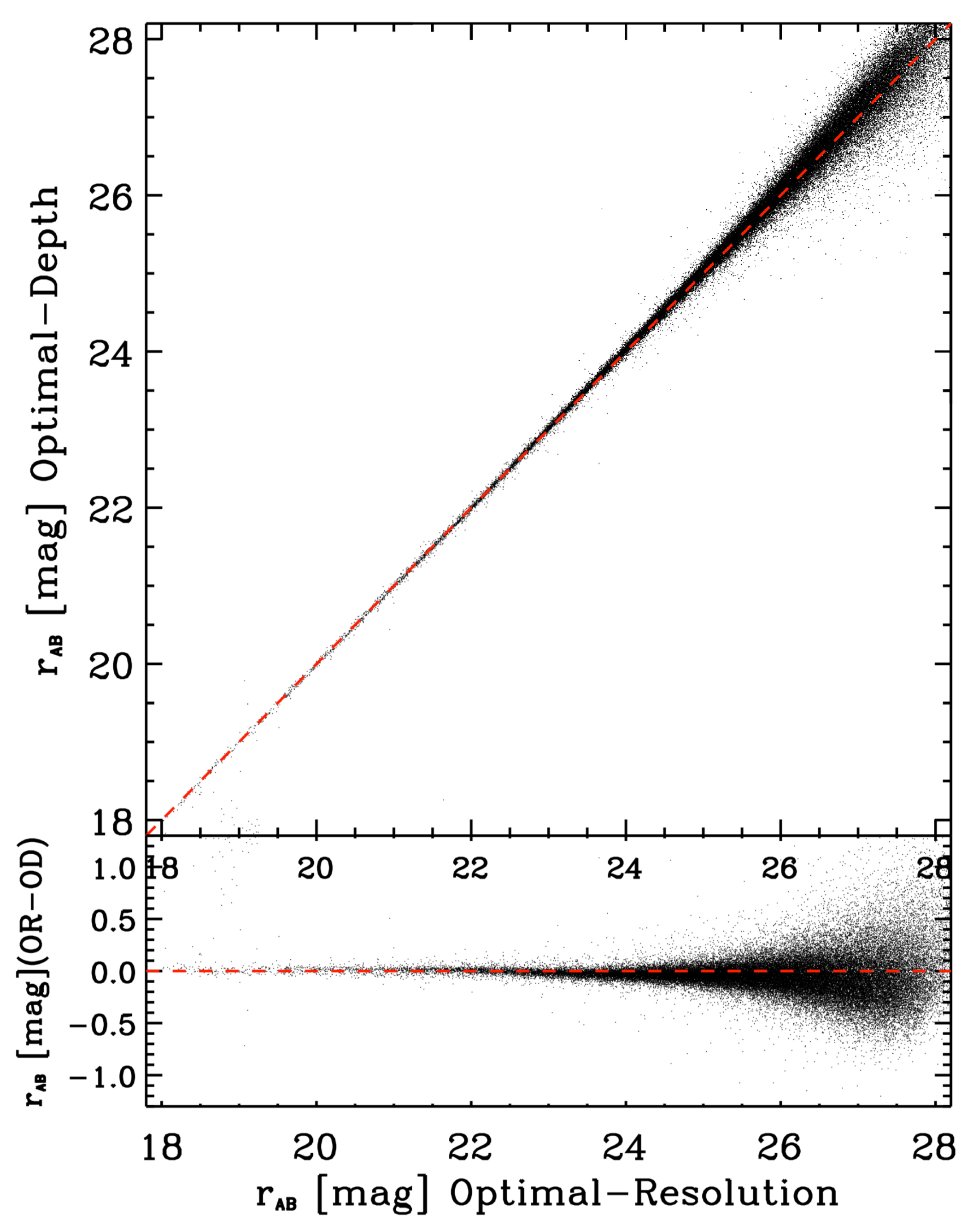}
\caption{\noindent\small 
Comparison of total \rb{} magnitudes measured by \sextractor, using dual image
mode, in the optimal resolution (OR) and the optimal depth (OD) mosaics. The bottom panel shows the OR - OD magnitude difference objects in the two catalogs. On average, there is little excess light in the optimal depth images than observed in the corresponding optimal resolution image. This aligns with what is observed in the surface brightness profile analysis where $\lesssim$10--14\% of galaxies are observed to have excess \rb{} flux in their outskirts. }
\label{figure:mvsmr}
\end{figure}

This fraction of galaxies with excess light in the optimal depth images is also represented in Figure \ref{figure:mvsmr}. These galaxies are most apparent in the bottom panel where the magnitude difference between the optimal resolution and optimal depth images for galaxies brighter than 21.5 mag. This panel shows a population of $\sim$19 galaxies with optimal resolution -- optimal depth magnitudes greater than 1.0 mag, which corresponds to the number of surface brightness profiles ranked as ``confident''. There are $\sim$30 galaxies that exist with a magnitude difference greater than 0.5 mag, which would match up with $\sim$50\% of the ``potential'' profiles as having significant excess light in in the optimal depth surface brightness profiles. 

In addition to the surface brightness profile analysis, the total contribution of light in galaxy outskirts to the EBL was calculated by integrating the normalized galaxy counts up to the optimal resolution and optimal depth completion limits of $\sim$27 mag following the methods in \citet{Driver2016, Carleton2022} and \citet{Windhorst2022}. The contribution to the EBL from the outskirts of galaxies is represented by the difference between the total energy from the optimal depth and optimal resolution integrated counts. This analysis showed that the total EBL contribution is 0.19\,$\rm{nW\,m^{-2}\,sr^{-1}}$ (3.52 - 3.33\,$\rm{nW\,m^{-2}\,sr^{-1}}$), which is $\sim5$\% of the total EBL in the Sloan $r$-band. \textit{Since these three independent methods of searching for additional light in the outskirts of galaxies provide similar results, it can confidently be stated that only a small fraction ($\sim$ 5-14\%) of extra light in galaxy outskirts is available to contribute towards missing EBL light.}

\figsetstart
\figsetnum{9}
\figsettitle{Surface brightness profiles with \rab{}$\simeq$21.5~mag}

\figsetgrpnum{9.1}
\figsetgrptitle{SB profiles 1-20}
\figsetplot{Fig9.1.pdf}
\figsetgrpnote{Radial surface brightness profiles for the 360 brightest objects with \rab{}$\lesssim$21.5~mag. The blue data points show the surface brightness profile for the optimal resolution image, while the red points show the corresponding profile for the optimal depth image. The total integrated \rab magnitudes from the profiles are listed in the lower left corner. The blue arrow represents the half-light radius measured with \sextractor{}. The galaxy inset images show the optimal resolution (left) and optimal depth (right) images. }
\figsetgrpend

\figsetgrpnum{9.2}
\figsetgrptitle{SB profiles 21-40}
\figsetplot{Fig9.2.pdf}
\figsetgrpnote{Radial surface brightness profiles for the 360 brightest objects with \rab{}$\lesssim$21.5~mag. The blue data points show the surface brightness profile for the optimal resolution image, while the red points show the corresponding profile for the optimal depth image. The total integrated \rab magnitudes from the profiles are listed in the lower left corner. The blue arrow represents the half-light radius measured with \sextractor{}. The galaxy inset images show the optimal resolution (left) and optimal depth (right) images. }
\figsetgrpend

\figsetgrpnum{9.3}
\figsetgrptitle{SB profiles 41-60}
\figsetplot{Fig9.3.pdf}
\figsetgrpnote{Radial surface brightness profiles for the 360 brightest objects with \rab{}$\lesssim$21.5~mag. The blue data points show the surface brightness profile for the optimal resolution image, while the red points show the corresponding profile for the optimal depth image. The total integrated \rab magnitudes from the profiles are listed in the lower left corner. The blue arrow represents the half-light radius measured with \sextractor{}. The galaxy inset images show the optimal resolution (left) and optimal depth (right) images. }
\figsetgrpend

\figsetgrpnum{9.4}
\figsetgrptitle{SB profiles 61-80}
\figsetplot{Fig9.4.pdf}
\figsetgrpnote{Radial surface brightness profiles for the 360 brightest objects with \rab{}$\lesssim$21.5~mag. The blue data points show the surface brightness profile for the optimal resolution image, while the red points show the corresponding profile for the optimal depth image. The total integrated \rab magnitudes from the profiles are listed in the lower left corner. The blue arrow represents the half-light radius measured with \sextractor{}. The galaxy inset images show the optimal resolution (left) and optimal depth (right) images. }
\figsetgrpend

\figsetgrpnum{9.5}
\figsetgrptitle{SB profiles 81-100}
\figsetplot{Fig9.5.pdf}
\figsetgrpnote{Radial surface brightness profiles for the 360 brightest objects with \rab{}$\lesssim$21.5~mag. The blue data points show the surface brightness profile for the optimal resolution image, while the red points show the corresponding profile for the optimal depth image. The total integrated \rab magnitudes from the profiles are listed in the lower left corner. The blue arrow represents the half-light radius measured with \sextractor{}. The galaxy inset images show the optimal resolution (left) and optimal depth (right) images. }
\figsetgrpend

\figsetgrpnum{9.6}
\figsetgrptitle{SB profiles 101-120}
\figsetplot{Fig9.6.pdf}
\figsetgrpnote{Radial surface brightness profiles for the 360 brightest objects with \rab{}$\lesssim$21.5~mag. The blue data points show the surface brightness profile for the optimal resolution image, while the red points show the corresponding profile for the optimal depth image. The total integrated \rab magnitudes from the profiles are listed in the lower left corner. The blue arrow represents the half-light radius measured with \sextractor{}. The galaxy inset images show the optimal resolution (left) and optimal depth (right) images. }
\figsetgrpend

\figsetgrpnum{9.7}
\figsetgrptitle{SB profiles 121-140}
\figsetplot{Fig9.7.pdf}
\figsetgrpnote{Radial surface brightness profiles for the 360 brightest objects with \rab{}$\lesssim$21.5~mag. The blue data points show the surface brightness profile for the optimal resolution image, while the red points show the corresponding profile for the optimal depth image. The total integrated \rab magnitudes from the profiles are listed in the lower left corner. The blue arrow represents the half-light radius measured with \sextractor{}. The galaxy inset images show the optimal resolution (left) and optimal depth (right) images. }
\figsetgrpend

\figsetgrpnum{9.8}
\figsetgrptitle{SB profiles 141-160}
\figsetplot{Fig9.8.pdf}
\figsetgrpnote{Radial surface brightness profiles for the 360 brightest objects with \rab{}$\lesssim$21.5~mag. The blue data points show the surface brightness profile for the optimal resolution image, while the red points show the corresponding profile for the optimal depth image. The total integrated \rab magnitudes from the profiles are listed in the lower left corner. The blue arrow represents the half-light radius measured with \sextractor{}. The galaxy inset images show the optimal resolution (left) and optimal depth (right) images. }
\figsetgrpend

\figsetgrpnum{9.9}
\figsetgrptitle{SB profiles 161-180}
\figsetplot{Fig9.9.pdf}
\figsetgrpnote{Radial surface brightness profiles for the 360 brightest objects with \rab{}$\lesssim$21.5~mag. The blue data points show the surface brightness profile for the optimal resolution image, while the red points show the corresponding profile for the optimal depth image. The total integrated \rab magnitudes from the profiles are listed in the lower left corner. The blue arrow represents the half-light radius measured with \sextractor{}. The galaxy inset images show the optimal resolution (left) and optimal depth (right) images. }
\figsetgrpend

\figsetgrpnum{9.10}
\figsetgrptitle{SB profiles 181-200}
\figsetplot{Fig9.10.pdf}
\figsetgrpnote{Radial surface brightness profiles for the 360 brightest objects with \rab{}$\lesssim$21.5~mag. The blue data points show the surface brightness profile for the optimal resolution image, while the red points show the corresponding profile for the optimal depth image. The total integrated \rab magnitudes from the profiles are listed in the lower left corner. The blue arrow represents the half-light radius measured with \sextractor{}. The galaxy inset images show the optimal resolution (left) and optimal depth (right) images. }
\figsetgrpend

\figsetgrpnum{9.11}
\figsetgrptitle{SB profiles 201-220}
\figsetplot{Fig9.11.pdf}
\figsetgrpnote{Radial surface brightness profiles for the 360 brightest objects with \rab{}$\lesssim$21.5~mag. The blue data points show the surface brightness profile for the optimal resolution image, while the red points show the corresponding profile for the optimal depth image. The total integrated \rab magnitudes from the profiles are listed in the lower left corner. The blue arrow represents the half-light radius measured with \sextractor{}. The galaxy inset images show the optimal resolution (left) and optimal depth (right) images. }
\figsetgrpend

\figsetgrpnum{9.12}
\figsetgrptitle{SB profiles 221-240}
\figsetplot{Fig9.12.pdf}
\figsetgrpnote{Radial surface brightness profiles for the 360 brightest objects with \rab{}$\lesssim$21.5~mag. The blue data points show the surface brightness profile for the optimal resolution image, while the red points show the corresponding profile for the optimal depth image. The total integrated \rab magnitudes from the profiles are listed in the lower left corner. The blue arrow represents the half-light radius measured with \sextractor{}. The galaxy inset images show the optimal resolution (left) and optimal depth (right) images. }
\figsetgrpend

\figsetgrpnum{9.13}
\figsetgrptitle{SB profiles 241-260}
\figsetplot{Fig9.13.pdf}
\figsetgrpnote{Radial surface brightness profiles for the 360 brightest objects with \rab{}$\lesssim$21.5~mag. The blue data points show the surface brightness profile for the optimal resolution image, while the red points show the corresponding profile for the optimal depth image. The total integrated \rab magnitudes from the profiles are listed in the lower left corner. The blue arrow represents the half-light radius measured with \sextractor{}. The galaxy inset images show the optimal resolution (left) and optimal depth (right) images. }
\figsetgrpend

\figsetgrpnum{9.14}
\figsetgrptitle{SB profiles 261-280}
\figsetplot{Fig9.14.pdf}
\figsetgrpnote{Radial surface brightness profiles for the 360 brightest objects with \rab{}$\lesssim$21.5~mag. The blue data points show the surface brightness profile for the optimal resolution image, while the red points show the corresponding profile for the optimal depth image. The total integrated \rab magnitudes from the profiles are listed in the lower left corner. The blue arrow represents the half-light radius measured with \sextractor{}. The galaxy inset images show the optimal resolution (left) and optimal depth (right) images. }
\figsetgrpend

\figsetgrpnum{9.15}
\figsetgrptitle{SB profiles 281-300}
\figsetplot{Fig9.15.pdf}
\figsetgrpnote{Radial surface brightness profiles for the 360 brightest objects with \rab{}$\lesssim$21.5~mag. The blue data points show the surface brightness profile for the optimal resolution image, while the red points show the corresponding profile for the optimal depth image. The total integrated \rab magnitudes from the profiles are listed in the lower left corner. The blue arrow represents the half-light radius measured with \sextractor{}. The galaxy inset images show the optimal resolution (left) and optimal depth (right) images. }
\figsetgrpend

\figsetgrpnum{9.16}
\figsetgrptitle{SB profiles 301-320}
\figsetplot{Fig9.16.pdf}
\figsetgrpnote{Radial surface brightness profiles for the 360 brightest objects with \rab{}$\lesssim$21.5~mag. The blue data points show the surface brightness profile for the optimal resolution image, while the red points show the corresponding profile for the optimal depth image. The total integrated \rab magnitudes from the profiles are listed in the lower left corner. The blue arrow represents the half-light radius measured with \sextractor{}. The galaxy inset images show the optimal resolution (left) and optimal depth (right) images. }
\figsetgrpend

\figsetgrpnum{9.17}
\figsetgrptitle{SB profiles 321-340}
\figsetplot{Fig9.17.pdf}
\figsetgrpnote{Radial surface brightness profiles for the 360 brightest objects with \rab{}$\lesssim$21.5~mag. The blue data points show the surface brightness profile for the optimal resolution image, while the red points show the corresponding profile for the optimal depth image. The total integrated \rab magnitudes from the profiles are listed in the lower left corner. The blue arrow represents the half-light radius measured with \sextractor{}. The galaxy inset images show the optimal resolution (left) and optimal depth (right) images. }
\figsetgrpend

\figsetgrpnum{9.18}
\figsetgrptitle{SB profiles_341-360}
\figsetplot{Fig9.18.pdf}
\figsetgrpnote{Radial surface brightness profiles for the 360 brightest objects with \rab{}$\lesssim$21.5~mag. The blue data points show the surface brightness profile for the optimal resolution image, while the red points show the corresponding profile for the optimal depth image. The total integrated \rab magnitudes from the profiles are listed in the lower left corner. The blue arrow represents the half-light radius measured with \sextractor{}. The galaxy inset images show the optimal resolution (left) and optimal depth (right) images. }
\figsetgrpend

\figsetend


\noindent
\vspace*{-\baselineskip}

\section{SUMMARY AND CONCLUSIONS} \label{sec:summaryr}

838 \rb{} exposures ($\sim$ 28 hours) were obtained between December 2012 and January 2014 of GOODS-N in order to examine the trade-off between optimal image resolution and depth. Following the seeing sorted stacking method detailed in \citet{Ashcraft2018}, the best seeing images were stacked to create optimal depth and optimal resolution mosaics. The optimal depth mosaic was found to be necessary for the study of low surface brightness regions presented in this work. The sacrifice in resolution was complemented by increased surface brightness sensitivity down to \sbabr$\leq31$~mag~arcsec$^{-2}$.

The main challenge for photometric measurements was to overcome the natural confusion limit for separating objects, which occurs once faint objects are closer than $\sim1\farcs0$ in ground based images \citep{Windhorst2011}. In order to accurately deblend sources in the mosaics, \sextractor{} in dual image mode was utilized in order to deblend and detect objects using the optimal resolution mosaic. Within the optimal depth and optimal resolution mosaics, objects can be detected to \mab$\sim$29 and \mab$\sim$28.5 magnitudes, respectively. 

A collection of 30 candidate interacting galaxies were investigated along with their low-surface brightness features. For the majority of these systems, the galaxies were not resolved sufficiently for detailed morphological classification. For the sample galaxies that did fall within the \hst{} footprint, the higher resolution images were used for interaction classification following the methods of \citet{Elmegreen2007} and the diffuse flux/tidal tails were described. Future studies can utilize the LBT \Ub and \rb{} mosaics along with additional filters to continue to study the properties of these 30 interacting galaxies. Specifically, the colors of the diffuse plumes and tidal tails can be measured to gain more insight into the age and properties of these stellar populations.

Lastly, surface brightness profiles were measured for the 360 brightest galaxies with \rab{}$\lesssim$ 21.5 mag in both the optimal resolution and optimal depth mosaics. Galaxies with magnitudes \mab$\lesssim$ 21.5 provide a representative sample of galaxies that could contribute significantly to the Extragalactic Background Light in the $r$-band \citep{Driver2016}. These surface brightness profiles show marginal differences between the optimal resolution and optimal depth mosaics to surface brightnesses of \sbabr$\sim$31 mag arcsec$^{-2}$. Only 19/360 galaxies confidently exhibited excess flux in the optimal depth radial profiles, while another 32/360 galaxies were categorized as ``potentially'' having excess flux in the outskirts. 

As a result, on average, we conclude that only $\sim$5-14\% of extra light in the outskirts of galaxies are likely to contribute to the Extragalactic Background Light out to surface brightness levels of $\sim$31 mag arcsec$^{-2}$. The EBL contribution from the outskirts of galaxies was found to be $\sim$5\% of the EBL determined from the integrated galaxy counts. We find that while there is some contribution to the EBL from diffuse light in galaxy outskirts, there is not enough of a contribution in the $r$-band to close the discrepancy between direct EBL measurements \citep{Puget1996, Hauser1998, Matsumoto2005, Matsumoto2018, Lauer2021, Korngut2022} and predicted values \citep{Driver2011, Driver2016, Andrews2018}.

\begin{acknowledgments}
\vspace{-0.3cm}
The Arizona State University authors acknowledge the twenty two Native Nations that have inhabited this land for centuries. Arizona State University’s four campuses are located in the Salt River Valley on ancestral territories of Indigenous peoples, including the Akimel O’odham (Pima) and Pee Posh (Maricopa) Indian Communities, whose care and keeping of these lands allows us to be here today. We acknowledge the sovereignty of these nations and seek to foster an environment of success and possibility for Native American students and patrons.

The LBT is an international collaboration among institutions in the United
States, Italy, and Germany. LBT Corporation partners are The University of
Arizona on behalf of the Arizona university system; Istituto Nazionale di
Astrofisica, Italy; LBT Beteiligungsgesellschaft, Germany, representing the
Max-Planck Society, the Astrophysical Institute Potsdam, and Heidelberg
University; The Ohio State University; and The Research Corporation, on behalf
of The University of Notre Dame, University of Minnesota, and University of
Virginia. R. A. Windhorst acknowledges support from NASA JWST Interdisciplinary
Scientist grants NAG5-12460, NNX14AN10G and 80NSSC18K0200 from GSFC.

The authors acknowledge support from UVCANDELS grant HST-GO-15647 provided by NASA through the Space Telescope Science Institute, which is operated by the Association of Universities for Research in Astronomy, Inc., under NASA contract NAS 5-2655, and from NASA JWST Interdisciplinary Scientist grants NAG5-12460, NNX14AN10G and 80NSSC18K0200 from GSFC.

\end{acknowledgments}

\newpage
\bibliography{References}
\bibliographystyle{aasjournal}

\appendix

\section{Tidal Tails and Mergers}
\label{ap:mergers}

Below, we briefly describe the full sample of different tidal tail and merging galaxy
candidates. Galaxies which reside within the \hst{} FOV are shown first. 

\begin{enumerate}


\begin{figure}
\includegraphics[width=\columnwidth]{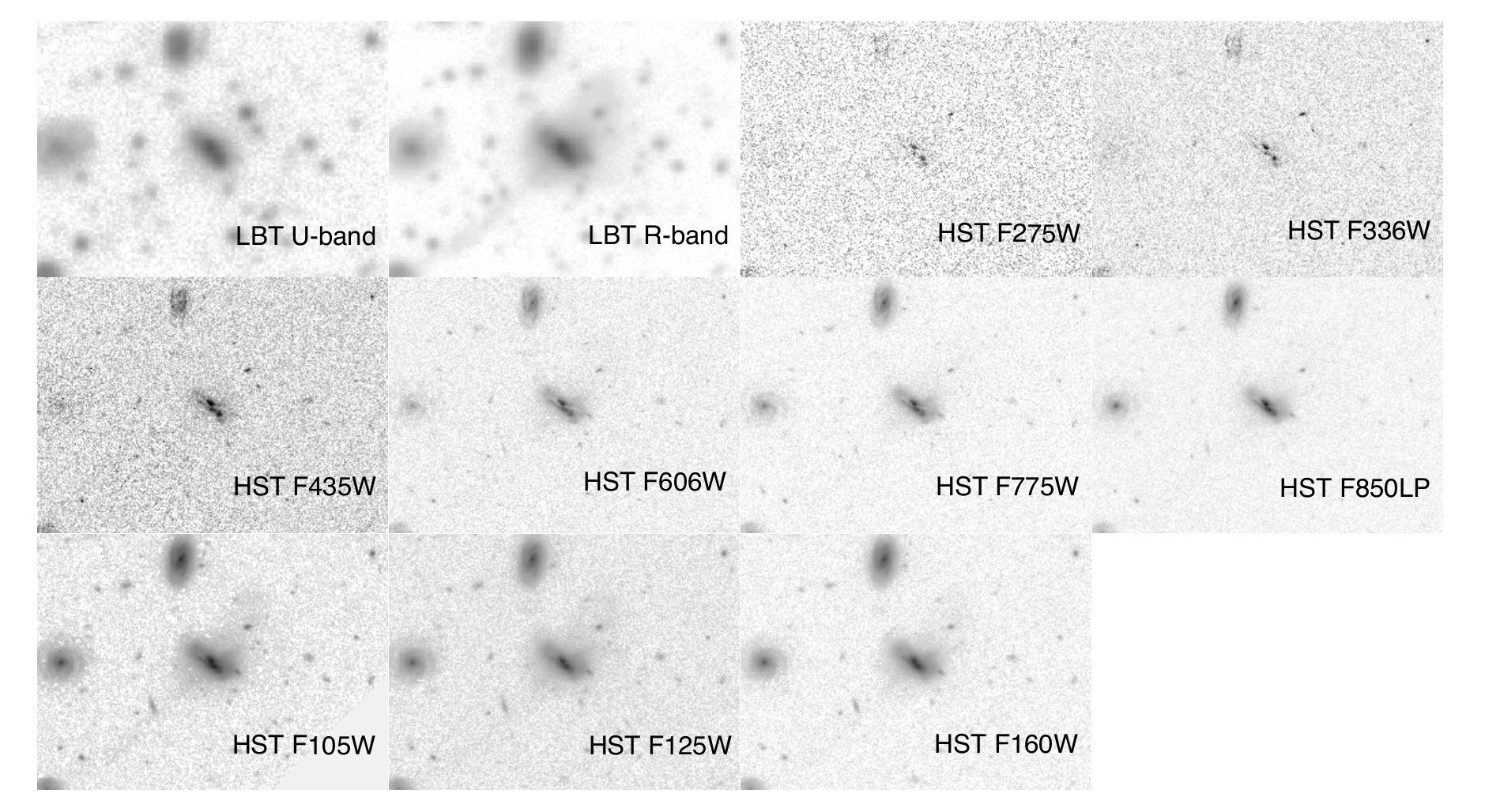}
\noindent\rule{\textwidth}{1pt}
\includegraphics[width=\columnwidth]{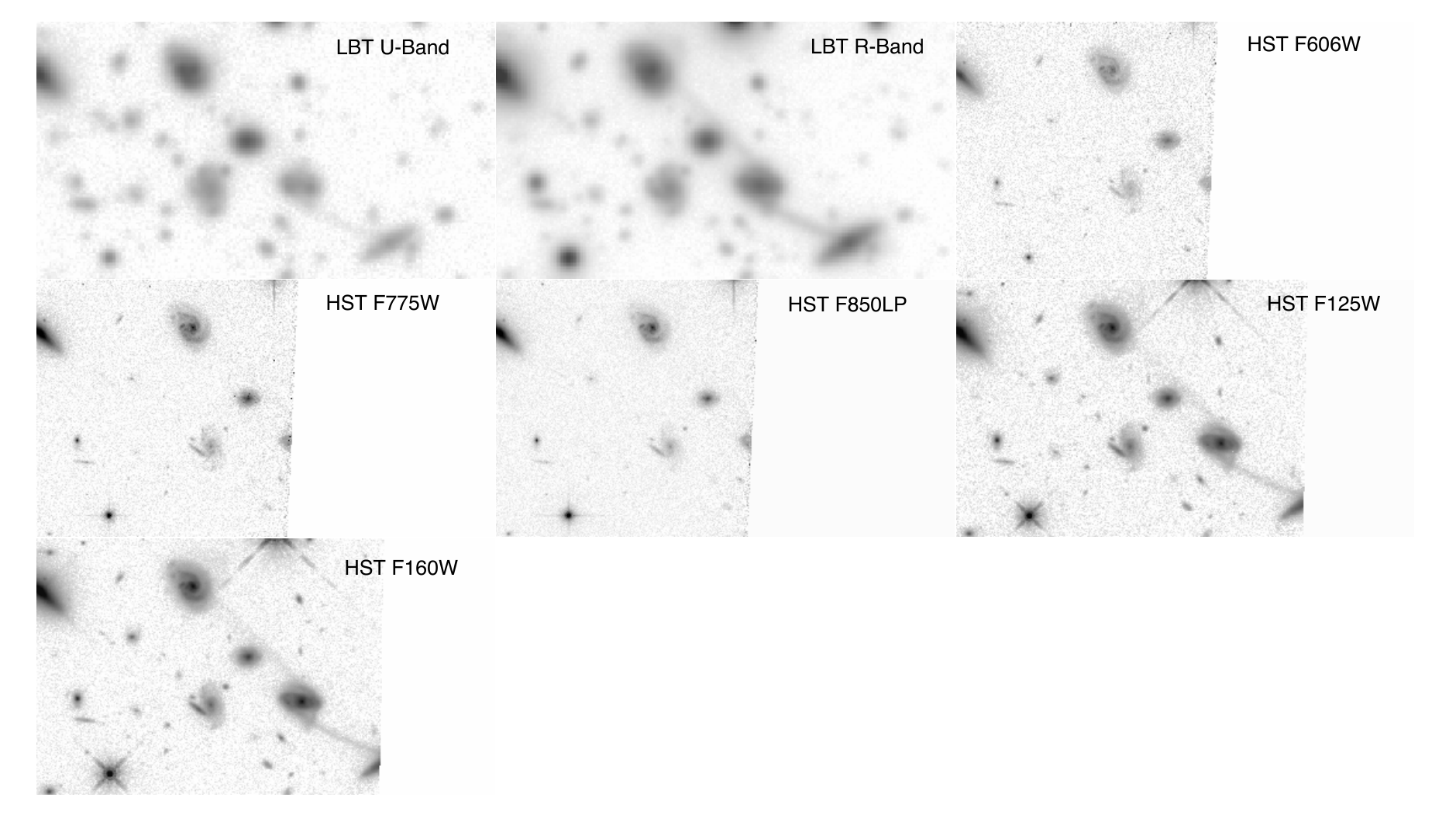}
\caption{\small\noindent
Galaxy 1 (top): A galaxy with long extended tidal tail and diffuse flux seen in
LBT \rb{} and has a $z_{spec}$\,=\,0.457. The galaxy is visible in \hst{}
images from F275W-F160W, with only a double lobe being visible in the UV. The
diffuse flux and tidal tail is visible in the NIR wavelengths, especially
bright in the F105W. Galaxy 2 (bottom): Multiple spiral galaxies interacting
with two long tidal tails visible linking the galaxies. The galaxies all have
either spectroscopic or photometric $z\sim$\,0.375. This system is only
partially visible in \hst{} imaging with best views being in the J and H-bands.
Only one of the tidal streams is clearly visible in the \Uz-band.}
\label{figure:gal1}
\end{figure}



\item In the $r$-band, there is a diffuse plume above the galaxy and a long
tidal tail downward to the left in the image (top galaxy in
Fig.\,\ref{figure:gal1}). This galaxy is inside the \hst{} FOV, but the tidal
debris is not clearly visible without smoothing the data, and is most prominent
in the F160W. It has a clear dust lane through the center perpendicular to the
tidal tails visible in the \hst{} optical wavelengths. In the F160W image, the
dust lane disappears, so that two distinct cores become visible. Multiple
brighter clumps of this galaxy are seen in the \hst{} F336W and F275W images.
Since it is detected in both x-ray \citep{Wang2016} and radio \citep{Biggs2006},
it may have outflows or cooling flows associated with an AGN. Other studies
have classified this galaxy as a starburst, high-excitation narrow-line radio
galaxy, and merging, and measured a redshift of $z_{spec}$\,=\,0.457 \citep{Wirth2004}.


\item Galaxy 2 shows a clear recent interaction with at least two galaxies,
possibly a third, with measured redshifts, z$\simeq$\,0.375 (bottom images in
Fig.\,\ref{figure:gal1}). There is a tidal tail between the two galaxies in the
bottom right, which is also visible in the LBT \Uz-band. Another tidal tail in
the opposite direction probably links the central galaxy to a third spiral
galaxy in the upper left. This tidal tail barely is detectable in the \Ub
mosaic, which could be due to multiple reasons including an older stellar
population in the interaction. Unfortunately, this system of galaxies is on the
edge of the \hst{} FOV and only the less dominant tidal tail is visible in the
\hst{} WFC3 IR images. 


\item An elliptical galaxy with a diffuse debris tail visible in the \rb{} and
partly in the \Uz-band mosaic (top galaxy in Fig.\,\ref{figure:gal3}). It has a
$z_{spec}$\,=\,0.530 \citep{Moran2007}. The diffuse flux above the main
galaxy has a similar photometric redshift to the main galaxy, and can been seen
in the \hst{} V to H-bands. There appears to be a tail of debris coming from
the bottom of the main elliptical galaxy, and the brightest clump in the stream
has a $z_{spec}$\,=\,0.533 \citep{Moran2007}.


\begin{figure}
\includegraphics[width=\columnwidth]{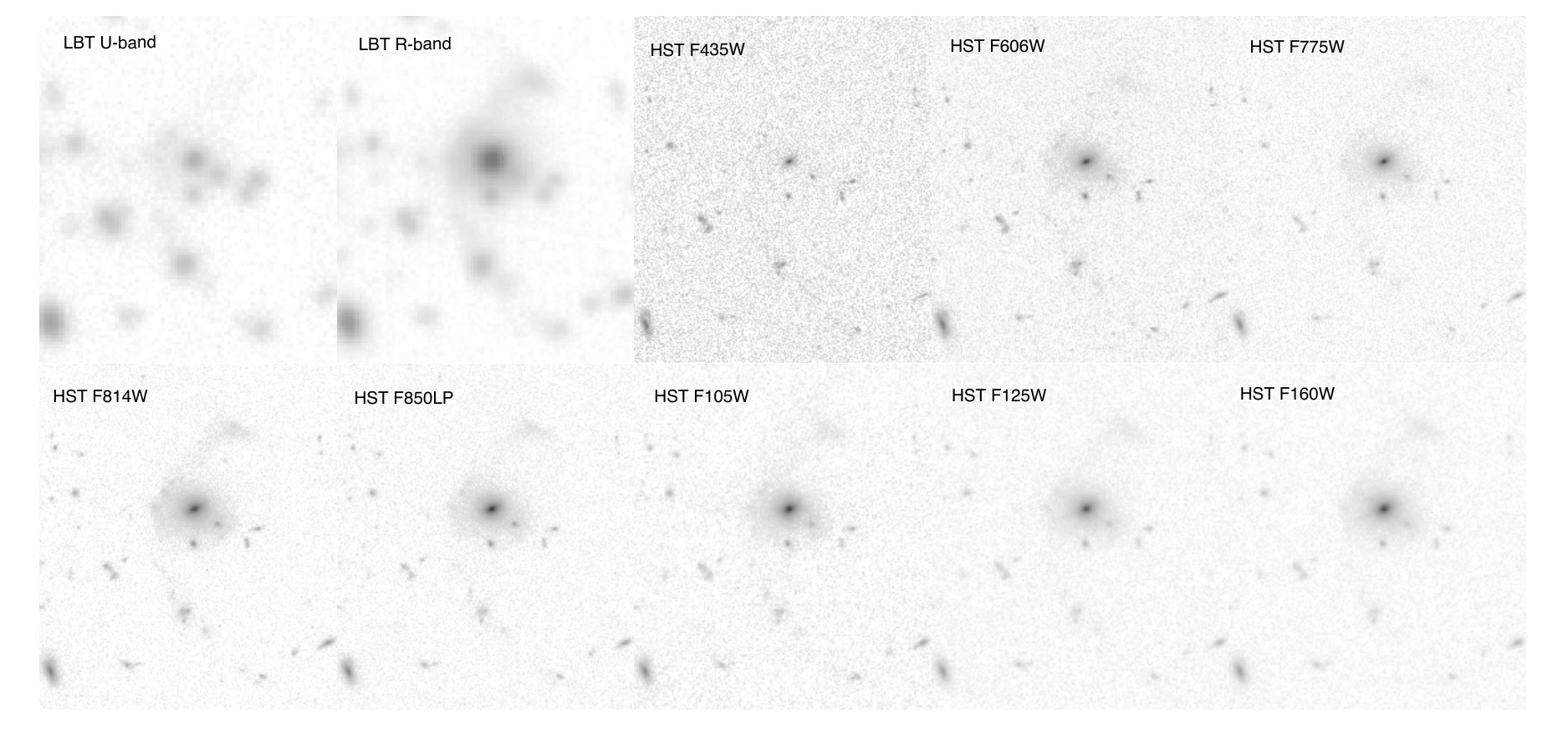}
\noindent\rule{\textwidth}{1pt}
\includegraphics[width=\columnwidth]{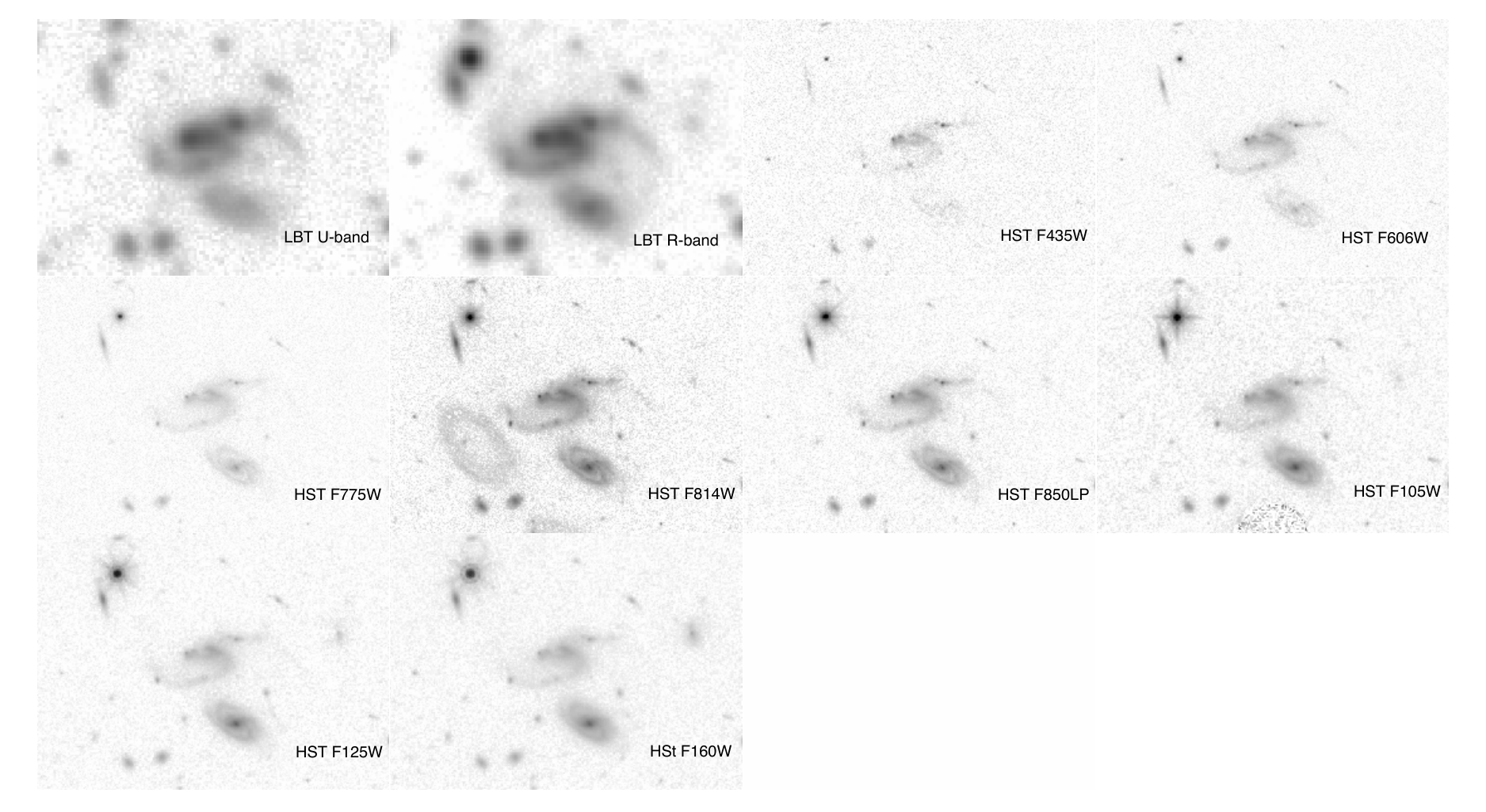}
\caption{\small\noindent 
Galaxy 3 (top): Elliptical galaxy with diffuse flux seen prominently above the
galaxy in \rb{} and only slightly in \Uz-band, and has a $z_{spec}$\,=\,0.530.
A smaller galaxy, out of the FOV in the figure, has the same spectroscopic
redshift, and could be associated with the tidal interaction. In addition,
there is a long tidal stream with clumps below the main galaxy. Galaxy 4
(bottom): Obvious merging system with multiple tidal arms and a
$z_{spec}$\,=\,0.299. From the the \hst{} images, there appears to be no
distinct core for the merging galaxy.}
\label{figure:gal3}
\end{figure}



\item Reminiscent of the nearby system, Stephan's Quintet, there are at least
three similar size spiral galaxies involved with this interaction (bottom set
of images in Fig.\,\ref{figure:gal3}). Two of the galaxies are already merging,
while a third appears to be gravitationally influencing one of the tidal tails. In the
\hst{} imaging, the tidal arms are visible in all bands, and the system has
$z_{spec}$= 0.299 \citep{Wirth2004}.


\item Similar to M51, galaxy 5 (top galaxy in Fig.\,\ref{figure:gal5}) has a
diffuse tidal stream on the bottom right linking it to a smaller companion
galaxy and with a $z_{spec}$\,=\,0.253 \citep{Wirth2004}. The \hst{} images
show a dust lane through the center of the galaxy and parallel to the diffuse
stream. In the centers of both the main and the companion galaxies, no
detectable flux is present in both the F336W and F275W filters.


\begin{figure}
\includegraphics[width=\columnwidth]{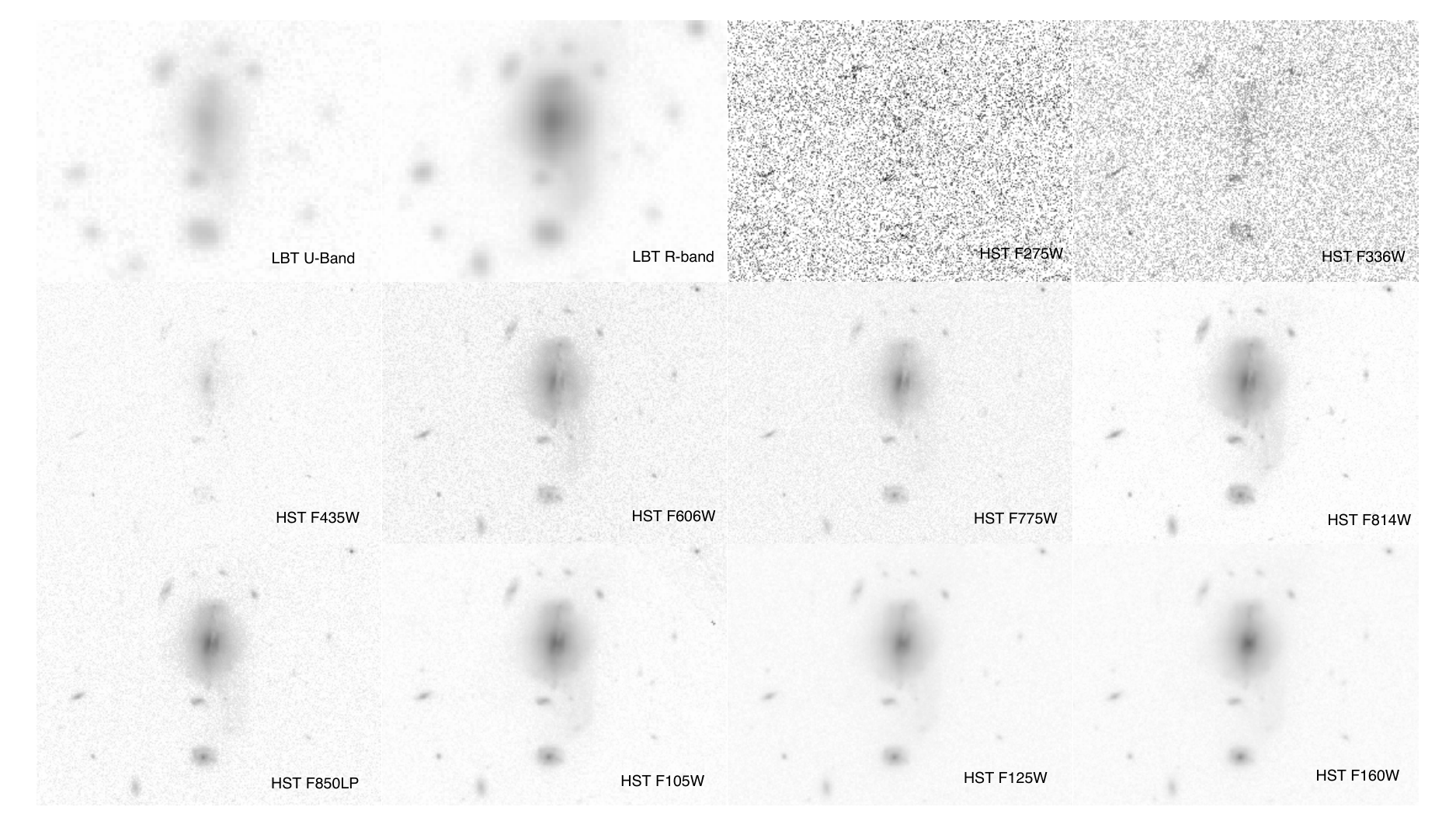}
\noindent\rule{\textwidth}{1pt}
\includegraphics[width=\columnwidth]{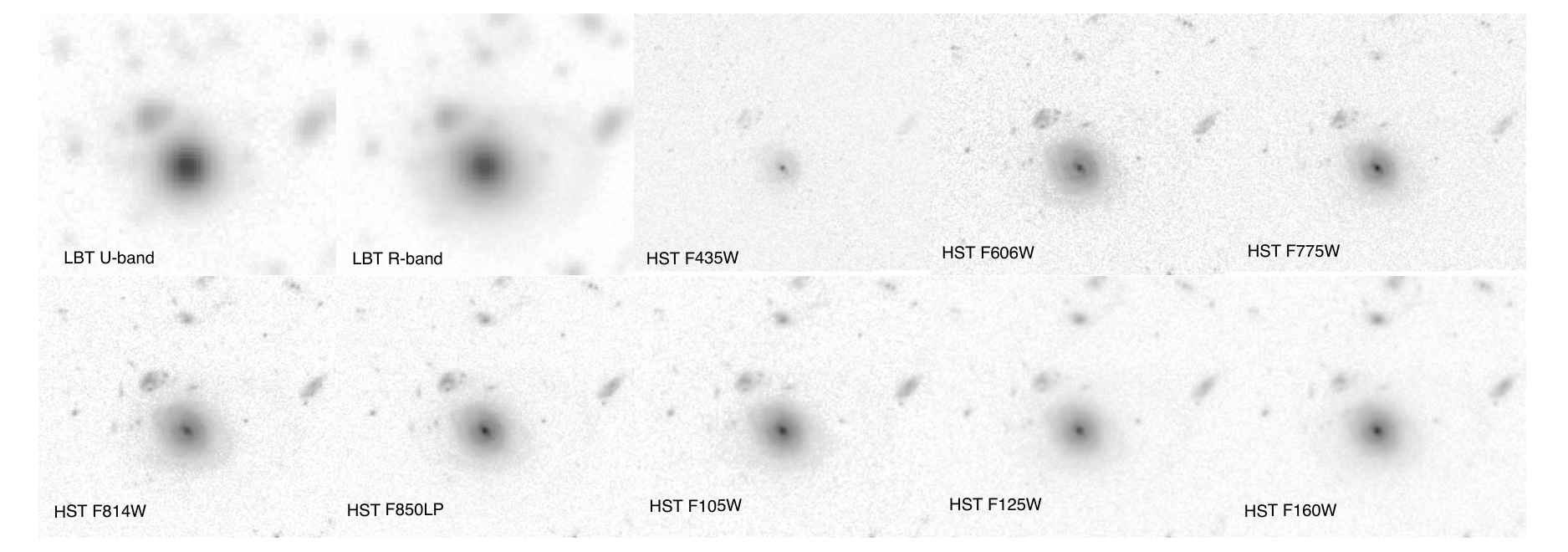}
\caption{\small\noindent 
Galaxy 5 (top): There is diffuse debris on the bottom right side toward a
smaller companion galaxy, similar to a M51 type system. This is one of only two
galaxies in our sample with both F275W and F336W imaging. It has
$z_{spec}$\,=\,0.253 \citep{Wirth2004}. Galaxy 6 (bottom): There is smooth
diffuse tidal debris to the right of the galaxy that appears to make a loop
from back to front, but is only seen in the LBT \rb{} and not the \Uz-band. The
\hst{} imaging shows a very bright core surrounded by a disk. It has
$z_{spec}$\,=\,0.306 \citep{Wirth2004}.}
\label{figure:gal5}
\end{figure}



\item There is smooth diffuse tidal debris visible in the LBT \rb{} to the
right of the galaxy that appears to make a loop around the galaxy, but is not
detectable in the LBT \Ub (bottom galaxy in Fig.\,\ref{figure:gal5}). From the
\hst{} imaging, it appears as a face on disk galaxy with a very bright center
and possible bar. The tidal debris is only marginally seen without smoothing.
It has been classified as a broad-line AGN and a measured $z_{spec}$\,=\,0.306
\citep{Wirth2004}.


\item An edge-on disk galaxy, which has been identified as an AGN (top galaxy
in Fig.\,\ref{figure:gal7}). It has extra diffuse flux upward and to the left,
which is seen in both \rb{} and $U$-band. In addition, it has a possible
approximately equal size grazing companion. The main galaxy has
$z_{spec}$\,=\,0.637 \citep{Barger2008}, but the potential companion galaxy
to the right is not spectroscopically confirmed to be at the same redshift.
However, the small galaxies to the left have $z_{spec}$\,=\,0.632
\citep{Barger2008}.


\begin{figure}[t]
\includegraphics[width=\columnwidth]{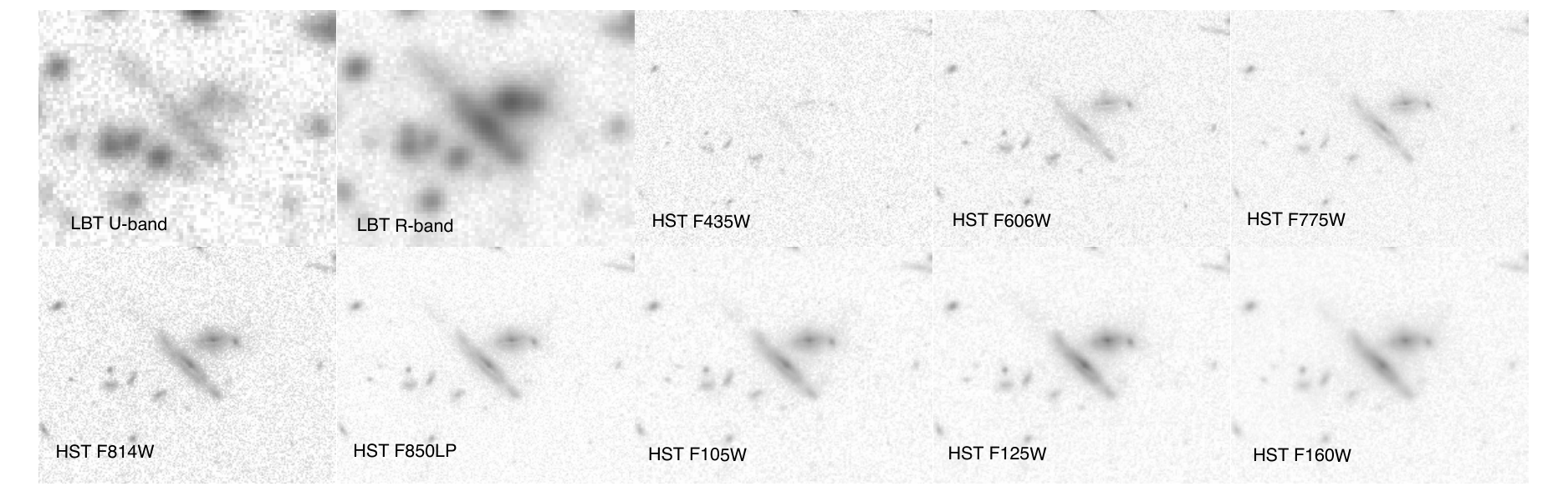}
\noindent\rule{\textwidth}{1pt}
\includegraphics[width=\columnwidth]{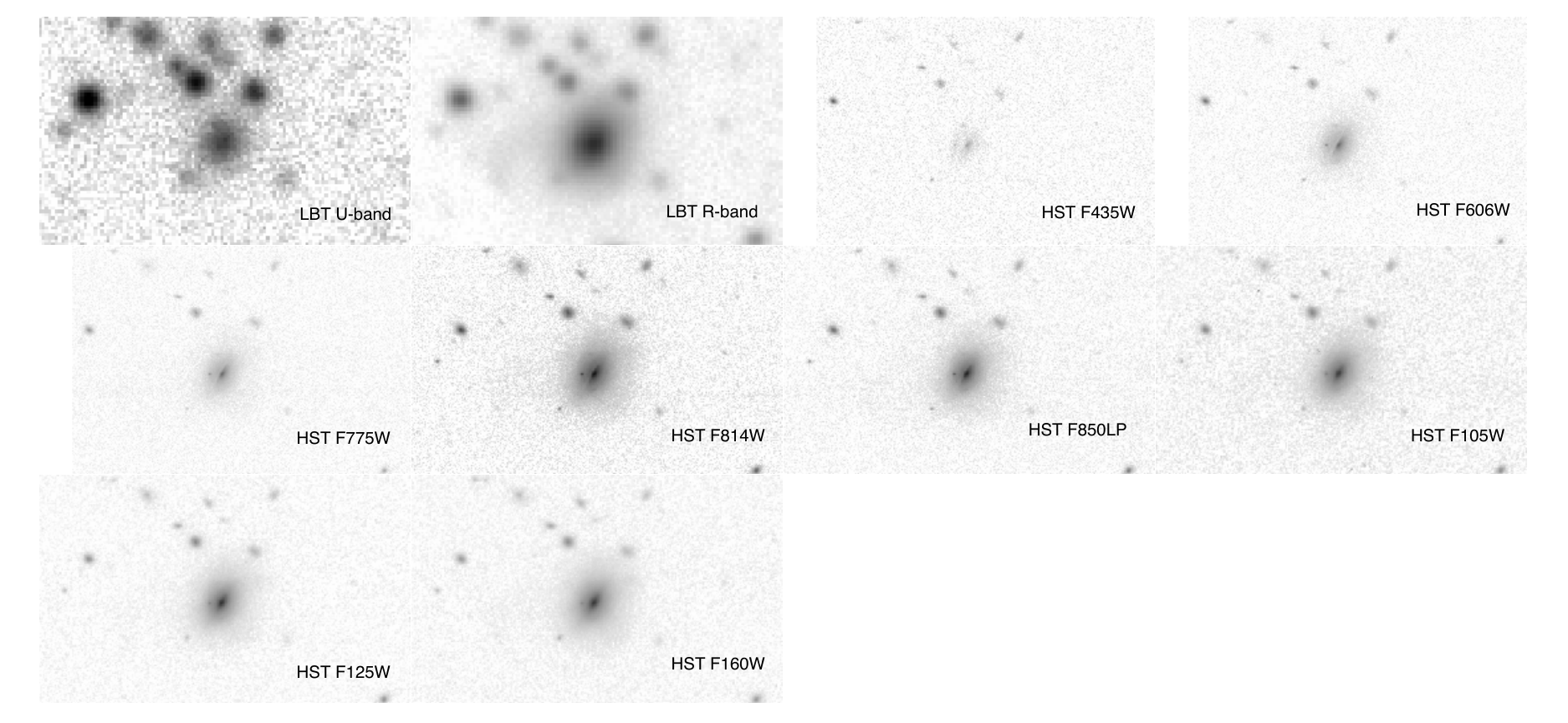}
\caption{\small\noindent 
Galaxy 7 (top): Galaxy 7 is an edge-on disk galaxy with an AGN. It has extra
diffuse flux upward and to the left, which is seen in both LBT \rb{} and
$U$-band, and a possible equal size grazing companion. Across most of the
\hst{} images, the diffuse flux is visible. The main galaxy has
$z_{spec}$\,=\,0.637 \citep{Barger2008}. Galaxy 8 (bottom): There is smooth,
diffuse debris on the left side of the galaxy only seen in the LBT \rb{} mosaic
and is only marginally detectable in some of the redder \hst{} images. The
\hst{} images show a bright core with a possible disk, but no spiral structure.
It has $z_{spec}$\,=\,0.440 \citep{Wirth2004}.}
\label{figure:gal7}
\end{figure}



\item There is smooth, diffuse debris to the left of the central galaxy seen in
the LBT \rb{}, while the \Ub only shows background objects (bottom galaxy in
Fig.\,\ref{figure:gal7}). The galaxy has $z_{spec}$\,=\,0.440 \citep{Wirth2004}.
It has also been categorized as a S0/spheroid galaxy, which is confirmed
by the \hst{} images, as well as a high-excitation narrow-line radio galaxy
\citep{Wirth2004}. The redder \hst{} filters also show the diffuse flux in
the same location as the LBT $r$-band.


\item Appearing as an elliptical galaxy, but
it clearly has rings with recent star forming clumps visible in the \hst{}
images (top galaxy in Fig.\,\ref{figure:gal9}). The small bright object to its
lower left has the same redshift $z_{spec}$\,=\,0.377 \citep{Wirth2004}. It
appears to be a smaller galaxy being tidally disrupted and absorbed by the
larger galaxy. In the LBT $r$-band, there appears to be even more diffuse tidal
arms that are not easily visible in the \hst{} imaging. Auxiliary data includes
a detection in the radio by the VLA \citep{Morrison2010}. The bright galaxy to the
upper left is a foreground interloper.

 
\begin{figure}
\includegraphics[width=\columnwidth]{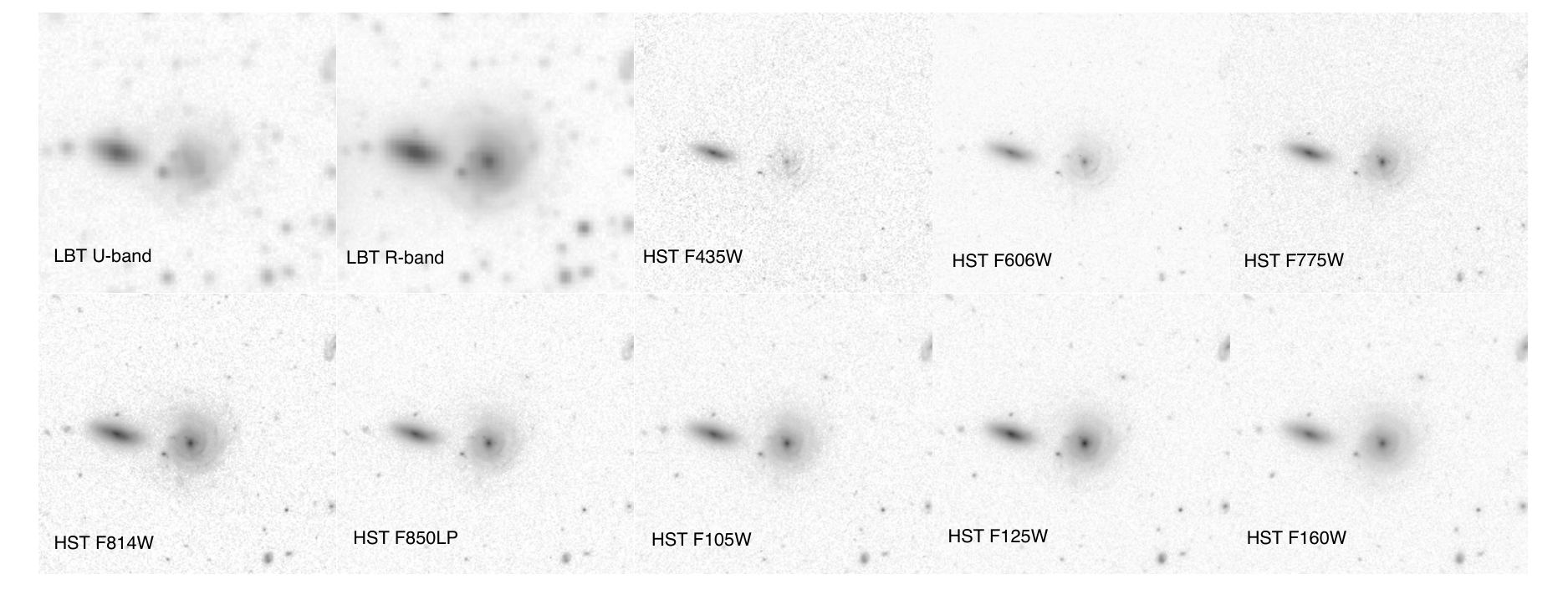}
\noindent\rule{\textwidth}{1pt}
\includegraphics[width=\columnwidth]{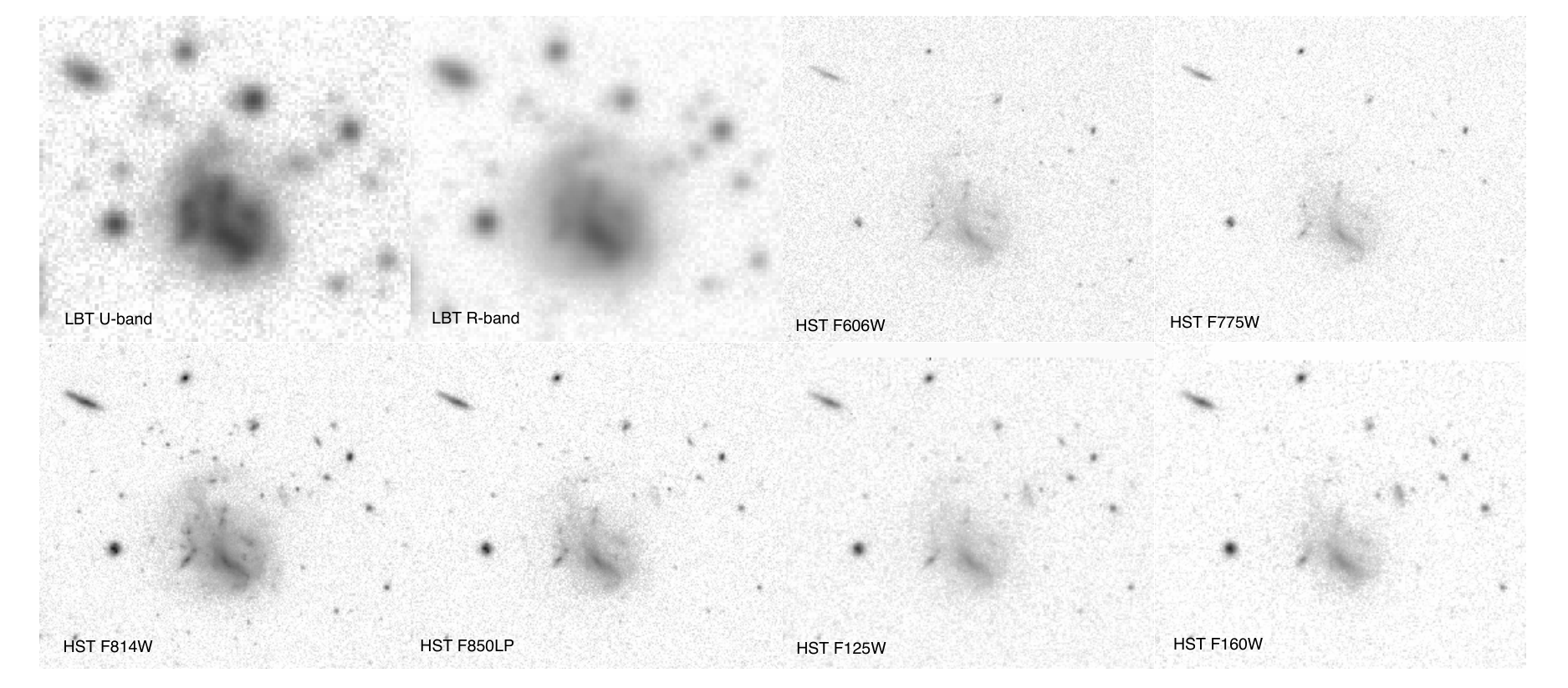}
\caption{\small\noindent 
Galaxy 9 (top): A bright face-on spiral galaxy with a small companion galaxy in
the lower left of the image, which is being absorbed by the main galaxy. The
core of the companion galaxy is seen in all the \hst{} images, while the
brighter galaxy to the upper left is a foreground interloper. Both the main
spiral and companion galaxy have confirmed spectroscopic redshifts of
$z_{spec}$\,=\,0.377 \citep{Wirth2004}. Galaxy 10 (bottom): A clearly merging
system, with signs of remnants of at least two distinct galaxies. There is one
dominant arm curving from the top of the galaxy downward to the right. Several
clumps of possible recent star formation are visible within the tidal arm.}
\label{figure:gal9}
\end{figure}



\item At least two merging galaxies with the central regions of each galaxy
still visible, but distorted (bottom galaxy in Fig.\,\ref{figure:gal9}). From
the LBT observations, there appears to be a tail coming from top and curving
around to the right. From the \citet{Elmegreen2007} designations, this galaxy
could be categorized as a shrimp type based on its shape and features visible
in the different imaging. There is no available redshift for this galaxy.


\item This system appears to have 3 elliptical galaxies with a long diffuse
stream toward the upper left corner linking to a possible 4th galaxy in the
system (top system of galaxies in Fig.\,\ref{figure:gal11}). With redshifts of
$z\,=$\,0.798 \citep{Treu2005}, these red galaxies are almost not detectable
in the LBT \Uz-band. The redder \hst{} filters show the same diffuse stream
visible in the LBT $r$-band at low surface brightness levels.


\begin{figure}
\includegraphics[width=\columnwidth]{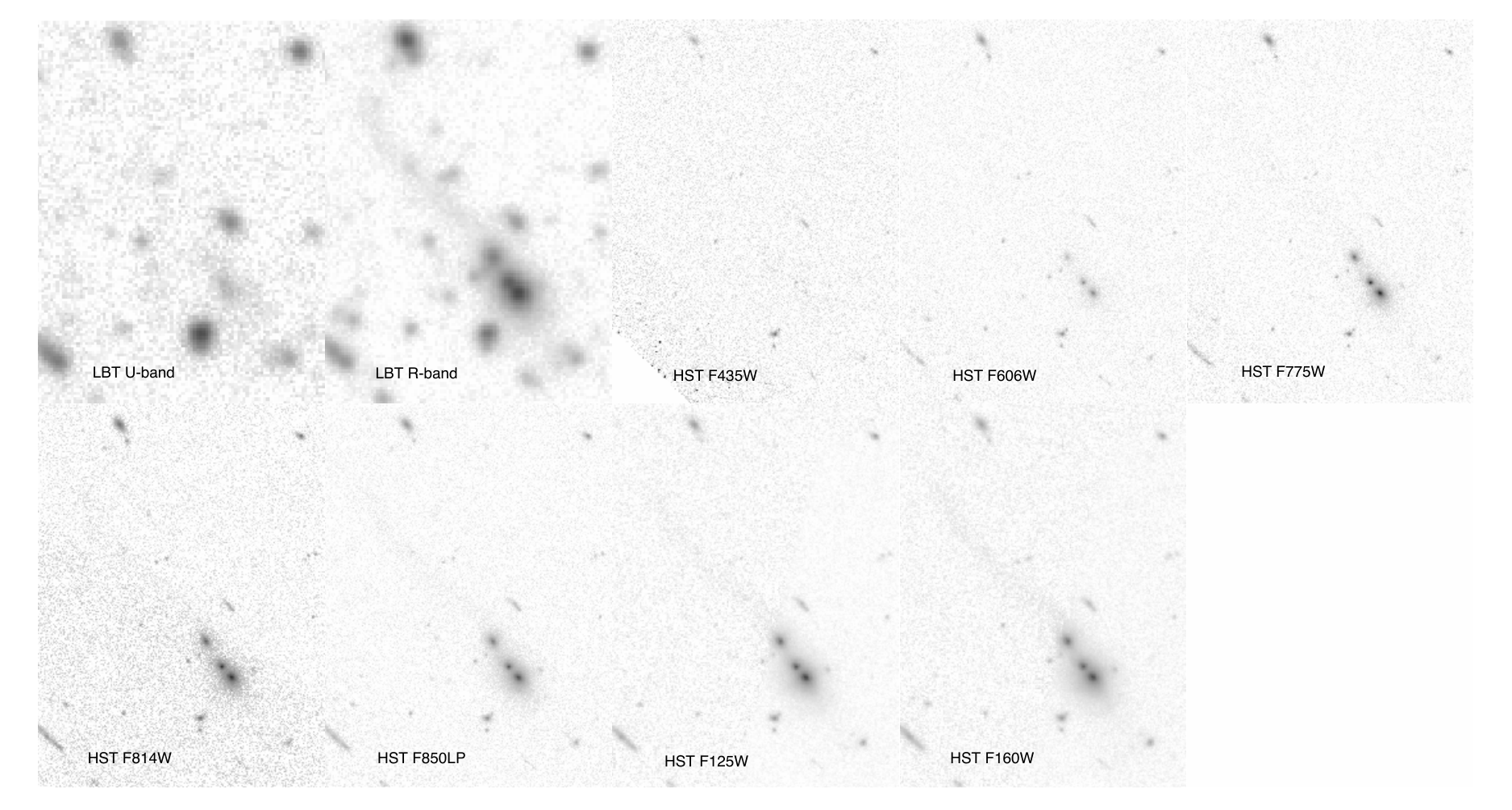}
\noindent\rule{\textwidth}{1pt}
\includegraphics[width=\columnwidth]{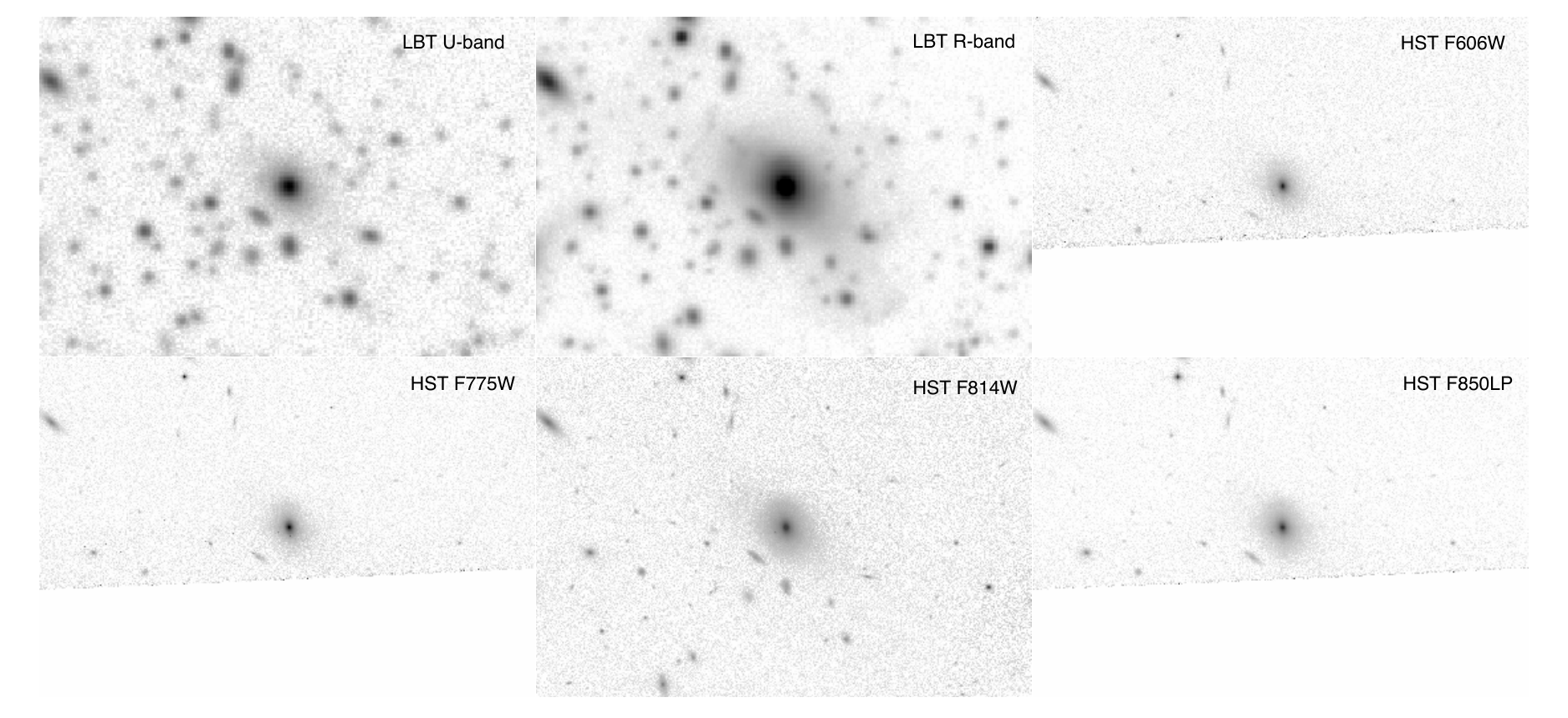}
\caption{\small\noindent 
Galaxy 11 (top): Three distinct elliptical galaxies are visible in the \hst{}
imaging, along with a diffuse stream linking to a possible 4th galaxy. The
tidal stream is only detectable in the LBT \rb{}, as well as the redder \hst{}
filters. For the bluest filters with data (LBT \Ub and \hst{} $b$-band), there
is only a very limited amount of flux detected even for the main galaxies. This
system has a redshift of $z\,=$\,0.798 \citep{Treu2005}. Galaxy 12 (bottom):
Another elliptical galaxy; however, there is only one galaxy with a diffuse
plume surrounding it and no evidence of a tidal stream of flux. The most
prominent area of diffuse flux is outside the \hst{} FOV, except for the I-band
(F814W). (The HST F814W image also shows a filter ghost).}
\label{figure:gal11}
\end{figure}



\item An elliptical galaxy surrounded by diffuse plumes with the most prominent
plume to the lower left (bottom galaxy in Fig.\,\ref{figure:gal11}). It is on
the edge of \hst{} FOV, and does not have any redshift information. Only the
\hst{} I-band (F814) image shows the entire region around the galaxy and there
does appear to be diffuse flux around the galaxy in this filter. However, it
would be easy to miss without re-binning or smoothing the \hst{} data.


\item This is a good example of where \hst{} imaging was critical to confirm
that it is an interacting system (Fig.\,\ref{figure:gal13}). Also, it is the
highest redshift galaxy in our sample with $z_{spec}$=\,0.937 \citep{Barger2008}.
The \hst{} image shows a second possible distinct core at the top of
the galaxy, or a starburst clump. The core near the center is not visible in
the bluer filters, which indicates a dusty galaxy. There are multiple clumpy
objects around the main galaxy, which could be associated with the main merging
system. \\

The following objects are completely outside the \hst{} FOV. In
Figs.\,\ref{figure:gal14} -- \ref{figure:gal29} only the optimal depth LBT
images in the \Ub on the left and the \rb{} on the right are shown. Some of the
objects are in the footprint for other surveys including Herschel,
\emph{Spitzer}, VLA, and Chandra.


\begin{figure*}
\centering
\includegraphics[width=0.96\txw]{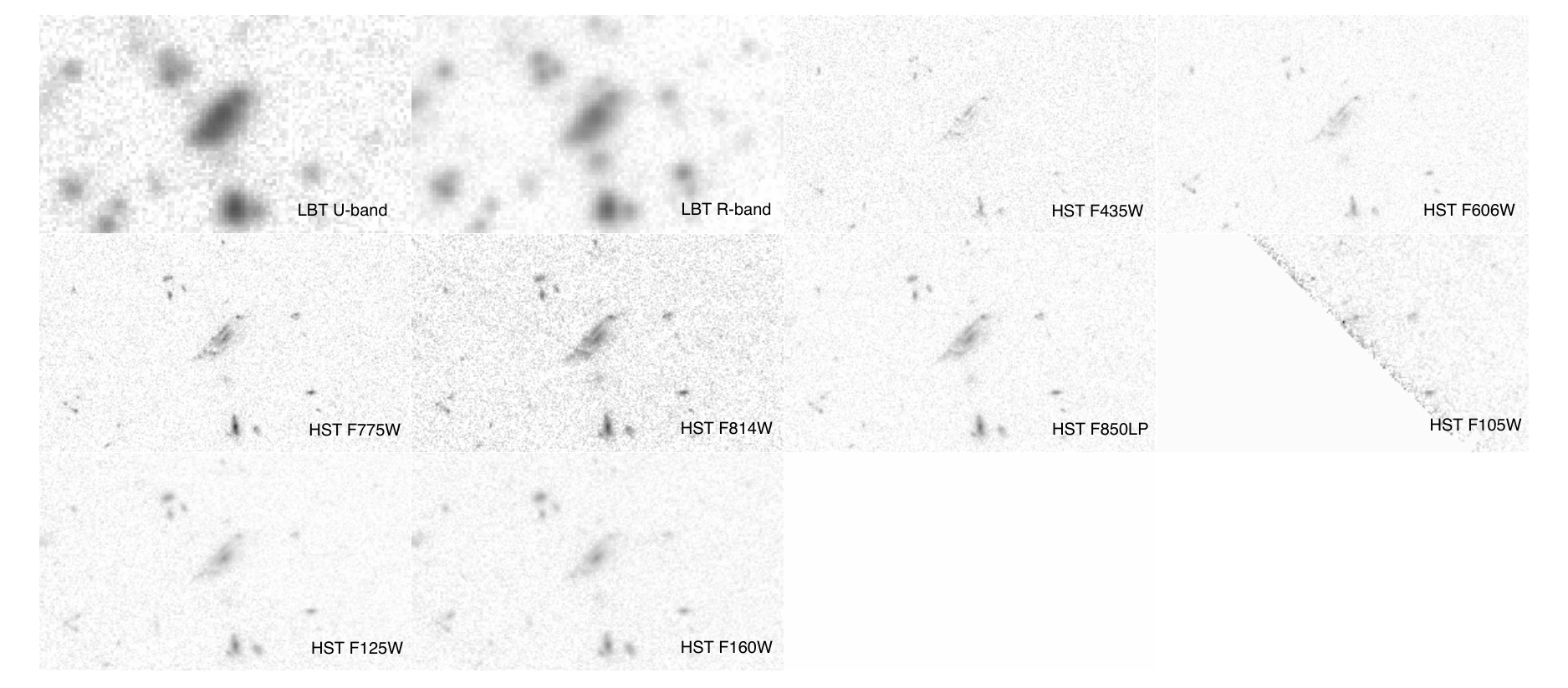}
\caption{\small\noindent Galaxy 13: 
An irregular galaxy with signatures of being in a merger, including a possible
second core, or a recent star-forming clump. It is the highest redshift galaxy
in our sample and has $z_{spec}=$\,0.937 \citep{Barger2008}, which is why
\hst{} imaging was key in confirming that it has morphology of a recent
interaction.}
\label{figure:gal13}
\vspace{0.9em}
\end{figure*}



\item A galaxy with long tidal streams reminiscent of the Antennae galaxy, but
the central region is unresolved (top galaxy in Fig.\,\ref{figure:gal14}). The
diffuse tidal streams are not seen seen in the \Uz-band, except for a few faint
spots along the tails. At a redshift of $z_{spec}$\,=\, 0.277 \citep{Casey2012}, 
the \Ub corresponds to NUV flux, which means there is little evidence of
recent star formation in these tidal streams. The galaxy was detected by both
Herschel (infrared; \citet{Casey2012}) and VLA (radio; \citet{Biggs2006}).


\begin{figure}
\includegraphics[width=\columnwidth]{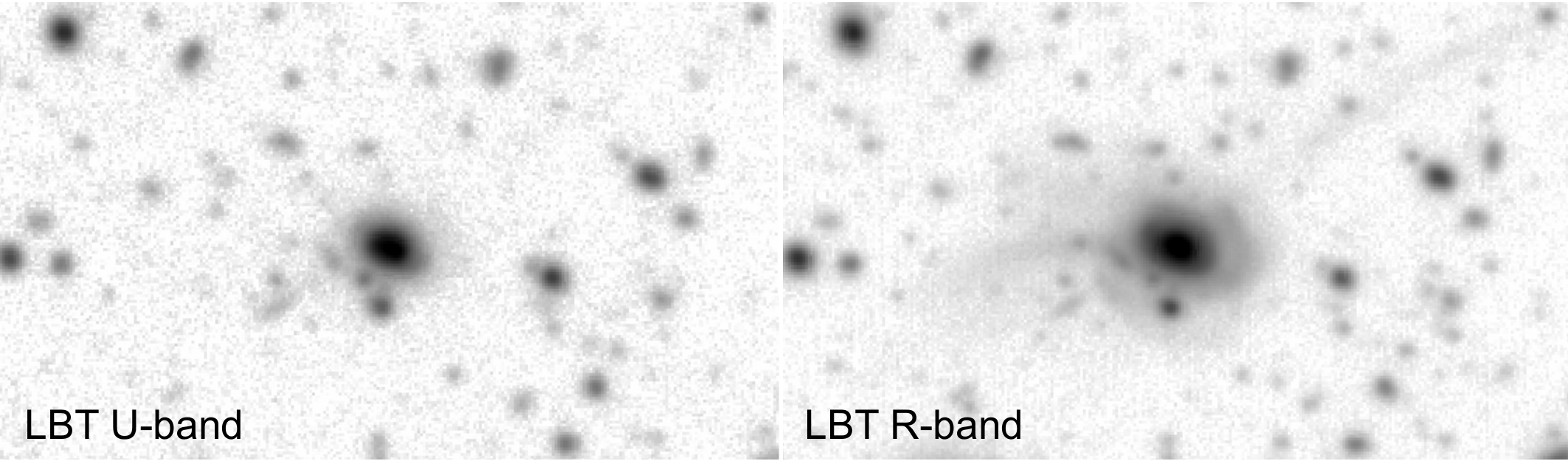}
\noindent\rule{\textwidth}{1pt}
\includegraphics[width=\columnwidth]{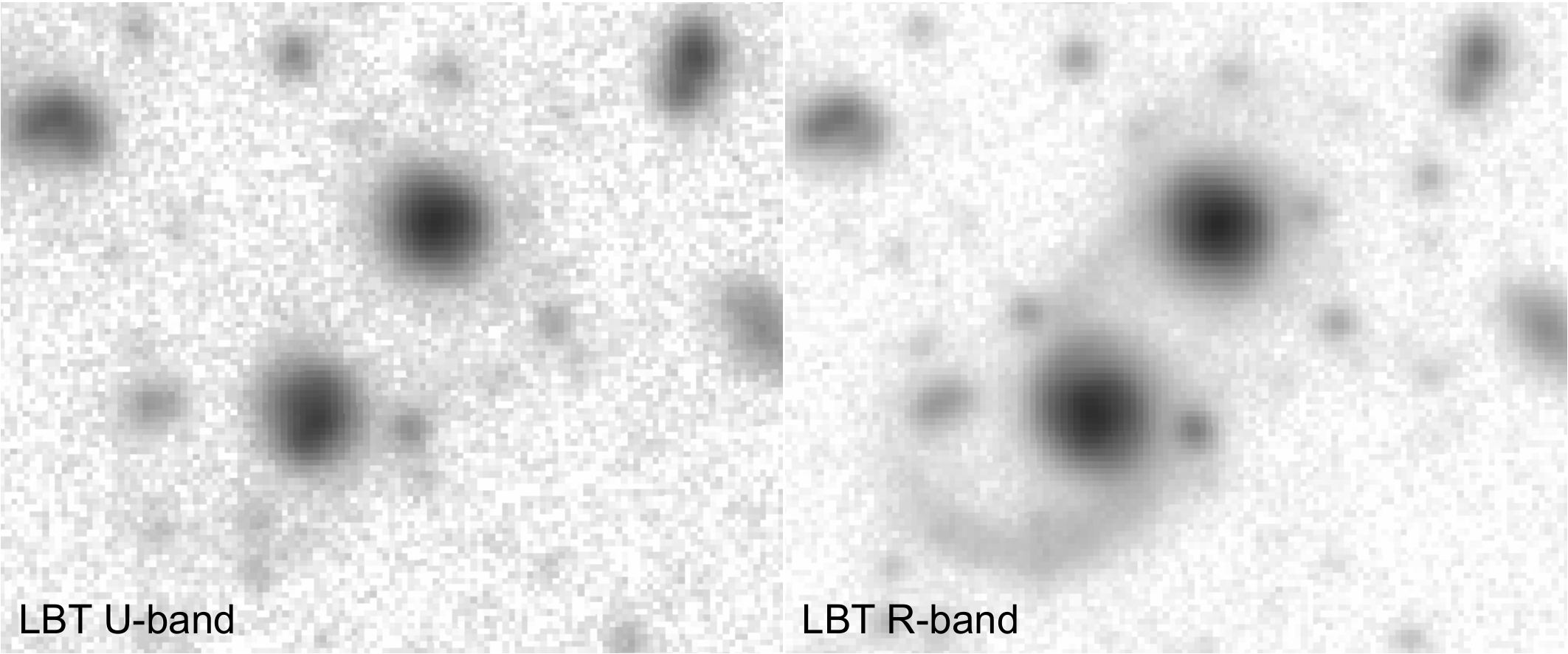}
\noindent\rule{\textwidth}{1pt}
\includegraphics[width=\columnwidth]{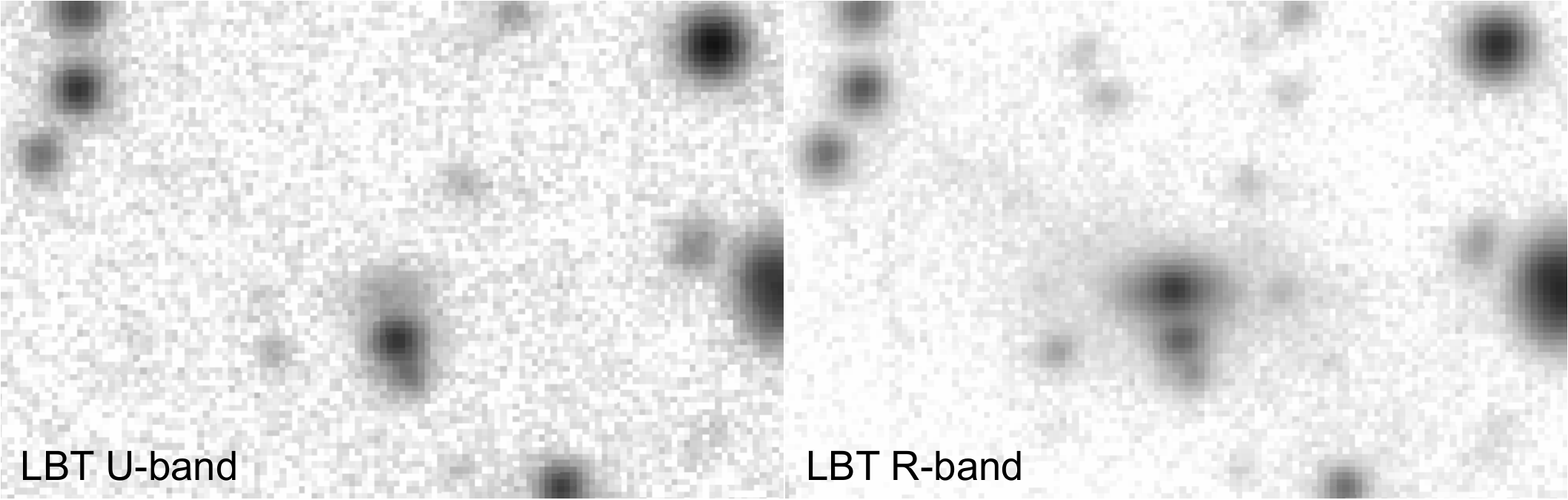}
\caption{\small\noindent 
Galaxy 14 (top): A merging system with a long extended tidal tail to the right
and a smaller tail with diffuse flux to the left. It is only significantly
detected in the LBT \rb{} and has $z_{spec}$\,=\,0.277 \citep{Casey2012}.
Galaxy 15 (middle): Two galaxies of equal size with multiple extended streams,
including a bridge between the galaxies, and a redshift of $z_{spec}$\,=\,0.456
\citep{Casey2012} is measured for the top galaxy. The diffuse light in the
streams is not easily visible in the \Uz-band, but there appears to be
potential star forming clumps. Galaxy 16 (bottom): A red galaxy with diffuse
debris to the upper left only detected in \rb{} and a possible redshift of
$z_{phot}$\,=\,0.48 \citep{Rafferty2011}.}
\label{figure:gal14}
\end{figure}



\item Two galaxies with extended flux in tidal tails seen in the LBT \rb{}, but
only the brightest clumps are in the tails are detectable in the \Uz-band
(middle system of galaxies in Fig.\,\ref{figure:gal14}). The system was
detected in the infrared by Herschel and \emph{Spitzer}, and the top galaxy is
associated with a radio source (VLA; \citet{Morrison2010}). The bottom large galaxy is interacting
with the smaller galaxy to the left, while the top galaxy has
$z_{spec}$=\,0.456 \citep{Casey2012}.


\item A very red galaxy, probably elliptical/S0 type galaxy, which is only
barley visible in the LBT $U$-band. This galaxy is surrounded by diffuse plumes
and has a stream to the upper left (bottom galaxy in Fig.\,\ref{figure:gal14}).
From the literature, it has a photometric redshift of $z_{phot}$=\,0.48 \citep{Rafferty2011}.


\item Two large irregular galaxies just outside the \hst{} FOV (top galaxy
system in Fig.\,\ref{figure:gal17}). There are multiple star forming clumps
around the two galaxies, which are most likely physically associated based on
their colors. The larger galaxy on the right side appears to have a M51 like
bridge with the small galaxy to it's right. They are at a low redshift of
$z_{spec}$\,=\,0.0366 \citep{Wirth2004}.


\begin{figure}
\includegraphics[width=\columnwidth]{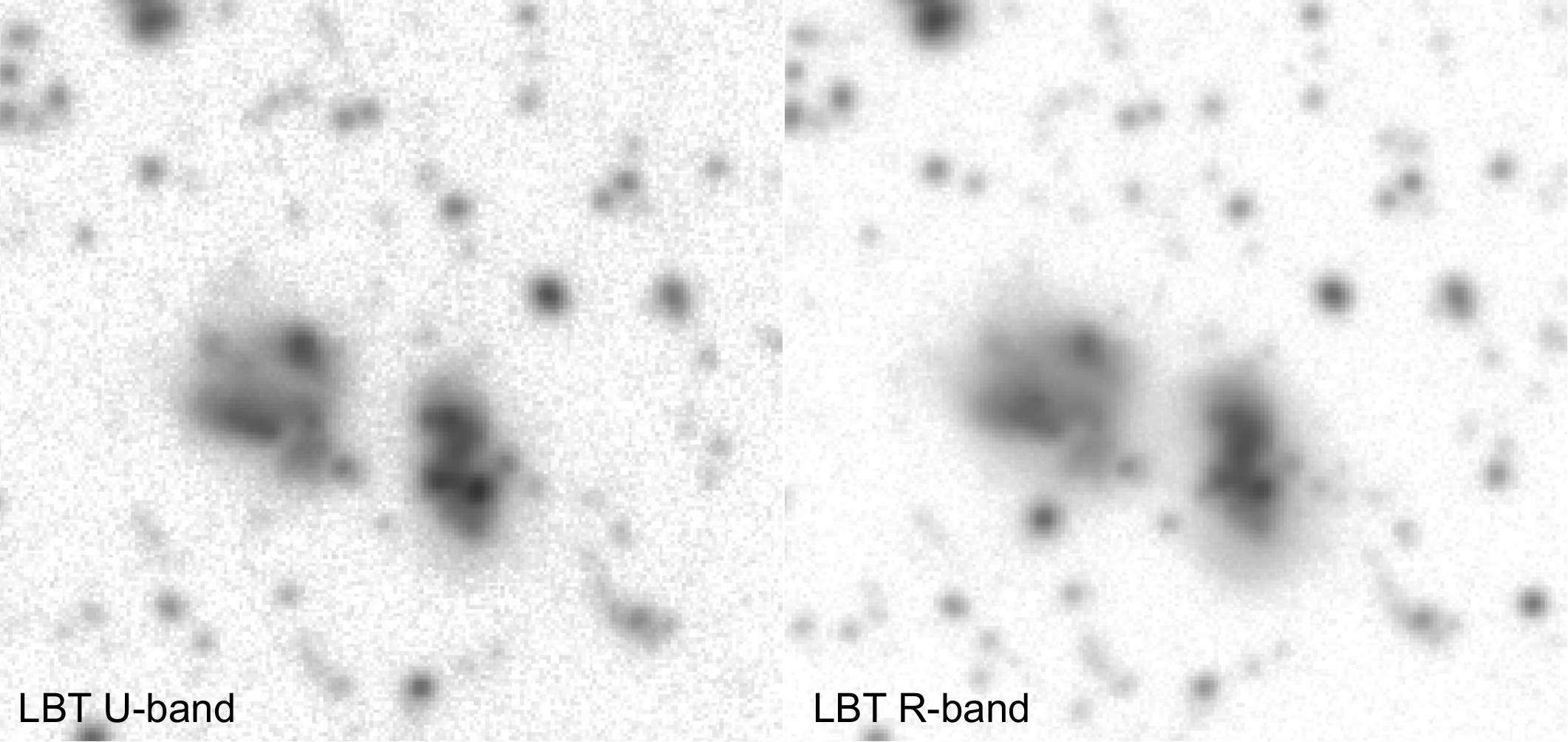}
\noindent\rule{\textwidth}{1pt}
\includegraphics[width=\columnwidth]{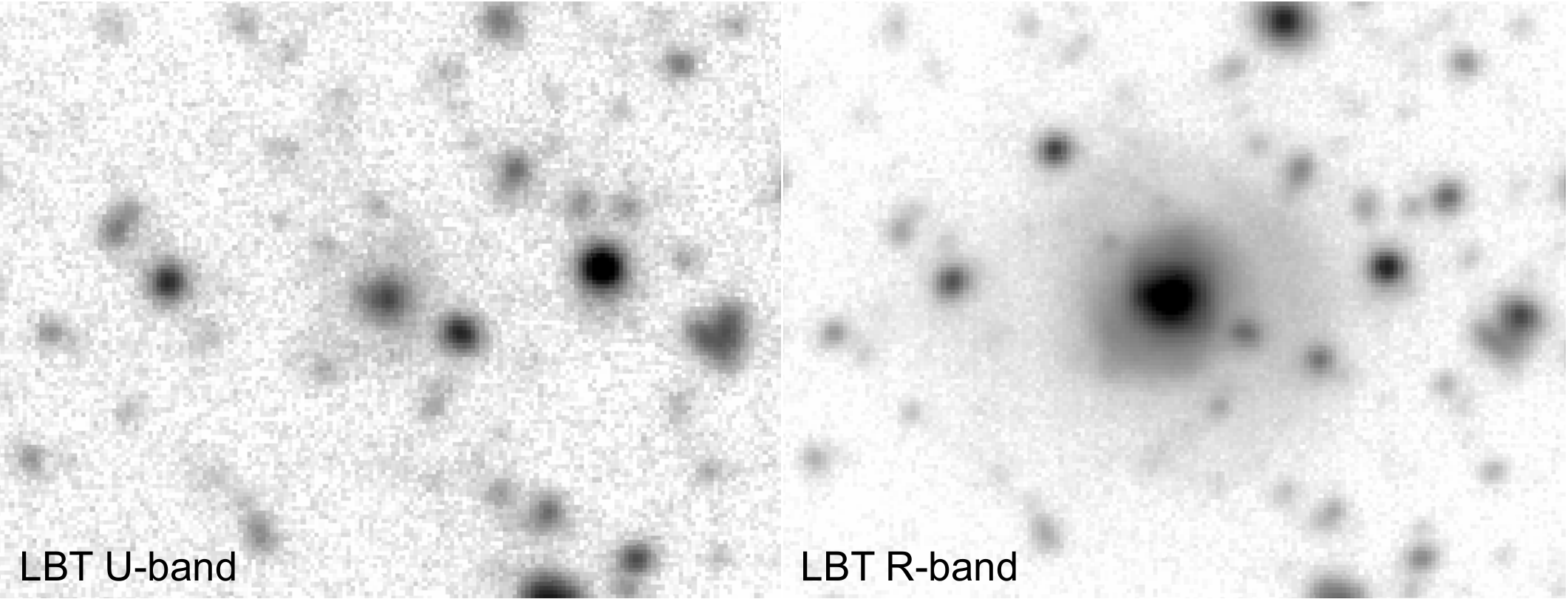}
\noindent\rule{\textwidth}{1pt}
\includegraphics[width=\columnwidth]{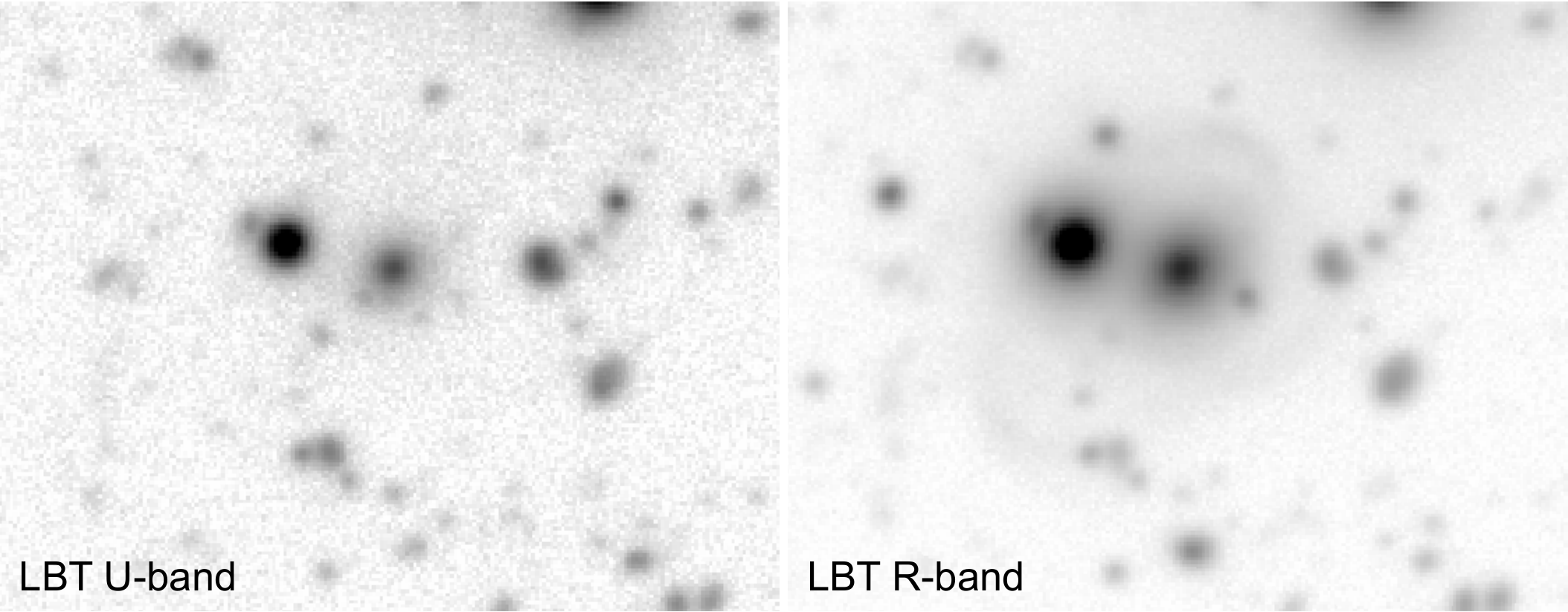}
\caption{\small\noindent 
Galaxy 17 (top): Two large distorted galaxies with antenna like main regions
and $z_{spec}$\,=\,0.0336 \citep{Wirth2004}. Both look similar in the \rb{}
and \Uz-band, and the galaxy on the right is also interacting with a smaller
galaxy to the lower right. Galaxy 18 (middle): A large diffuse shell of flux is
seen only in the \rb{} and as a $z_{spec}$\,=\,0.560 \citep{Albareti2017}.
Galaxy 19 (bottom): A galaxy with long extended diffuse plumes and loops around
the galaxy, distinctly visible in $r$-band, with one stream of flux partially
detectable in \Uz-band.}
\label{figure:gal17}
\end{figure}



\item In the \rb{} there is a large extended shell of diffuse light around the
galaxy (middle galaxy in Fig.\,\ref{figure:gal17}), which is not seen in the
$U$-band. This galaxy has $z_{spec}$\,=\,0.560 \citep{Albareti2017}. The
diffuse shell could be a signature of a recent merger.


\item A galaxy with long extended diffuse plumes and loops around the galaxy
that is only seen in the LBT \rb{} (bottom galaxy in Fig.\,\ref{figure:gal17}).
There is one tidal stream in the bottom left, which has detectable flux in the
corresponding \Ub area.


\item Another galaxy with rings of diffuse light surrounding it (top galaxy in
Fig.\,\ref{figure:gal20}). The outer most ring is only seen in the LBT
$r$-band, and could be a sign of a recent merger. There is no indication of
interaction with any other galaxies nearby it, and no available redshift
information. 


\begin{figure}
\includegraphics[width=\columnwidth]{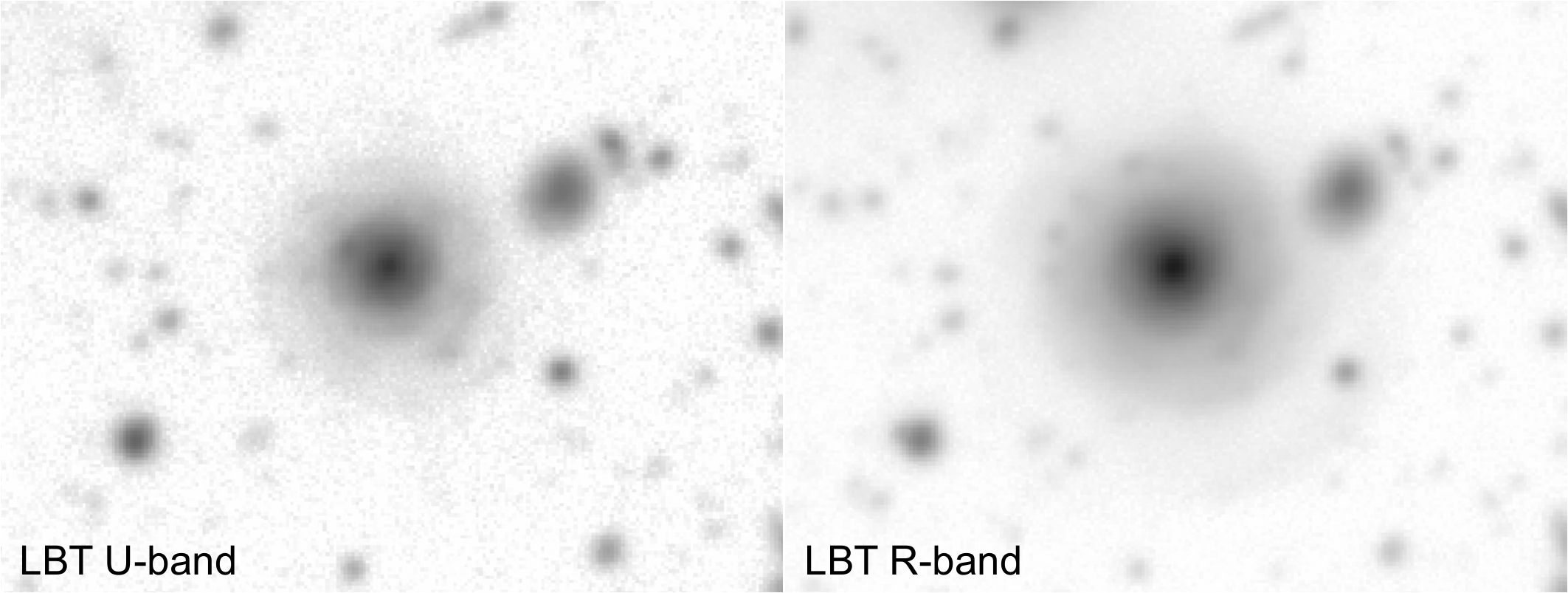}
\noindent\rule{\textwidth}{1pt}
\includegraphics[width=\columnwidth]{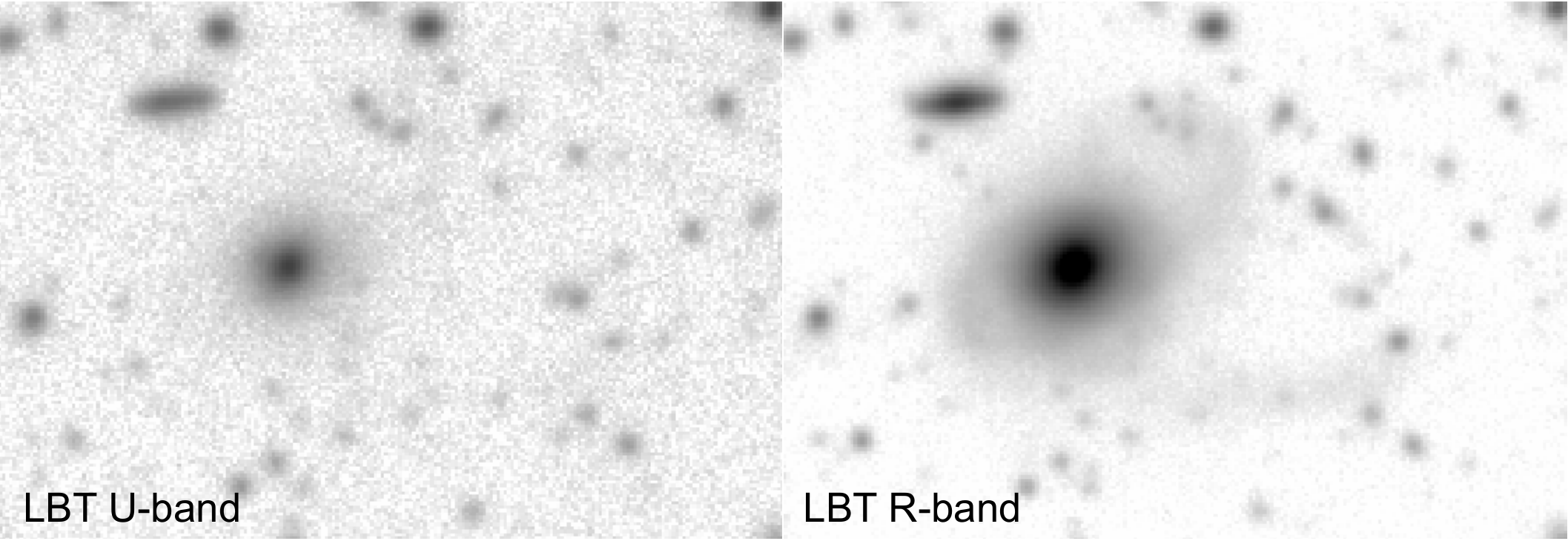}
\caption{\small\noindent 
Galaxy 20 (top): A galaxy with rings of diffuse light around it, with the outer
most ring only noticeable in the LBT $r$-band. Galaxy 21 (bottom): An
elliptical galaxy with one stream interacting with a much smaller nearby
galaxy, and a large diffuse plume above the galaxy. Only potential star-forming
clumps along the tidal stream are visible in the \Uz-band. }
\label{figure:gal20}
\end{figure}



\item An elliptical galaxy with one stream interacting with a much smaller
nearby galaxy, and a large diffuse plume above the galaxy (middle galaxy in
Fig.\,\ref{figure:gal20}). This extended diffuse light is not detected in the
\Uz-band, except for discreet clumps along the tidal arm, which are most likely
associated with recent star-formation.


\begin{figure}
\includegraphics[width=\columnwidth]{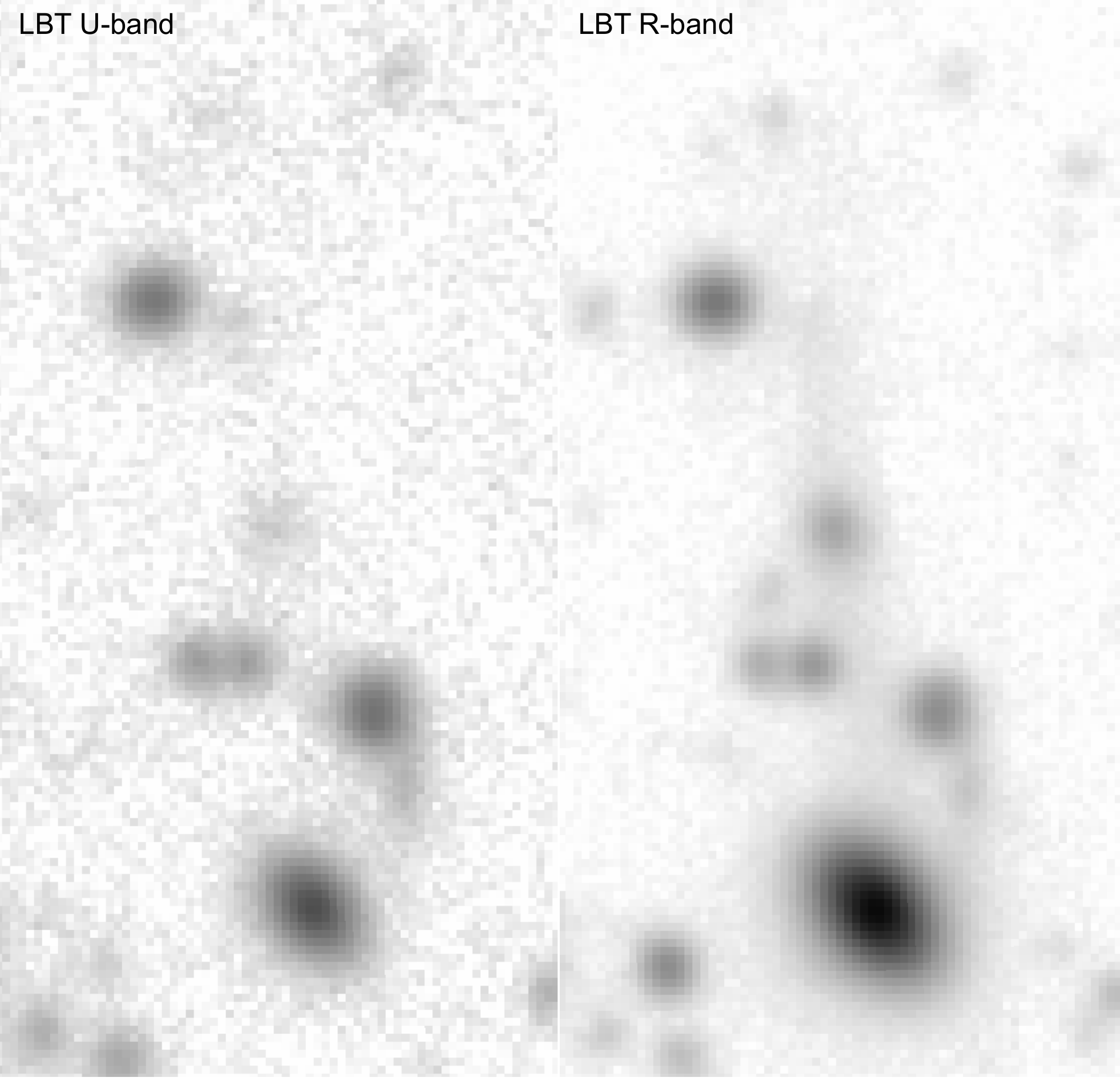}
\caption{\small\noindent 
Galaxy 22: In the $r$-band, there is a diffuse plume surrounding the central galaxy, along
with a couple of bubble like features at the bottom of the galaxy toward
another similar size galaxy. Only the main bubble feature is detected in the
\Uz-band.}
\label{figure:gal22}
\end{figure}



\item Diffuse plumes surround the galaxy in the center of the image, as well as
a couple of bubbles, or loop features at the bottom (bottom galaxy in
Fig.\,\ref{figure:gal22}). In the \Uz-band, the majority of the diffuse flux is
not detected with only the brighter bubble feature visible. For galaxies within
this field, there is no redshift information. The main galaxy could be
interacting with multiple galaxies around it, including the similar size
elliptical below it, and a galaxy farther to the right following one of the
tidal streams.


\item An interesting system with potentially multiple galaxies interacting (top
system of galaxies in Fig.\,\ref{figure:gal22}). There is a larger elliptical
at the of bottom of the image with a bridge to a galaxy above it, and a tidal
stream extending even farther above it. Without redshift information, it is unknown if the various objects along the tidal tail are gravitationally associated with the interaction.


\begin{figure}
\includegraphics[width=0.97\columnwidth]{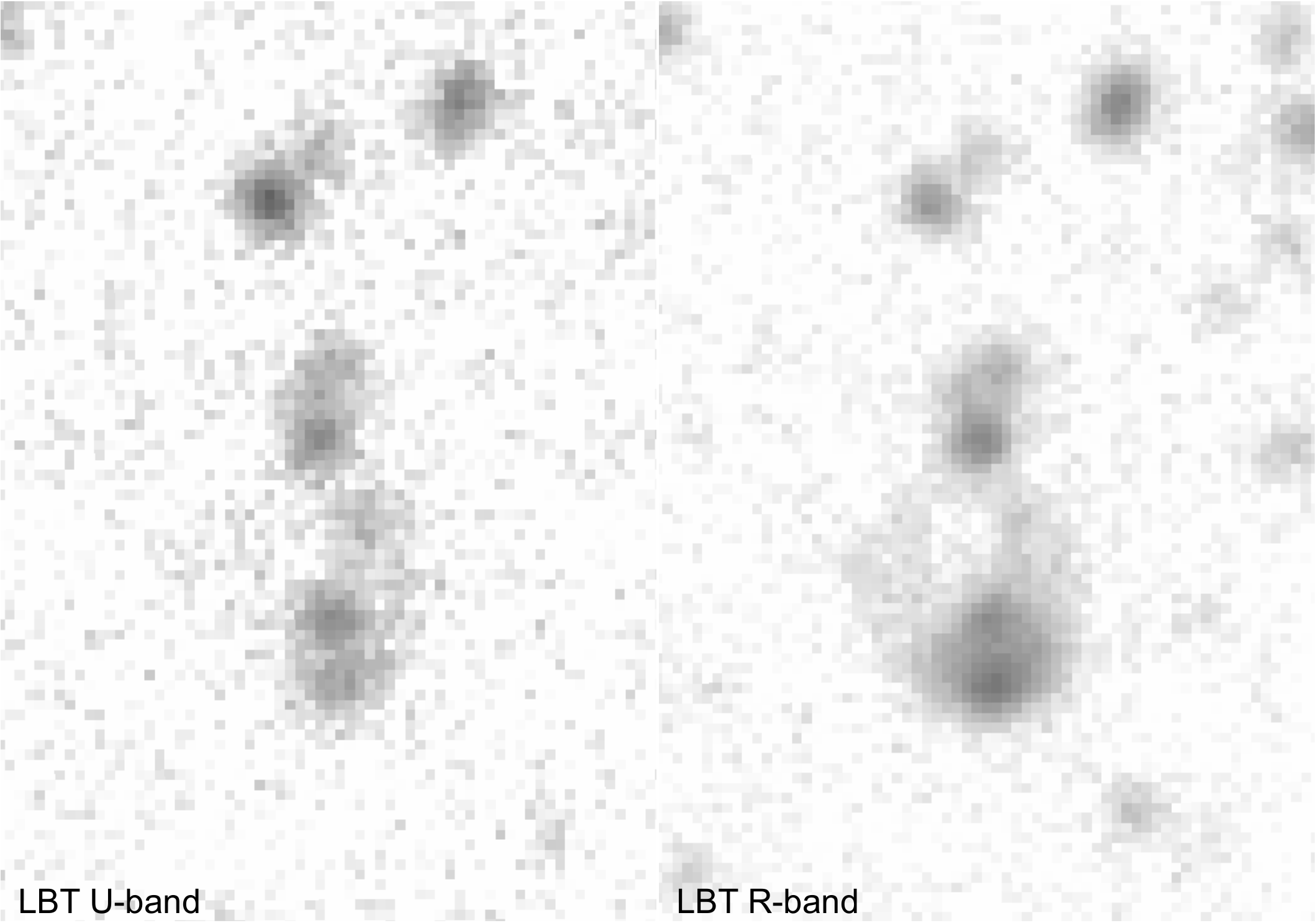}
\noindent\rule{\textwidth}{1pt}
\includegraphics[width=0.97\columnwidth]{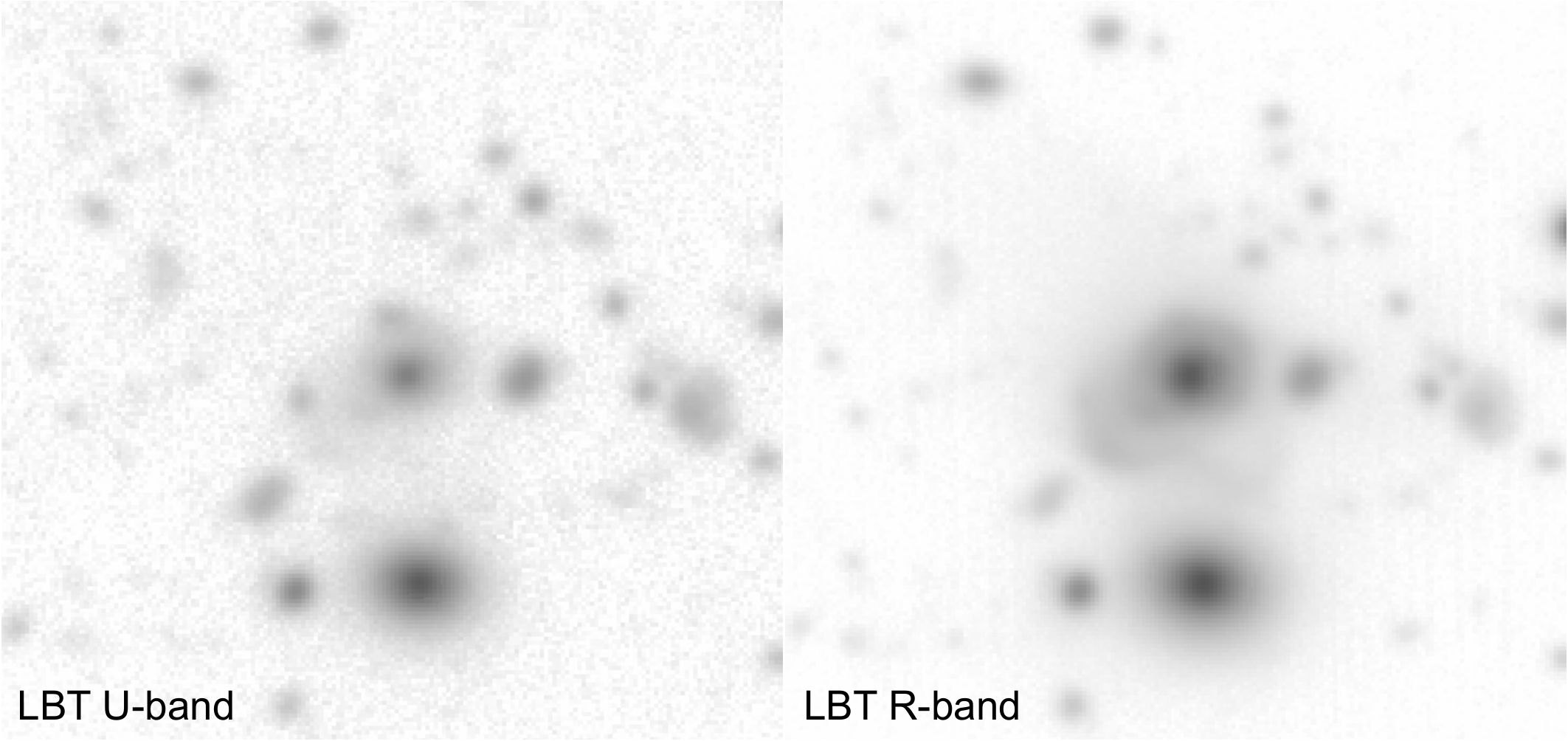}
\caption{\small\noindent
Galaxy 23 (top): A long diffuse stream is running from the bright galaxy at the
bottom to the top of the image, which is not as clearly defined in the
\Uz-band. There is a bridge visible in both the \rb{} and the \Ub linking the
main galaxy to a slightly smaller one. Galaxy 24 (bottom): A galaxy with
extended flux along the top and right portion seen in LBT \rb{} and partly
visible in the \Uz-band. It appears to be a face-on disk with multiple
star-forming clumps and possibly interacting with galaxy above it.}
\label{figure:gal22}
\end{figure}



\item A face-on disk galaxy with multiple star-forming clumps and an extended
flux feature on the top and left side of galaxy (bottom galaxy in
Fig.\,\ref{figure:gal22}). While most noticeable in the $r$-band, there are a
few brighter spots visible in \Uz-band. Another galaxy without any redshift
information.


\item An assembly type with two diffuse bridges linking two brighter clump
cores (top galaxy in Fig.\,\ref{figure:gal23}). The left bridge connecting the
cores is much fainter in the \Ub compared to the rest of the components of the
system. There is no redshift information available for this system.


\begin{figure}
\includegraphics[width=\columnwidth]{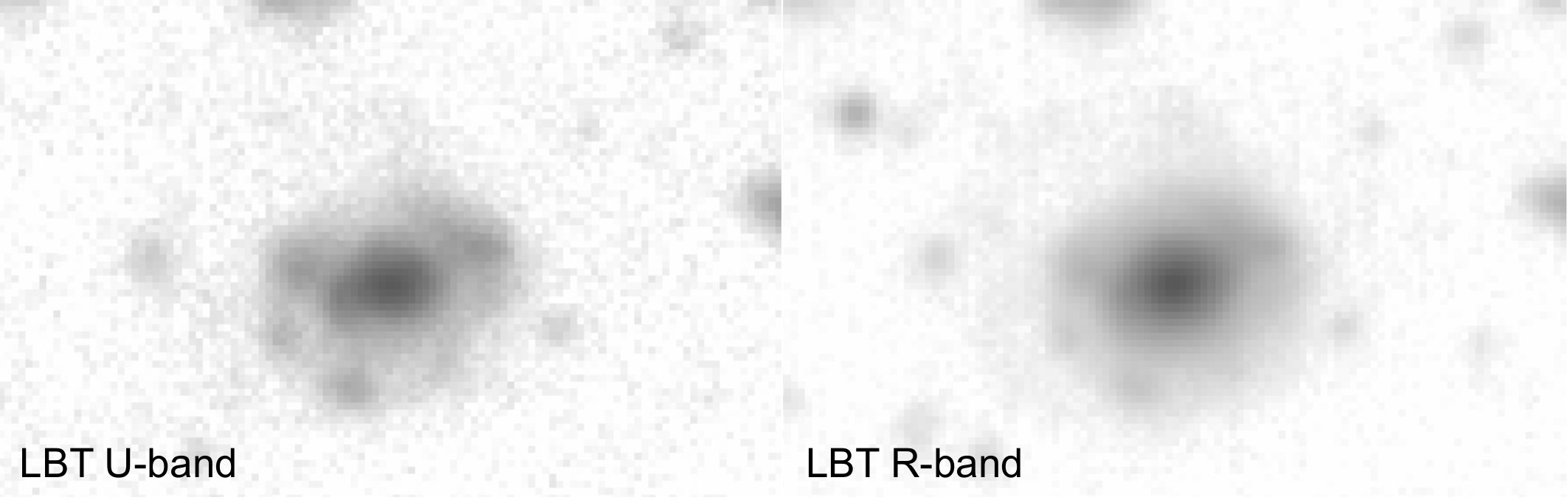}
\noindent\rule{\textwidth}{1pt}
\includegraphics[width=\columnwidth]{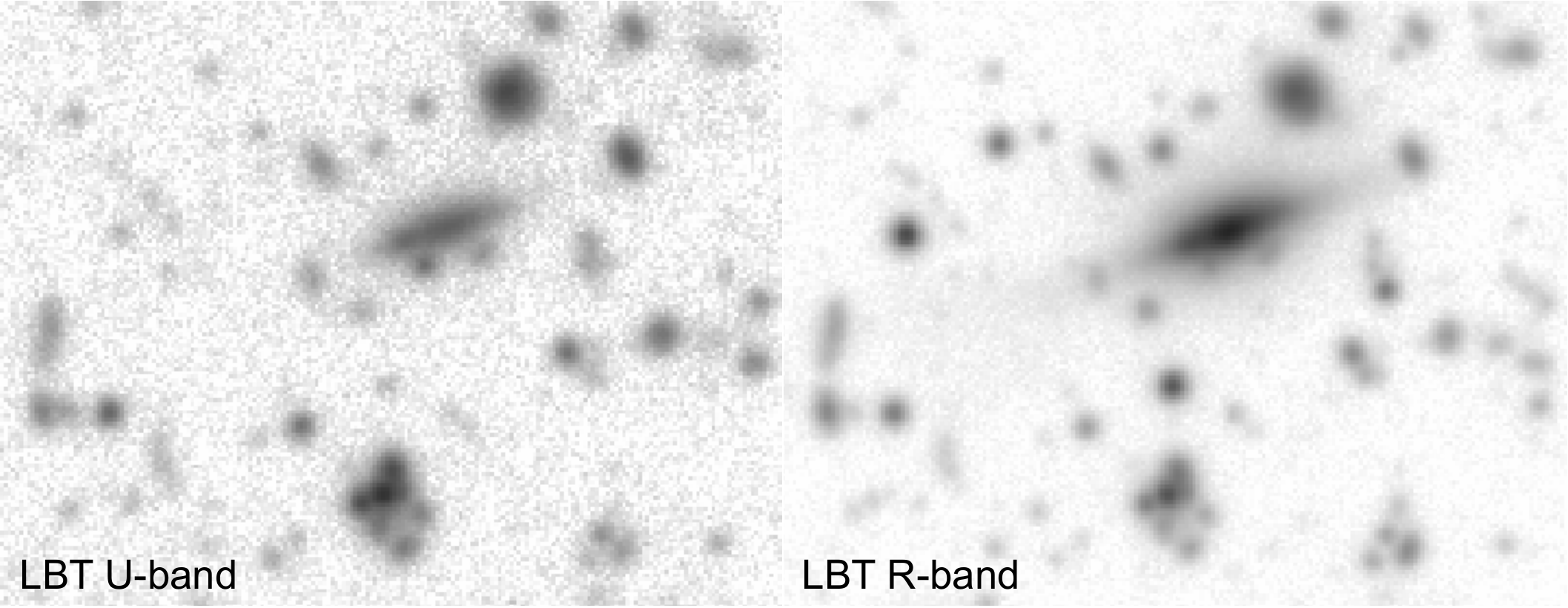}
\caption{\small\noindent 
Galaxy 25 (top): An assembly type system with two diffuse bridges linking two
brighter clump cores. All except the bridge on the left connecting the two
cores are detected in both \rb{} and \Uz-band. Galaxy 26 (bottom): An edge-on
spiral galaxy with extended flux toward the left seen in the \rb{}, but not in
the \Uz-band. At the bottom of the image, there is a galaxy consisting of
several clumps.}
\label{figure:gal23}
\end{figure}


\item An edge-on spiral galaxy with extended flux on the left side, which is
only detected in the $r$-band (bottom galaxy in Fig.\,\ref{figure:gal23}).
There are other galaxies in the image with irregular shapes, and signatures of
recent interactions or mergers, including a galaxy consisting of clumps at the
bottom of the image. There is no redshift information available for any of the
objects in the image.


\begin{figure}
\includegraphics[width=\columnwidth]{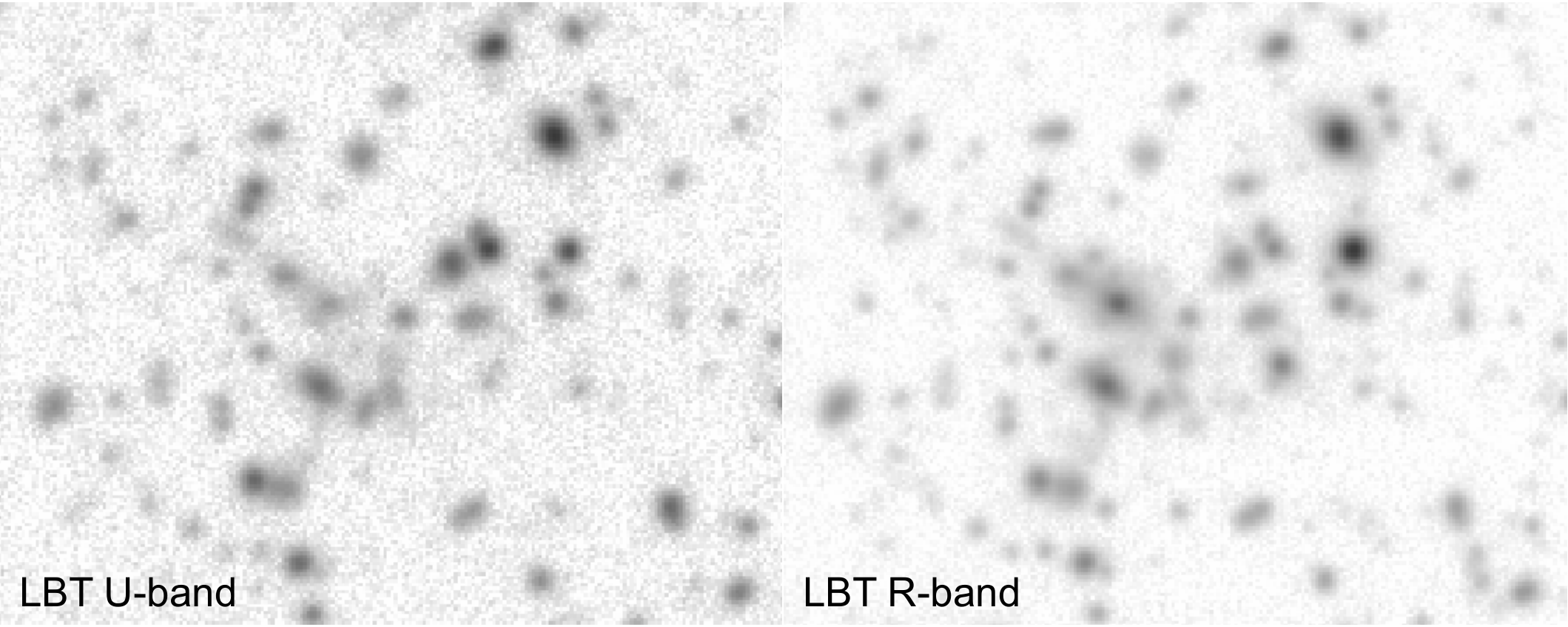}
\noindent\rule{\textwidth}{1pt}
\includegraphics[width=\columnwidth]{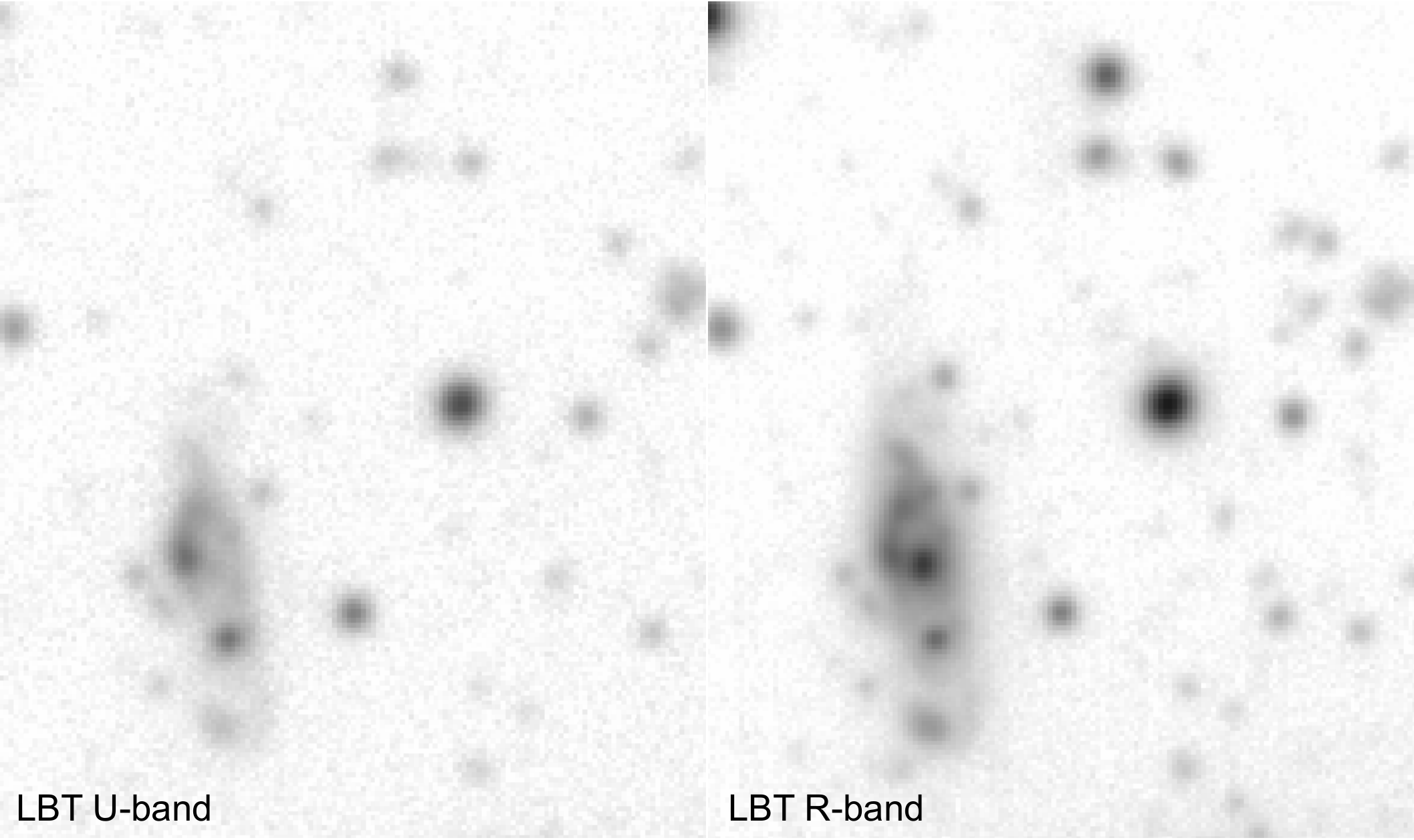}
\caption{\small\noindent 
Galaxy 27 (top): A dense area with multiple galaxies interacting and a clear
tidal stream between two galaxies near center. Most of the galaxies and
features within the field are detected in both \rb{} and \Uz-band. The QSO in
the system has a $z_{spec}$\,=\,0.334. Galaxy 28 (bottom): A system in the
process of merging with potential double nuclei and a diffuse stream toward the
top of the image. In the \Uz-band, the central nucleus is not visible. }
\label{figure:gal26}
\end{figure}



\item A dense area of galaxies with multiple possible interactions (top system
of galaxies in Fig.\,\ref{figure:gal26}). The most obvious one is a bridge
between two galaxies near the center and is visible in both the \rb{} and the
$U$-band. There is also a known QSO in the field with a $z_{spec}$\,=\,0.334
\citep{Albareti2017}. The remainder of the objects in the field do not have
known redshifts.


\item A potential double nuclei merging system with diffuse flux above the
galaxy in \rb{} (bottom galaxy in fig.\,\ref{figure:gal26}). While most of the
galaxy is detected in \Uz-band, remarkably, the bright core in the center not
visible in the \Uz-band. Visually there appears to be links to nearby objects,
but there is no redshift information to confirm.


\item An elliptical galaxy with diffuse plumes seen in the \rb, but in the
\Uz-band the plumes are less distinct (top galaxy in Fig.\,\ref{figure:gal29}).
There is no redshift information for this galaxy.


\begin{figure}
\includegraphics[width=\columnwidth]{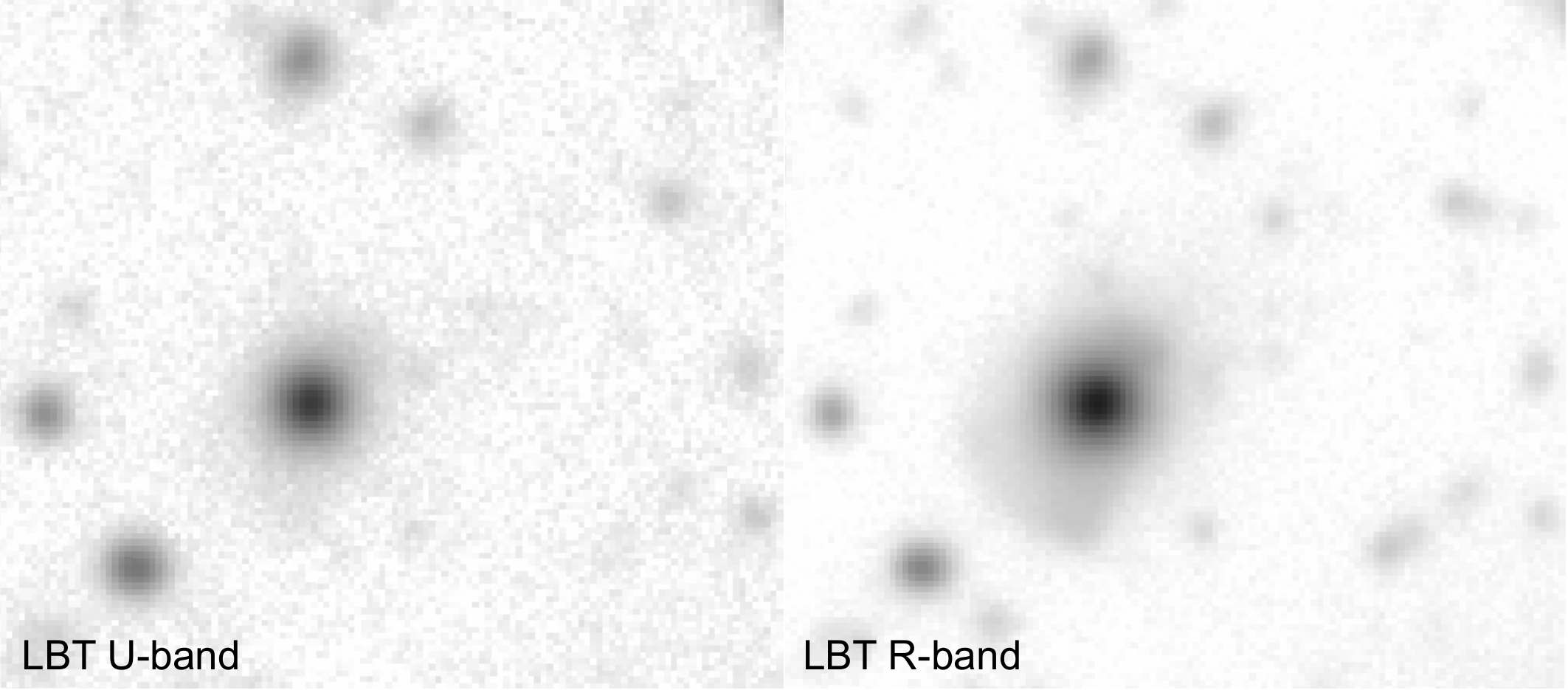}
\noindent\rule{\textwidth}{1pt}
\includegraphics[width=\columnwidth]{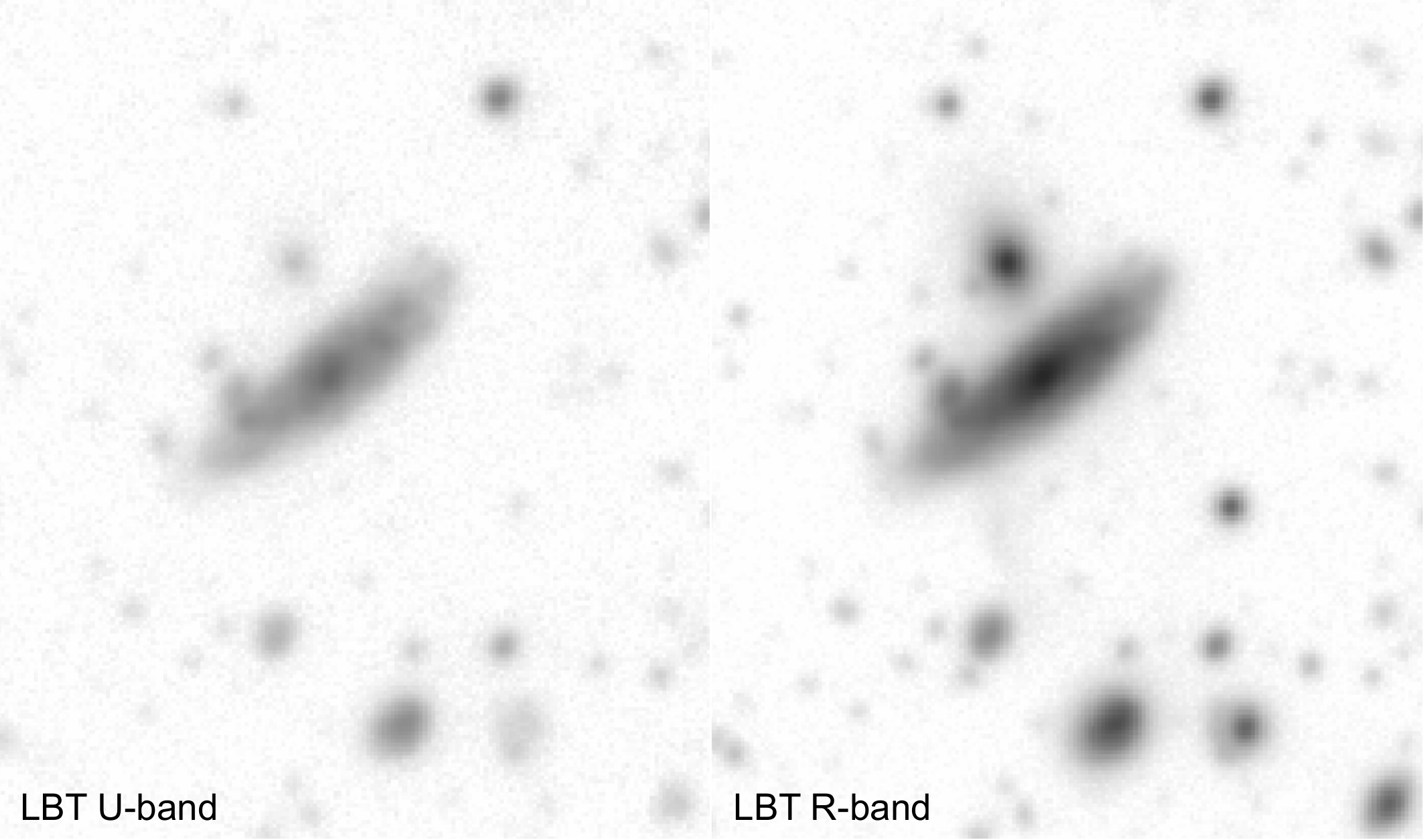}
\caption{\small\noindent 
Galaxy 29 (top): An interesting extended plume of diffuse light toward the
bottom left seen only in the $r$-band. Galaxy 30 (bottom): An angled spiral
galaxy, with a tidal bridge extending down toward the smaller galaxy below it,
which is not detected in the \Uz-band. }
\label{figure:gal29}
\end{figure}



\item An angled spiral galaxy with diffuse flux extending down toward the
smaller galaxy below it (bottom galaxy in Fig.\,\ref{figure:gal29}). The
possible diffuse bridge linking the two galaxies is only detected in the
$r$-band. However, without redshift information, it is not possible to know for
sure if these objects are truly interacting.

\end{enumerate}


\end{document}